\documentclass[11pt]{article}
\usepackage{fullpage}
\usepackage[utf8]{inputenc}
\pdfoutput=1

\usepackage[margin=1in]{geometry}

\usepackage{amssymb}
\usepackage{amsmath}
\usepackage{mathtools}
\usepackage{latexsym}
\usepackage{amsfonts}

\usepackage{amsthm}
\usepackage{thmtools, thm-restate}

\newtheorem{theorem}{Theorem}[section]

\newtheorem{lemma}[theorem]{Lemma}
\newtheorem{corollary}[theorem]{Corollary}
\theoremstyle{remark}

\theoremstyle{definition}

\newtheorem{example}[theorem]{Example}

\usepackage[numbers]{natbib}

\usepackage{xcolor} 
\usepackage{tikz} % for graph / drawing
\usetikzlibrary{arrows.meta,shapes,snakes}
\usepackage{cancel} % for crossover
\usepackage{enumitem} % to adjust spacing 
\usepackage{soul} 
\usepackage[labelfont=bf]{caption}
\usepackage{subcaption} % for subfigure
\usepackage{blkarray} % for latin matrix labels
\usepackage{ulem}   % for waving underline, but need
\usepackage[breaklinks=true]{hyperref} % clickable reference to lemmas...
\hypersetup{
	colorlinks=true,
	citecolor=green!80!black,
	linkcolor=blue,  
}
\usepackage{breakurl}

\usepackage{algorithm}
\usepackage{algorithmic}
\renewcommand{\algorithmiccomment}[1]{\bgroup\hfill//~#1\egroup}

\def\A{{\cal A}}
\def\B{{\cal B}}
\def\C{{\cal C}}

\def\L{{\cal L}}
\def\I{{\cal I}}
\def\N{{\mathbb N}}
\def\q{{\mathbf q}}
\def\M{{\cal M}}
\def\Rot{{\cal R}}
\def\SR{{\mathcal{SR}}}
\def\Z{{\mathbb Z}}

\def\SS{{\cal S}}

\let\bar\overline
\def\proof#1{{\textit {#1}}}
\def\endproof{}

\def\Halmos{\mbox{\quad$\square$}}

\author{Yuri Faenza \\ yf2414@columbia.edu \and Xuan Zhang \\ xz2569@columbia.edu}
\title{Legal Assignments and fast EADAM with consent \\ via classical theory of stable matchings}
\date{Department of Industrial Engineering and Operations Research,\\ Columbia University, New York, NY 10027
}

\begin{document}
%%%%%%%%%%%%%%%%

\maketitle
\begin{abstract} Gale and Shapley's \emph{stable assignment problem}~\citep{GS} has been extensively studied, applied, and extended. In the context of school choice, mechanisms often aim at finding an assignment that is more favorable to students. We investigate two extensions introduced in this framework  -- \emph{legal assignments}~\citep{Morr} and the \emph{EADAM} algorithm~\citep{Kesten} --  through the lens of classical theory of stable matchings. In any instance, the set ${\cal L}$ of legal assignments is known to contain all stable assignments. We prove that ${\cal L}$ is exactly the set of stable assignments in \emph{another} instance. Moreover, we show that essentially all optimization problems over  ${\cal L}$ can be solved within the same time bound needed for solving it over the set of stable assignments. A key tool for this latter result is an algorithm that finds the student-optimal legal assignment. We then generalize our algorithm to obtain the assignment output of EADAM with any given set of consenting students without sacrificing the running time, hence largely improving in both theory and practice over known algorithms. Lastly, we show that the set ${\cal L}$ can be much larger than the set of stable matchings, connecting legal matchings with certain concepts and open problems in the literature.
\end{abstract}


\section{Introduction}

Stable matchings and stable assignments are fundamental paradigms in operations research and the design of matching markets.  Since the seminal work of~\citet{GS}, stable assignments have received widespread attention for their mathematical elegance and broad applicability (see, e.g., \citet{GI,RS,SU}). Those two facets are tightly connected. For instance, a detailed understanding of the lattice structure of stable matchings led to many fast algorithms for e.g., enumerating all stable matchings~\citep{Gusfield} and finding a  stable matching that maximizes some linear profit function~\citep{ILG}. In turns, these algorithmic results propelled the application of stable matchings to many markets, such as college admission, assignment of residents to hospitals~\citep{Roth.Medical}, and kidney transplant~\citep{Roth.Kidney}. 
 
One of the most important applications of matching theory, the school choice problem, considers the assignment of high school students to public schools. After the pioneering work of~\citet{NYC03}, many school districts, such as New York City and Boston, adopted the student-optimal stable mechanism for its \emph{fairness} (no priority violation or stability) and \emph{strategy-proofness} (for students). The mechanism asks students to report their (strict) preferences of the schools and schools to report their priorities\footnote{Priorities are preferences with ties, as schools usually rank students based on categorical information such as demographics, test scores, etc.} (preferences with ties) over the students. It then randomly breaks ties in the latter to obtain an instance of the stable assignment problem and performs Gale-Shapley's algorithm\footnote{In some literature, Gale-Shapley's algorithm is referred to as \emph{deferred acceptance}. In this paper, we stick to Gale-Shapley.} to obtain the student-optimal stable assignment. Gale-Shapley's algorithm embodies many desirable qualities an algorithm can have: it is simple, elegant, runs in time linear in the size of the instance, and outputs an assignment that satisfies the aforementioned strong properties. In our simulations, on random instances of the size of the New York City school system, it terminates on average in less than 3 minutes (see Figure~\ref{fig:algo-compare-large}). 

However, in this setting, public schools are often perceived as commodities, and only students' welfare matters. Hence, enforcing stability implies a loss of \emph{efficiency}. \citet{APR} demonstrated the magnitude of such efficiency loss with empirical data from the New York City school system, showing that over 4,000 eighth graders could have improved their assignments if stability constraints were relaxed.  Striving to regain this loss in welfare for the students, many alternative concepts and mechanisms have therefore been introduced and extensively studied (see, e.g., \citet{abdulkadirouglu2015expanding, erdil2008s, Kesten, kloosterman2016efficient, Morr}). 

Those mechanisms lead therefore to solutions outside the well-structured set of stable assignments. As a consequence, ad-hoc structural studies and algorithms must be presented. Unfortunately, properties of the former and performance of the latter rarely match theory of and algorithms for stable assignments~\citep{Kesten, kloosterman2016efficient, Tang.Yu}. For instance, Kesten's \emph{Efficiency Adjusted Deferred Acceptance Mechanism} (\emph{EADAM})~\citep{Kesten} (one of the main focuses of the present paper), in our experiments, cannot terminate after 24 hours of computation, on average, on random instances of similar size as the New York City high school system. This algorithmic inefficiency harms the applicability of such mechanisms to real-world instances, especially if policy designers want to run them multiple times either as a subroutine in a more complex mechanism or to test the effects of different tie-breaking rules~\cite{ashlagi2016matters,erdil2008s}. 

The goal of this paper is to show how certain concepts, introduced in the literature to regain the loss of welfare caused by stability constraints, can be fully understood through the lens of classical theory of stable assignments. Moreover, we show that this better understanding leads to theoretically and practically faster algorithms, as well as extensions and new connections within this classical theory. We believe that our results can stimulate further applications of those concepts, as well as future theoretical research. The two topics that we study in depth are \emph{legal assignments}~\citep{Morr} and \emph{EADAM with consent}~\citep{Kesten}. Let us therefore introduce them next.

\smallskip \noindent\textbf{Legal Assignments.} Legality gives an alternative interpretation of fairness, in an attempt to eliminate the tension between stability and efficiency. The stability condition prohibits, in the assignment output, the existence of a student-school pair that prefer each other to their assigned partners. Such pairs are called \emph{blocking pairs}. Therefore, stability makes sure that no student is harmed and thus no student has the justification to take a legal action against the public school system. However,~\citet{Morr} observed that legal standing, as interpreted by the United States Supreme Court, is not exactly the same as prohibiting blocking pairs. Specifically, in order for a student to have a legal standing, not only must he be harmed (i.e., forming a blocking pair with a school), this harm also must be \emph{redressable}.  That is, there must be an assignment that is accepted as feasible under which the student is assigned to the school.

With this interpretation, an institution is safe from legal actions if the set ${\cal L}$ of assignments that are considered feasible has the property that if a student-school pair blocks an assignment from ${\cal L}$, then this pair is not matched in any assignment from ${\cal L}$ (\emph{internal stability}). On the other hand, in order to justify the exclusion of an assignment $M$ from the set ${\cal L}$, there must be a pair that blocks $M$ and is matched in some assignment from ${\cal L}$ (\emph{external stability}). Following Morrill, we call a set ${\cal L}$ with those properties \emph{legal}. Note that every legal set contains the set of stable assignments. We illustrate this concept with an example.
	
\begin{example} \label{ex:legal-block}
    Here and throughout the paper, one side of the bipartition is called \emph{students} and the other is called \emph{schools}. In this example, we also assume that each school can admit at most one student. Consider the instance with preference lists given below. 
    $$\begin{array}{ l c c c c l c c c   }
    	\text{student 1:} & A & B & C  & & \text{school A:} & 2 & 3 & 1  \\
    	\text{student 2:} & B & A & & & \text{school B:} & 1 & 2 \\
    	\text{student 3:} & A & C & & \hspace{2cm}~ & \text{school C:} & 3 & 1
    \end{array} $$

    Below, we list all five maximal matchings. Note that it is sufficient to consider only maximal matchings because if a matching is not maximal, it cannot be in a legal set. We also list the blocking pairs each matching admits. In this instance, $M_1$ is the only stable matching. 
    $$\begin{array}{lll}
		\# \hspace{2cm}~ & \text{maximal matching} \hspace{2cm}~ & \text{blocking pairs} \\
		\hline
		M_1 & 1B,2A,3C & \emptyset \\
		M_2 & 1A,2B,3C & 3A \\
		M_3 & 1B,3A & 2A \\
		M_4 & 1C,2B,3A & 1B \\
		M_5 & 1C,2A & 1B, 2B, 3C
    \end{array}$$

    We now construct a digraph with each maximal matching as a vertex. We add an arc $(a, b)$ if and only if the matching represented by $a$ blocks the matching represented by $b$, where we say a matching $M_1$ \emph{blocks} $M_2$ if $M_1$ contains an edge that is a blocking pair for $M_2$. 

    \begin{figure}[ht]
    	\centering
    	\begin{tikzpicture}[scale = 1]
    	\node[rectangle,draw] (1) at (2,1.3) {$M_2$};
    	\node[rectangle,draw] (2) at (0.5,0) {$M_1$};
    	\node[rectangle,draw] (3) at (4.4,-.4) {$M_4$};
    	\node[rectangle,draw] (4) at (2,-1.5) {$M_5$};
    	\node[rectangle,draw] (5) at (4,1) {$M_3$};
    	\draw[solid, -{Latex[length=2mm]}, ultra thick] (1) to (4);
    	\draw[solid, -{Latex[length=2mm]}, ultra thick] (2) to (3);
    	\draw[solid, -{Latex[length=2mm]}, ultra thick] (2) to (4);
    	\draw[solid, -{Latex[length=2mm]}, ultra thick] (2) to (5);
    	\draw[solid, -{Latex[length=2mm]}, ultra thick] (3) to (1);
    	\draw[solid, -{Latex[length=2mm]}, ultra thick] (3) to [out=240, in=-10] (4);
    	\draw[solid, -{Latex[length=2mm]}, ultra thick] (3) to (5);
    	\draw[solid, -{Latex[length=2mm]}, ultra thick] (4) to (3);
    	\draw[solid, -{Latex[length=2mm]}, ultra thick] (5) to (1);
    	\draw[solid, -{Latex[length=2mm]}, ultra thick] (5) to (4);
    	\draw[thick, dashed, rotate around={45:(1.2,.65)}] (1.2,0.65) ellipse (2cm and .8cm);
    	\draw[thick, dotted, rotate around={45:(3.4,-.4)}] (3.4,-.4) ellipse (2.5cm and 1.3cm);
    	\node[] () at (1.2,1) {\large $\L$};
    	\end{tikzpicture}	
    \end{figure}

    By the definition of legality, we claim that, in this instance, $\L=\{M_1,M_2\}$ is a legal set. This is because $M_1$ and $M_2$ do not block each other, while all other matchings are blocked by at least one of $M_1$ and $M_2$. 
    \hfill{$\diamondsuit$}
\end{example}

From the example, one can see that relaxing stability to legality allows us to extend the set of feasible assignments, while maintaining a certain level of fairness. As we will show in Section~\ref{SEC:LATIN}, the increase in the size of the feasible set can be very significant.
\citet{Morr} observed that, in the present setting, legality coincides with the concept of \emph{von Neumann-Morgenstern stability}~\citep{vNM} in game theory under an appropriate definition of dominance. This has been investigated in the one-to-one case by~\citet{WJ}. (We defer a more detailed discussion of von Neumann-Morgenstern stability and of the work of~\citet{WJ} to Section~\ref{sec:literature}). \citet{Morr} also showed that every one-to-many instance has a unique legal set ${\cal L}$. Moreover, assignments in ${\cal L}$ form a lattice under the classical dominance relation. By standard arguments, this implies the existence of a student-optimal legal assignment, which~\citet{Morr} showed is \emph{Pareto-efficient} and can be found using Kesten's EADAM algorithm~\cite{Kesten}.

\smallskip \noindent\textbf{EADAM with Consent.} EADAM similarly aims at reducing the loss in efficiency due to stability constraints without creating legal concerns. Starting from the student-optimal stable assignment, EADAM algorithm iteratively asks for certain students' consent to allow the removal of certain schools from their preference lists, and then re-runs Gale-Shapley's algorithm. This removes the possibility of certain student-school pairs acting as blocking pairs. We defer a detailed description of the algorithm to Section~\ref{SEC:CONNECT-TO-EADAM}, and illustrate here a number of properties (showed in~\citet{Kesten} and~\citet{Tang.Yu}) that make EADAM attractive for school choice.

If a student is asked to give consent, whether or not he consents, his assignment will not change and thus, no student has the incentive to not consent and no student is harmed under EADAM. Moreover, EADAM outputs an assignment that is \emph{constrained efficient}. That is, this assignment does not violate any nonconsenting students' priorities (i.e., no nonconsenting student is part of a blocking pair), but any other assignment that is weakly preferred by  all students does. When all students consent, the output is therefore Pareto-efficient. 

Although EADAM is not strategy-proof (i.e., a student can misstate his preference list in order to be assigned to a better school),~\citet{Kesten} remarked that violation of strategy-proofness does not necessarily imply easy manipulability in practice (see, e.g., \citet{roth1999redesign}), as agents usually do not have complete information about the preferences of other agents in the market and are thus unlikely to engage in potentially profitable strategic behaviors~\citep{roth1999truncation}. Moreover, as the market size increases, the possibilities agents have in manipulation vanish rapidly~\citep{kojima2009incentives}. \citet{Kesten} also proved that any mechanism that improves over the student-optimal stable mechanism either violates some nonconsenting students' priority or is not strategy-proof.

Although both EADAM and legal assignments have been further analyzed and extended by several authors (see, e.g., \citet{DGY, EM, kloosterman2016efficient, AAB, TZ}), our knowledge of those two concepts is far from complete. In particular, the knowledge that legal assignments form a lattice gives little information on how to exploit it for algorithmic purposes, e.g., how to find the legal assignment that maximizes some linear profit function\footnote{Indeed, even though Birkhoff's representation theorem~\cite{birkhoff1937rings} implies that there is a bijection between the elements of a distributive lattice and the closed sets of an associated poset, it is not clear how to use this information algorithmically. A typical example are~\emph{strongly stable matchings}, which have been known for a long time to form a distributive lattice~\citep{ManTie}, but only recently was this structure exploited for algorithmic purposes~\citep{Strong}.}. Moreover, little is known on how to exploit the structure of legal assignments to obtain the output of EADAM when not all students consent, since the assignment output by the algorithm may not be legal.

%
%
% ------------------------------------------
%
%

\subsection{Our Contribution} 

Our first contribution addresses the structure of legal assignments. We prove in Section~\ref{sec:disguised} that the set of legal assignments coincides with the set of stable assignments in a subinstance of the original one. That is, we can describe the set of legal assignments exactly as a set of stable assignments in a subinstance. The proof we present in Section~\ref{sec:disguised} builds on results by~\citet{Morr}. However, a self-contained, not much harder proof that only relies on classical concepts from the theory of stable assignment is also possible, and it is presented in Appendix~\ref*{app:self-contained}. This latter proof studies the fixed points of a certain function defined in~\citet{Morr}, but our approach greatly simplifies the overall treatment. In particular, \citet{Morr} proved many properties of legal assignments from scratch and showed that the concept of legal assignments is well-defined through its connection with EADAM. In our proof, we first showed that legal assignments are stable assignments in a subinstance. Afterwards, all structural properties shown in~\citet{Morr} follow immediately. We defer a more detailed comparison between those proofs to Appendix~\ref*{app:self-contained}. We also show, by building on our approach and on results by~\citet{EM}, that legal assignments coincide with the set of stable assignments in a subinstance for the more general case where school preferences are represented by \emph{substitutable} choice functions that satisfy the \emph{law of aggregate demand}. We defer details to Section~\ref{sec:disguised} and to the Appendix. 

As our second contribution, in Section~\ref{sec:G-L-efficient} we show how to obtain the aforementioned subinstance in time linear in the number of edges of the input. Hence, in order to solve an optimization problem over the set of legal assignments (e.g., to find the already mentioned school-optimal, or other assignments of interest such as the \emph{egalitarian}, \emph{profit-optimal}, \emph{minimum regret}), one can resort to the broad literature on algorithms developed for the same problem on the set of stable assignments (see e.g., \citet{Manlove} for a collection of those results).  Since the worst-case running time of those algorithms is at least linear in the number of the edges, the complexity of the related problems over the set of legal assignments does not exceed their complexity over the set of stable assignments.  To achieve this second contribution, we rely on the concept of meta-rotations~\cite{metarot} and develop a symmetric pair of algorithms in Section~\ref{SEC:OUR-ALGO}, which we name \texttt{student-rotate-remove} and \texttt{school-rotate-remove}, that respectively find the school-optimal and student-optimal legal assignments.

Our third contribution is a fast algorithm for EADAM with consent. Algorithmic results above imply that, when all students consent, EADAM can be implemented as to run with the same time complexity as that of Gale-Shapley's. However, when only some students consent, the output of EADAM may no longer be legal (see Example~\ref{ex:eadam-seadam}). We show in Section~\ref{SEC:CONNECT-TO-EADAM} how to modify \texttt{school-rotate-remove} to produce the output of EADAM, again within the same time bound as Gale-Shapley's. Computational tests on random instances performed in Section \ref{sec:computation-compare} confirm that our algorithm run significantly faster in practice. 

As our last contribution, we show that when relaxing stability to legality, we can greatly increase the number of feasible matchings. We show one-to-one instances that have only one stable matching, but exponentially many (in the number of agents) legal matchings. This is achieved by an exploration of the connection between \emph{Latin marriages} introduced by \citet{Ben.et.al} and legal matchings. We defer details to Section \ref{SEC:LATIN}.

Our algorithm implementations for (1) finding student-optimal and school-optimal legal assignments and obtaining the legal subinstance; and (2) EADAM with consent can be found online\footnote{(1). \url{ https://github.com/xz2569/LegalAssignments}. (2). \url{https://github.com/xz2569/FastEADAM}. }.

%
%
% ------------------------------------------
%
%

\subsection{Literature Review} \label{sec:literature}

There is a vast amount of literature on mechanism design for the school choice problem, balancing the focus among strategy-proofness, efficiency and stability. From a theoretical prospective,~\citet{EH} shows that under certain acyclicity conditions on the priority structure, the student-optimal stable assignment is also Pareto-efficient for the students. \citet{Kesten} interprets these cycles as the existence of interrupting pairs and proposes the EADAM mechanism, which improves efficiency by obtaining students' consent to waive their priorities.

Extending upon Kesten's framework, many researchers offer new perspectives. \citet{Tang.Yu} propose a simplified version of EADAM, which repeatedly runs Gale-Shapley's algorithm after fixing the assignments of underdemanded schools. \citet{BK} shows an algorithm which iteratively runs Gale-Shapley's algorithm after fixing the assignments of the set of last proposers. \citet{BK} also shows that when restricting to the one-to-one setting, his algorithm finds the student-optimal matching in the \emph{von Neumann-Morgenstern (vNM) stable set}. \emph{vNM stable set} is a concept proposed by~\citet{vNM} for cooperative games. The definition of vNM stable set requires an irreflexive dominance relation among outcomes in the set. 

For the stable assignment problem, the definition of legal assignments in~\citet{Morr} corresponds to vNM stable set under the dominance relation \emph{dom}, where assignment $M_1$ \emph{dom} $M_2$ if $M_1$ blocks $M_2$. Under this dominance relation, results from~\citet{E} and~\citet{WJ} show existence and uniqueness of the vNM stable set in the one-to-one setting. \citet{Morr} further proves the existence and uniqueness results in the one-to-many setting, as well as the fact that the vNM stable set has a lattice structure. \citet{Morr} is superseded by~\citet{EM}, where the concept of legality and the above mentioned results are generalized to the setting where schools' preferences are specified by substitutable choice functions that satisfy the law of aggregate demand. To the best of our understanding, results from~\citet{EM} do not have any implication in the stable assignment setting, other than those that already follow from~\citet{Morr}, mentioned above. Interestingly,~\citet{EM} also investigate a different dominance relation \emph{dom$'$} (which they call ``vNM-blocks'') and observe that \emph{dom} and \emph{dom$'$} lead to different vNM stable sets. 

\citet{WJ} presents an algorithm that finds the man- and woman-optimal matchings in the vNM stable set (under the dominance relation \emph{dom} defined above) in the one-to-one case, and shows that the vNM stable set coincides with the set of stable matchings in another instance. When restricted to the one-to-one case, our algorithms from Section~\ref{SEC:OUR-ALGO} essentially projects to that of~\citet{WJ}, as~\citet{WJ} also obtains, e.g., the woman-optimal legal matching by starting from the woman-optimal stable matching and iteratively finding rotations and eliminating edges. However, our approach is different because, unlike~\citet{WJ}, we show that legal assignments are stable assignments in a subinstance before and independently of the algorithm for finding them. Even when restricted to the one-to-one case, this allows for a more direct derivation and, we believe, a more intuitive understanding of the algorithm, and a simpler and shorter proof overall. Moreover, as~\citet{WJ} points out, his results do not have either structural or algorithmic implications for the vNM stable set in the one-to-many setting, and he actually poses as an open question to construct an algorithm to produce such assignments.

Our results answer this open question and allow us to also characterize legal assignments in the more general setting of~\citet{EM}. We remark that, although there is a standard reduction from one-to-many instances to one-to-one instances~\citep{GI,RS} such that the set of stable assignments of the former and the set of stable matchings of the latter correspond, this one-to-one mapping fails for the set of legal assignments (see Example~\ref{ex:legal-donot-commute}). So we need to directly tackle the one-to-many setting. 

%
%
% ------------------------------------------
%
%

\section{Basics} \label{sec:basic}

We introduce here basic notions and facts. We point readers to the book by~\citet{GI} for a more comprehensive introduction on stable marriage and stable assignment problems.

For $n \in \N$, we denote by $[n]$ the set $\{1,\dots,n\}$. All (di)graphs in this paper are simple. All paths and cycles in (di)graphs are therefore uniquely determined by the sequence of nodes they traverse, and are denoted using this sequence, e.g., $a_0,b_0,a_1,b_1, \cdots$. The edge connecting two nodes $a,b$ in an undirected graph is denoted by $ab$. For a graph $G$, we denote by $V(G)$ and $E(G)$ its set of vertices and set of edges respectively. For $v\in V(G)$, we let $\deg_G(v)$ denote the degree of $v$ (i.e., number of adjacent vertices of $v$) in $G$. For a graph $G(V,E)$ and $F\subseteq E$, we denote by $G[F]:=G(V,F)$. A \emph{singleton} of a graph is a node of degree $0$. For sets $S,S'$, $S\triangle S'$ denotes their symmetric difference.

%
%
% ------------------------------------------
%
%

\subsection{The Stable Assignment Problem}

An instance of the \emph{stable assignment} problem is a triple $(G, <, \q)$ with $G=(A\cup B, E)$, where $G$ is a bipartite graph with bipartition $(A,B)$, $<$ denotes the set $\{<_v\}_{v \in A\cup B}$, with $<_v$ being a strict ordering of the neighbors of $v$ in $G$, and $\q = \{q_b\}_{b\in B} \in \N^{B}$ denotes the maximum number of vertices in $A$ that can be assigned to each $b\in B$. $q_b$ is called the \emph{quota} of $b$. Elements of $A$ are referred to as \emph{students} and elements of $B$ are referred to as \emph{schools}. For $x, y, y' \in A \cup B$ with $xy, xy' \in E$ we say $x$ \emph{strictly prefers} $y$ to $y'$ if $y>_x y'$, and we say that $x$ \emph{(weakly) prefers} $y$ to $y'$ and write $y\geq_x y'$ if $y>_x y'$ or $y=y'$. For all $xy \in E$, we assume $y>_x \emptyset$. When ${\bf q}$ is the vector all of $1$'s, we speak of an instance of the \emph{stable marriage} problem, and denote it by $(G,<)$. In this case, elements of $A$ are referred to as \emph{men} and elements of $B$ are referred to as \emph{women}. 

An \emph{assignment} $M$ for an instance $(G, <, \q)$ is a collection of edges of $G$ such that: at most one edge of $M$ is incident to $a$ for each $a \in A$; at most $q_b$ edges of $M$ are incident to $b$ for each $b \in B$. For $x \in A \cup B$, we write $M(x)=\{y: xy\in M\}$. When $M(x)=\{y\}$, we often think of $M(x)$ as an element instead of a set and write $M(x)=y$. For $ab \in E$ and an assignment $M$, we call $ab$ a \emph{blocking pair} for $M$ if $b >_a M(a)$, and either $a >_b a'$ for some $a'\in M(b)$ or $|M(b)|<q_b$. In this case, we say that $ab$ \emph{blocks} $M$, and similarly, we say that $M'$ blocks $M$ for every assignment $M'$ containing edge $ab$\footnote{This notion of an assignment blocking another assignment is not standard, and is adopted from~\citet{Morr}.}. An assignment is \emph{stable} if it is not blocked by any edge of $G$. 

We let $\M(G, \q)$ be the set of assignments of $(G, <, \q)$, and let $\SS(G, <, \q)$ be the set of stable assignments of $(G,<,\q)$. For a subgraph $G'$ of $G$, we denote by $(G',<,\q)$ the stable assignment instance whose preference lists are those induced by $<$ on $G'$ and quotas are those obtained by restricting $\mathbf{q}$ to nodes in $G'$. If $ab \in M$ for some $M \in \SS(G,<,\mathbf{q})$, we say that $a$ is a \emph{stable partner} of $b$ and that $ab$ is a \emph{stable pair}. 

Every instance has at least one stable assignment. Algorithms proposed by~\citet{GS} output special stable assignments.

\begin{theorem}\label{thm:GS-optimal}
	The student-proposing (resp.~school-proposing) Gale-Shapley's algorithm outputs a stable assignment $M_0$ (resp.~$M_z$)  such that $M_0(a)\ge_a M'(a)$ (resp.~$M'(a)\ge_a M_z(a)$) for any $a \in A$ and any stable assignment $M'$.
\end{theorem}

%
%
% ------------------------------------------
%
%

\subsection{Reduction: Stable Assignments to Stable Marriages}\label{sec:redu}

A stable assignment instance $(G, <, \q)$ can be transformed into a stable marriage instance $(H_G, <_G)$ via the following well-known reduction~\citep{GI,RS}. For each school $b \in B$, create $q_b$ copies $b^1,\dots,b^{q_b}$ of $b$, and replace $b$ in the preference list of each adjacent $a\in A$ by the $q_b$ copies in exactly this order. The preference list of each $b^i$ is identical to the preference list of $b$. We call these copies \textit{seats} of the schools and denote their collection by $B_H$. With this reduction, we can construct a map $\pi: \M(G,\q)\rightarrow \M(H_G, \mathbf{1})$ that induces a bijection between $\SS(G,<,\q)$ and $\SS(H_G,<_G,\mathbf{1})$. Given $M \in \M(G,\q)$, assume for some $b\in B$, $M(b)=\{a_1,\dots,a_j\}$ and $a_1>_b a_2>_b \cdots>_b a_j$. Define $\pi(M)(b^i)=a_i$ for $i \in [j]$ and $\pi(M)(b^i)=\emptyset$ for $i=j+1, \dots,q_b$.  For the sake of shortness, we often abbreviate $M_H=\pi(M)$.

%
%
% ------------------------------------------
%
%

\section{Legal Assignments are Stable Assignments in Disguise} \label{sec:disguised}

For an instance $(G, <, \q)$ of the stable assignment problem and a set $\M' \subseteq \M(G,\q)$, define $\I(\M')$ as the set of assignments that are blocked by some assignment from $\M'$. We say a set $\M'$ has the \emph{legal property} if $\I(\M')=\M(G,\q) \setminus \M'$ and in this case, we say that $(\M', \I(\M'))$ is a \emph{legal partition} of $\M(G,\q)$. That is, no assignment from $\M'$ is blocked by any assignment from $\M'$ (\emph{internal stability}), and every assignment in $\M(G,\q) \setminus \M'$ is blocked by some assignment from $\M'$ (\emph{external stability}).

We devote this section to the proof of the following theorem.

\begin{theorem}\label{thr:main1}
	Let $(G,<,\q)$ be an instance of the stable assignment problem. There exists a unique set $\L\subseteq \M(G,\q)$ that has the legal property. This set coincides with the set of stable assignments in $(G_L,<,\q)$, where $G_L$ is a subgraph of $G$. Moreover, $$E(G_L) = \bigcup\{M: M\in \SS(G_L, <, \q)\} = \bigcup\{M: M\in \L\}.$$
\end{theorem}

Before going into the proof, let us first show an interesting example. As introduced in Section~\ref{sec:basic}, there is a one-to-one correspondence between stable assignments in $(G,<,\q)$ and stable matchings in the reduced instance $(H_G, <_G)$. One could think of proving Theorem~\ref{thr:main1} by showing the (simpler) results for the stable marriage instance $(H_G, <_G)$, and then deducing the set of legal assignments of $(G,<,\q)$ from the set of legal matchings of $(H_G, <_G)$. Unfortunately, the bijection between stable assignments and stable matchings does not extend to the legal setting, as the next example shows.

\begin{example}\label{ex:legal-donot-commute} 
    Consider an instance with $4$ students and $2$ schools, each with $2$ seats. Let $a_i$, $b_i$, $b_i^j$ represent students, schools, and seats respectively. The preference lists are given as follows. In this and all following examples, when it is clear whose preference list we are referring to, the subscript in $>$ is dropped.
	$$\begin{array}{clccl}
	a_1: & b_1 > b_2 & \qquad \qquad \qquad & b_1: & a_3 > a_4 > a_2 > a_1 \\
	a_2: & b_2> b_1 & & b_2: & a_2 > a_4 > a_3 > a_1 \\
	a_3: & b_2> b_1 & && \\
	a_4: & b_1> b_2 & &&
	\end{array}$$
	
	Since all preference lists are complete, we can restrict our attention to the $6$ assignments where all students are matched. One can easily verify that $M=\{ a_1b_1, a_2b_2, a_3b_2, a_4b_1 \}$ is the only stable assignment, and all other assignments are blocked by some edge in $M$. Thus, ${\cal L}=\{M\}$. Now, consider the reduced stable marriage instance. The preference lists can be expanded:
	$$\begin{array}{clccl}
	a_1: & b_1^1 > b_1^2 > b_2^1 > b_2^2 & \qquad \qquad \qquad & b_1^1: & a_3 > a_4 > a_2 > a_1 \\
	a_2: & b_2^1 > b_2^2 > b_1^1 > b_1^2 & & b_1^2: & a_3 > a_4 > a_2 > a_1 \\
	a_3: & b_2^1 > b_2^2 > b_1^1 > b_1^2 & & b_2^1: & a_2 > a_4 > a_3 > a_1  \\
	a_4: & b_1^1 > b_1^2 > b_2^1 > b_2^2 & & b_2^2: & a_2 > a_4 > a_3 > a_1
	\end{array}$$ 
	
	The corresponding matchings and their blocking pairs are:
	$$\begin{array}{lccclcl}
	& & \text{matchings} & & \text{blocking pairs} & & \text{classification} \\
	\hline
	M_{1H} & & \{ a_1b_1^2, a_2b_1^1, a_3b_2^2, a_4b_2^1 \} & & \underline{a_2b_2^1}, a_2b_2^2, a_4b_1^1, a_4b_1^2 & & \text{illegal} \\
	M_{2H} & & \{ a_1b_1^2, a_2b_2^1, a_3b_1^1, a_4b_2^2 \} & & a_4b_1^2 & & \text{legal}  \\
	M_{3H} & & \{ a_1b_1^2, a_2b_2^1, a_3b_2^2, a_4b_1^1 \} & & \text{none} & & \text{stable}  \\
	M_{4H} & & \{ a_1b_2^2, a_2b_1^2, a_3b_1^1, a_4b_2^1 \} & & \underline{a_2b_2^1}, a_2b_2^2, \underline{a_3b_2^2}, a_4b_1^2 & & \text{illegal} \\
	M_{5H} & & \{ a_1b_2^2, a_2b_1^2, a_3b_2^1, a_4b_1^1 \} & & \underline{a_2b_2^1}, a_2b_2^2 & & \text{illegal} \\
	M_{6H} & \quad \quad \qquad & \{ a_1b_2^2, a_2b_2^1, a_3b_1^1, a_4b_1^2 \} & \quad \quad \qquad & \underline{a_3b_2^2} & \qquad \quad \quad & \text{illegal}
	\end{array}$$ 
	
	$M_{3H}$ is the only stable matching. All other matchings except for $M_{2H}$ are blocked by some edge in $M_{3H}$ (underlined). Hence, one easily verifies that $\{M_{2H},M_{3H}\}$ has the legal property and $\pi^{-1}(M_{3H})= M$ but $\pi^{-1}(M_{2H})\neq M$. 
	\hfill{$\diamondsuit$}
\end{example}

The proof of Theorem~\ref{thr:main1} presented in this section relies on the following result by~\citet{Morr}.

\begin{theorem} \label{thm:Morr}
    Let $(G,<,\q)$ be an instance of the stable assignment problem. There exists a unique set $\L\subseteq \M(G,\q)$ that satisfies the legal property.
\end{theorem}

As the proof of Theorem~\ref{thm:Morr} from~\citet{Morr} is somehow involved, in Appendix~\ref*{app:self-contained}, we present an alternative proof of Theorem~\ref{thr:main1} not relying on Theorem~\ref{thm:Morr}.

For a stable assignment instance $(G,<,\q)$, we denote by $\L(G, <, \q)\subseteq \M(G,<,\q)$ the unique set that satisfies the legal property and call it the set of \emph{legal assignments} of instance $(G, <, \q)$. We say that an edge $e\in E(G)$ is \emph{legal} if it is contained in some assignment from $\L(G,<,\q)$, and is \emph{illegal} otherwise. 

For the rest of the section, we fix a stable assignment instance $(G,<,\q)$ with $G=(A\cup B, E)$ and let $\M:= \M(G,\q)$, $\L:=\L(G,<,\q)$, $\I:=\M\setminus\L$, $\bar E :=\bigcup \{M: M\in \L\}$, and $G_L := G[\bar E]$. Next lemmas show that illegal edges can be removed from $G$ without modifying the set of legal assignments.

\begin{lemma}\label{lem:remove-illegal}
	Let $e$ be an illegal edge and let $\widetilde G:=G[E \setminus \{e\}]$. Then $\L=\L(\widetilde G,<,\q)$.
\end{lemma}

\proof{Proof.}
    Let $\M^e:=\{ M\in \M: e\in M \}$ and $\widetilde{\M}:= \M(\widetilde{G}, \q)$. Note that $\widetilde \M = \M\setminus \M^e$ and ${\cal L}\subseteq \widetilde \M$, since $e$ is an illegal edge.  Hence, $(\L, \widetilde{\M} \setminus \L)$ is a partition of $\widetilde{\M}$. We show next that it is also a legal partition. To see this, first note that no two assignments $M_1, M_2\in \L$ block each other, since $\L$ is the set of legal assignments of the original instance. Next, for any assignment $M'\in \widetilde{\M} \setminus \L$, since $\widetilde{\M} \setminus \L\subseteq \I$, $M'$ must be blocked by some assignment in $\L$. Thus, together with the uniqueness of the legal partition given by Theorem~\ref{thm:Morr}, we conclude $\L(\widetilde{G}, <,\q) = \L$.
\hfill{\Halmos}\endproof

\begin{lemma} \label{lem:legal-remain-rm-edge}
     $\L = \L(G_L, <, \q)$.
\end{lemma}

\proof{Proof.}
	Let $e_1, e_2, \cdots, e_k$ be an ordering of the illegal edges and let $G^i := G[E\setminus \{e_1, e_2, \cdots, e_i\}]$. Observe that $G^k = G_L$.  By Lemma~\ref{lem:remove-illegal}, we have $\L(G^1, <, \q) = \L$ and thus, by definition of illegal edges, edges $e_2, \cdots, e_k$ remain illegal in instance $(G^1, <, \q)$. Therefore, applying Lemma~\ref{lem:remove-illegal} again to $(G^1, <, \q)$, we have $\L(G^2, <, \q) = \L(G^1, <, \q)= \L$. Iterating the process, we can conclude that $\L(G_L, <, \q) = \L(G^k, <, \q) = \L(G^{k-1}, <, \q) = \dots = \L$.
\hfill{\Halmos}\endproof

Once all illegal edges have been removed, we are left with a graph whose edges are all and only those appearing in some legal assignment. In such an instance, legality and stability coincide.

\begin{lemma} \label{lem:legal-equal-stable}
    $\SS(G_L,<,\q) = \L(G_L,<,\q)$.
\end{lemma}

\proof{Proof.}
	The direction $\SS(G_L,<,\q)\subseteq \L(G_L,<,\q)$ is clear, since a stable assignment is not blocked by any other assignment. For the other direction, let $M\in \L(G_L,<,\q)$. Then $M$ is not blocked by any assignment in $\L(G_L,<,\q)$. Since every edge in $E(G_L)$ appears in at least one assignment in $\L(G_L,<,\q)$, $M$ admits no blocking pair and thus is stable. This concludes the proof.
\hfill{\Halmos}\endproof

\medskip
\proof{Proof of Theorem~\ref{thr:main1}.} 
    Immediately from Theorem~\ref{thm:Morr}, Lemma~\ref{lem:legal-remain-rm-edge}, and Lemma~\ref{lem:legal-equal-stable}. 
\hfill{\Halmos}\endproof
\smallskip

The approach developed in this section can be extended to a more general setting studied in~\citet{EM}, where schools' preferences are represented by certain choice functions. In particular, Theorem~\ref{thr:main1} also holds in this setting. We defer details to Appendix~\ref*{app:choice-setting}.

We have shown that legal assignments are stable assignments in $G_L$. Since there might be an exponential number of legal assignments, one cannot expect to construct $G_L$ efficiently by explicitly listing all the legal assignments. Instead, the main tool we use is an efficient mechanism in identifying legal and illegal edges, which is developed in Section~\ref{SEC:OUR-ALGO}.

Before we introduce this algorithm, we need some more properties of stable assignments. 

%
%
% ------------------------------------------
%
%

\section{The structure of stable assignments} \label{sec:lattice-rotation}

In this section, we recall known results on structural properties of stable assignments and their algorithmic consequences. Throughout the section, we fix a stable assignment instance $(G, <, \q)$, with $G=(A\cup B, E)$. Given $M,M' \in \M(G, \q)$, we say $M$ (\emph{weakly}) \emph{dominates} $M'$, and write $M \succeq M'$, if $M(a)\ge_a M'(a)$ for every student $a \in A$. If moreover $M\neq M'$, we say that $M$ \emph{strictly dominates} $M'$ and write $M\succ M'$. $M \in \M(G,\q)$ is said to be \emph{Pareto-efficient} if there is no $M' \in \M(G,\q)$ such that $M'\succ M$. The following fact is well-known (see, e.g., \citet{GI}).

\begin{theorem}\label{thm:sa-lattice}
	$\SS(G,<,\q)$ endowed with the dominance relation $\succeq$ forms a distributive lattice. In particular, there exists a stable assignment $M_0$ (resp.~$M_z$) such that $M_0\succeq M$ (resp.~$M\succeq M_z$) for all $M\in \SS(G, <, \q)$, and it is called the \emph{student-optimal} (resp.~\emph{school-optimal}) stable assignment. 
\end{theorem}

Note that the student-optimal (resp.~school-optimal) stable assignment coincides with the one output by Gale-Shapley's algorithm with students (resp.~schools) proposing, as described in Theorem~\ref{thm:GS-optimal}. Hence, the notation describing those assignments coincide.

Next, we introduce the concept of rotations in the one-to-many setting. Informally speaking, a rotation exposed in a stable assignment $M$ is a certain $M$-alternating cycle $C$ such that $M\triangle C$ is again a stable assignment. $C$ has the property that every agent from one side of the bipartition prefers $M$ to $M\triangle C$, while every agent from the other side prefers $M\triangle C$ to $M$. We can interpret a rotation as a cycle of un-matches and re-matches with one side getting better and the other side getting worse. Hence, rotations provide a mechanism to generate one stable assignment from another, moving along the lattice.

Because of the different role played by the two sides of the bipartition, we distinguish between \emph{school-} and \emph{student-rotations}. In the following, we present them jointly by choosing $X$ to be one side of the bipartition and $Y$ the other\footnote{It is worth noticing that the definition of both student-rotation and school-rotation can be simplified, but in different ways. However, in order to keep the treatment compact, we give a unique presentation encompassing both.}. We extend the quota vector ${\bf q}$ to students by letting $q_a=1$ for each student $a \in A$.

For $M \in \SS(G,<,\q)$ and $x\in X$, let $s_M(x)$ be the first agent $y\notin M(x)$ on $x$'s preference list satisfying $x>_y x'$ for some $x' \in M(y)$. Note that since $M$ is stable, we must have $y'>_x y$ for all $y'\in M(x)$. If $y:=s_M(x)$ exists and if moreover $|M(y)|=q_y$, define $next_M(x)$ as the \emph{least preferred} among the current partners of  $y$, i.e., $next_M(x) \in M(y)$ and for all $x' \in M(y)$, $x' \geq_y next_M(x)$. If otherwise $|M(y)|<q_y$, then define $next_M(x)=\emptyset$. 

Given distinct $x_0,\dots, x_{r-1} \in X$ and $y_0,\dots,y_{r-1}\in Y$, a cycle $y_0, x_0, y_1, x_1, \dots, y_{r-1}, x_{r-1}$ of $G$ is an \emph{$X$-rotation} \emph{exposed} in $M$ if $x_iy_i \in M$ and $s_M(x_i)=y_{i+1}$ for all $i=0,\dots, r-1$ (here and later, indices are taken modulo $r$). Define the \emph{$X$-rotation digraph} $D_X$ (of $M$) with vertices $X\cup Y\cup \{\emptyset\}$ and arcs $(x, y)$ and $(y,x')$ if $s_M(x)=y$ and $next_M(x)=x'$. If $s_M(x)$ does not exist for some $x\in X$, then $x$ is a sink. Thus, note that sinks in $D_X$ are either agents in $X$ or $\emptyset$. One easily observes that $X$-rotations exposed in $M$ are in one-to-one correspondence with directed cycles in $D_X$.

Let $\rho:=y_0, x_0, \cdots, y_{r-1}, x_{r-1}$ be an $X$-rotation exposed in  $M$. The \emph{elimination} of $\rho$ maps $M$ to the assignment $M':=M/\rho$ with $M'(x)=M(x)$ for $x \in X \setminus \rho$ and $M'(x_i)=(M'(x_i)\setminus \{y_i\})\cup \{ y_{i+1}\}$ for $i=0,1,\cdots,r-1$. 

$X$-rotation (digraph) represents the student- or school- rotation (digraph) respectively when $X$ is the set of students or schools. When it is clear whether we are referring to students or schools, we drop the prefix $X$-. See Example~\ref{ex:rar} for an illustration of rotations and rotation digraphs in the context of our algorithm.  

The following lemmas~\cite{metarot} extend classical results on rotations in the one-to-one setting to our one-to-many setting. It shows that the set of stable assignments is complete and closed under the elimination of exposed rotations.

\begin{lemma} \label{thm:255-ext}
	Let $M \in \SS(G,<,\q)$, $\rho$ be an $X$-rotation exposed in $M$, and $M'=M/ \rho$. Then $M' \in \SS(G,<,\q)$. Moreover, $M\succ M'$ if $X$ is the set of students and $M\prec M'$ if $X$ is the set of schools. If there is no $X$-rotation exposed in $M$, $M$ is the $Y$-optimal stable assignment. In addition, every stable assignment can be generated by a sequence of $X$-rotation eliminations, starting from the $X$-optimal stable assignment, and every such sequence contains the same set of $X$-rotations.
\end{lemma}

\begin{lemma} \label{thr:256-ext}
	$xy\in E$ is a stable pair if and only if $x$ is assigned to $y$ in the $Y$-optimal stable assignment or, for some $X$-rotation $y_0, x_0, y_1, x_1, \cdots, y_{r-1}, x_{r-1}$ exposed in some stable assignment and some $i \in \{0,\dots, r-1\}$,  we have $x=x_i$ and $y=y_{i}$.
\end{lemma}

For an instance $(G,<,\q)$, we denote by $\Rot(G,<,\q)$ the set of student-rotations exposed in some of its stable assignments, and by $\SR(G,<,\q)$ the set of school-rotations exposed in some of its stable assignments. 

\begin{lemma} \label{lem:students-rotations-and-school-rotation}
	$|\Rot(G,<,\q)|=|\SR(G,<,\q)|$. There is a bijection $\sigma: \Rot(G,<,\q) \rightarrow \SR(G,<,\q)$ such that for each $M \in \SS(G,<, \q)$ and $\rho \in \Rot(G,<,\q)$ exposed in $M$, we have $M=(M/\rho)/\sigma(\rho)$.
\end{lemma}

%
%
% ------------------------------------------
%
%

\section{Algorithms for Student- and School-Optimal Legal Assignments} \label{SEC:OUR-ALGO}

Because of Theorem~\ref{thr:main1} and Theorem~\ref{thm:sa-lattice}, the concepts of student- and school-optimal legal assignments are well-defined. In this section, we show efficient routines for finding them. Throughout the section, we again fix a stable assignment instance $(G, <, \q)$ with $G=(A\cup B, E)$. We denote by $M^{\cal L}_0$ (resp. $M^{\cal L}_z$) the student-optimal (resp. school-optimal) legal assignment.

Suppose first we want to find the student-optimal legal assignment. The basic idea of the algorithm is the following: at each iteration, a legal assignment $M$ and a set of edges identified as illegal are taken as input, and one of the following three cases will happen: either (i) the set of identified illegal edges is expanded; or (ii) a legal assignment $M'\succ M$ is produced; or (iii)  $M$ is certified as the student-optimal legal assignment. If we are in case (i), then we can safely remove the newly found illegal edge (because of Lemma~\ref{lem:remove-illegal}) and proceed to the next iteration. If we are in case (ii), we replace $M$ with $M'$, and proceed to the next iteration. If we are in case (iii), we halt the algorithm and output the current assignment.

In order to distinguish between cases (i), (ii), and (iii) above, we rely on properties of the rotation digraph. In the following, $X$ can again be either the set of students or the set of schools. The proof of the following lemma uses some of the tools developed in the proof of Theorem~\ref{thr:main1} given in the appendix, and thus is deferred to Appendix~\ref*{app:proof-find-rot2}.

\begin{lemma}\label{lem:find-rot2}
	Let $M \in \SS(G,<, \q)$. If $x \in X$ is a sink in the $X$-rotation digraph $D_X$ of $M$ and $(x',y), (y,x) \in A(D_X)$, then $x'y$ is an illegal edge.
\end{lemma}

Hence, if the algorithm finds a sink in the school-rotation digraph, we are in case $(i)$ above. If the school-rotation digraph has a directed cycle, eliminating the corresponding school-rotation from $M$ brings us to case (ii)\footnote{If $D_B$ has both a sink and a directed cycle, the algorithm is free to choose between the two cases.}. Lastly, if $D_B$ has no arc, we conclude that we are in case (iii). The initial iteration starts with the set of identified illegal edges being empty, and $M$ being the student-optimal stable assignment. The algorithm that finds the school-optimal legal assignment proceeds similarly, with a legal assignment $M' \prec M$ generated in case (ii).

A formal description of our algorithm is given in Algorithm~\ref{alg:rar}. Proof of correctness is given in Theorem~\ref{thr:rotate-remove-finds-legal}, and we illustrate the algorithm in Example~\ref{ex:rar}. %{\color{red}In the following, whenever it is clear which rotation digraph, student- or school-, we are referring to, we drop the subscript in $D_X$.}

\begin{algorithm}[h!t]
	\caption{\texttt{$X$-rotate-remove} to find the $Y$-optimal legal assignment}\label{alg:rar}
	\begin{algorithmic}[1] 
		\REQUIRE $(G(A\cup B,E),<,\q)$
		\STATE{Find the $Y$-optimal stable assignment $M_Y$ of $(G,<,\q)$ via Gale-Shapley's algorithm.}  \label{step:find-optimal}
		\STATE{Let $G^0:=G$ and $M^0:=M_Y$.}
		\STATE{Set $i=0$ and let $D^0$ to be the $X$-rotation digraph of $M^0$ in $(G^0,<,\q)$.}
		\WHILE{$D^i$ still has an arc}	
		\STATE{Find (i) arcs $(x',y), (y,x) \in A(D^i)$ where $x$ is a sink in $D^i$ or (ii) a cycle $C^i$ of $D^i$.}
		\IF{(i) is found}\label{step:sink-found}
		\STATE{Define $G^{i+1}$ from $G^i$ by removing $x'y$, and set $M^{i+1} =M^i$.}
		\ELSIF{(ii) is found}\label{step:rotation-found}
		\STATE{Let $\rho^i$ be the corresponding $X$-rotation. Set $M^{i+1}=M^i / \rho^i$, and $G^{i+1}=G^i$.}
		\ENDIF
		\STATE{Set $i=i+1$ and let $D^i$ to be the $X$-rotation digraph of $M^i$ in $(G^i, <, \q)$.}
		\ENDWHILE
		\STATE{Output $M^{i}$.}		
	\end{algorithmic}
\end{algorithm}

\begin{example}\label{ex:rar}
	We apply \texttt{student-rotate-remove} and \texttt{school-rotate-remove} to the following instance with $6$ students and $3$ schools, where each school has a quota of $2$. 
	$$\begin{array}{clccl}
	a_1: & \boxed{b_2} >b_3 >b_1 & & b_1: &  a_1 >\boxed{a_4} >\boxed{a_3} >a_5 >a_2 >a_6\\
	a_2: & b_1 >\boxed{b_2} >b_3 & & b_2: & a_3 >\boxed{a_2} >a_6 >\boxed{a_1} >a_5 >a_4 \\
	a_3: & b_3 >\boxed{b_1} >b_2 & & b_3: & \boxed{a_6} >a_1 >\boxed{a_5} >a_2 >a_4 >a_3 \\
	a_4: & \boxed{b_1} >b_2 >b_3 & \qquad \qquad & & \\
	a_5: & \boxed{b_3} >b_2 >b_1 & & & \\
	a_6: & b_1 >\boxed{b_3} >b_2 & & & 
	\end{array}$$
	
    The student- and school-optimal stable assignments coincide, and are given by $\{a_1b_2$, $a_2b_2$, $a_3b_1$, $a_4b_1$, $a_5b_3$, $a_6b_3\}$ (squared entries above). This is the $M^0$ for both algorithms.
    
    \smallskip\noindent {\bf {Student-Rotate-Remove.}} On $a_1$'s preference list, $b_3$ is the first school after $M^0(a_1)$. In addition, $b_3$ prefers $a_1$ to $a_5$, who is $b_3$'s least preferred student among $M^0(b_3)$. Thus, $s_{M^0}(a_1) = b_3$ and $next_{M^0}(a_1)= a_5$. After working out $s_{M^0}(\cdot)$ and $next_{M^0}(\cdot)$ of all the students, we have the rotation digraph $D^0$ for the first iteration of \texttt{student-rotate-remove}:
	
	\begin{center}
		\begin{tikzpicture}[scale = 1]
		\node[]() at (0,.5) {$ $};
		\node[circle, draw] (1) at (0,0) {$a_6$};
		\node[circle, draw] (2) at (0,-1) {$a_3$};
		\node[rectangle, draw] (10) at (2,-1) {$b_1$};
		\node[rectangle, draw] (3) at (2,0) {$b_2$};
		\node[circle, draw] (4) at (4,0) {$a_1$};
		\node[rectangle, draw] (5) at (6,0) {$b_3$};
		\node[circle,draw](6) at (8,0) {$a_5$};
		\node[circle,draw](7) at (4,-1) {$a_2$};
		\node[circle,draw](8) at (6,-1) {$a_4$};
		\node[circle,draw](9) at (8,-1) {$\emptyset$};
		\draw[solid, thick, ->] (1) to (3);
		\draw[solid, thick, ->] (2) to (3);
		\draw[solid, thick, ->] (3) to (4);
		\draw[solid, thick, ->] (4) to (5);
		\draw[solid, thick, ->] (5) to (6);
		\end{tikzpicture}
	\end{center}
	
	Here, we find a case (i) with $x'=a_1$, $y=b_3$, and $x=a_5$. So we set $M^1 = M^0$, remove $x'y=a_1b_3$ from the instance, and update the rotation digraph $D^1$ for the next 
	iteration:
	
	\begin{center}
		\begin{tikzpicture}[scale = 1]
		\node[]() at (0,.6) {$ $};
		\node[circle, draw] (1) at (0,0) {$a_6$};
		\node[circle, draw] (2) at (0,-1) {$a_3$};
		\node[rectangle, draw] (10) at (2,-1) {$b_1$};
		\node[rectangle, draw] (3) at (2,0) {$b_2$};
		\node[circle, draw] (4) at (4,0) {$a_1$};
		\node[rectangle, draw] (5) at (6,0) {$b_3$};
		\node[circle,draw](6) at (8,0) {$a_5$};
		\node[circle,draw](7) at (4,-1) {$a_2$};
		\node[circle,draw](8) at (6,-1) {$a_4$};
		\node[circle,draw](9) at (8,-1) {$\emptyset$};
		\draw[solid, thick, ->] (1) to (3);
		\draw[solid, thick, ->] (2) to (3);
		\draw[solid, thick, ->] (3) to (4);
		\draw[solid, thick, ->] (4) to (10);
		\draw[solid, thick, ->] (10) to (2);
		\end{tikzpicture}
	\end{center}
	
	Now, we have a case (ii), with the corresponding student-rotation $\rho^1=b_2, a_1, b_1, a_3$. Eliminating $\rho^1$ from $M^1$, we have $M^2 = M^1/\rho^1 = \{a_1b_1, a_2b_2, a_3b_2, a_4b_1, a_5b_3, a_6b_3\}$. In the next iteration, the rotation digraph $D^2$ only contains sinks. Thus, the algorithm terminates and output $M^2$ as the school-optimal legal assignment.
	
	\smallskip \noindent {\bf {School-Rotate-Remove.}} The first student on $b_1$'s preference list that prefers $b_1$ to his assigned school under $M^0$ is $a_2$. Thus, $s_{M^0}(b_1) = a_2$ and $next_{M^0}(b_1)= b_2$. After working out $s_{M^0}(\cdot)$ and $next_{M^0}(\cdot)$ of all the schools, we have the rotation digraph $D^0$ for the first iteration:
	
	\begin{center}
		\begin{tikzpicture}[scale = 1]
		\node[]() at (0,.5) {$ $};
		\node[circle, draw] (1) at (0,0) {$b_3$};
		\node[rectangle, draw] (2) at (2, 0) {$a_3$};
		\node[circle, draw] (3) at (4,0) {$b_1$};
		\node[rectangle, draw] (4) at (6, 0) {$a_2$};
		\node[circle, draw] (5) at (8,0) {$b_2$};
		\node[rectangle, draw] (6) at (1, -1) {$a_6$};
		\node[rectangle, draw] (7) at (4, -1) {$a_5$};
		\node[rectangle, draw] (8) at (5, -1) {$a_4$};
		\node[rectangle, draw] (9) at (6, -1) {$a_1$};
		\node[circle, draw] () at (8, -1) {$\emptyset$};
		\draw[solid, thick, ->] (1) -- (2);
		\draw[solid, thick, ->] (2) -- (3);
		\draw[solid, thick, ->] (3) -- (4);
		\draw[solid, thick, ->] (4) -- (5);
		\end{tikzpicture}
	\end{center}
	
	Here, we find a case (i) with $x'=b_1$, $y=a_2$ and $x=b_2$. So we set $M^1 = M^0$, remove $x'y = a_2b_1$ from the instance, and update the rotation digraph $D^1$ for the next iteration:
	
	\begin{center}
		\begin{tikzpicture}[scale = 1]
		\node[]() at (0,.5) {$ $};
		\node[circle, draw] (1) at (0,0) {$b_3$};
		\node[rectangle, draw] (2) at (2, 0) {$a_3$};
		\node[circle, draw] (3) at (4,0) {$b_1$};
		\node[rectangle, draw] (4) at (6, 0) {$a_2$};
		\node[circle, draw] (5) at (8,0) {$b_2$};
		\node[rectangle, draw] (6) at (1, -1) {$a_6$};
		\node[rectangle, draw] (7) at (4, -1) {$a_5$};
		\node[rectangle, draw] (8) at (5, -1) {$a_4$};
		\node[rectangle, draw] (9) at (6, -1) {$a_1$};
		\node[circle, draw] () at (8, -1) {$\emptyset$};
		\draw[solid, thick, ->] (1) -- (2);
		\draw[solid, thick, ->] (2) -- (3);
		\draw[solid, thick, ->] (3) -- (6);
		\draw[solid, thick, ->] (6) to (1);
		\end{tikzpicture}
	\end{center}
	
	Now, we have a case (ii), with the corresponding school-rotation $\rho^1=a_6, b_3, a_3, b_1$. Eliminating $\rho^1$ from $M^1$, we have $M^2 = M^1/\rho^1 = \{a_1b_2, a_2b_2, a_3b_3, a_4b_1, a_5b_3, a_6b_1\}$. In the next iteration, the rotation digraph $D^2$ only contains sinks. Thus, the algorithm terminates and output $M^2$ as the student-optimal legal assignment.
	\hfill {$\diamondsuit$}
\end{example}

%
%
% ------------------------------------------
%
%

\subsection{Correctness of Algorithm~\ref{alg:rar}} \label{sec:algo-correct}

Using the lattice structure of the legal assignments, we can deduce the correctness of Algorithm~\ref{alg:rar}.

\begin{theorem}\label{thr:rotate-remove-finds-legal}
	Algorithm~\ref{alg:rar} finds the $Y$-optimal legal assignment.
\end{theorem}

\proof{Proof.}
	We focus on the statement with $Y$ being the set of students, the other follows analogously. We first show, by induction on the iteration $i$ of the algorithm, that $M^i\in \SS(G^i, <, \q)$ and $\L(G^i, <, \q) =\L$. This is obvious for $i=0$. Assume the claim is true for $i-1\ge 0$ and consider iteration $i$. If the condition at Step~\ref{step:sink-found} is satisfied, $M^i=M^{i-1}$ is unchanged and the edge removed from $G^{i-1}$ is illegal by Lemma~\ref{lem:find-rot2}. Hence, $M^i=M^{i-1}\in \SS(G^{i-1}, <, \q) \subseteq \SS(G^i, <, \q)$ and $\L(G^i, <, \q) = \L(G^{i-1}, <, \q)=\L$ by induction and Lemma~\ref{lem:remove-illegal}. If conversely the condition at Step~\ref{step:rotation-found} is satisfied, then $\rho^{i-1}$ is a school-rotation exposed in $M^{i-1}$, and $M^i=M^{i-1} / \rho^{i-1}\in \SS(G^{i-1}, <, \q)$ by induction and Lemma~\ref{thm:255-ext}. Moreover, since $G^{i} = G^{i-1}$, we have $\SS(G^i, <, \q)= \SS(G^{i-1}, <, \q)$ and $\L(G^i, <, \q) =\L(G^{i-1}, <, \q) = \L$.	

	In order to conclude the proof, observe that at the end of the algorithm, the school-rotation digraph -- call it $D^*$ -- only has sinks. We first claim that the assignment output -- call it $M^*$ -- strictly dominates every assignment in $\M(G^*,\q)$. Assume by contradiction that there is $M\in \M(G^*,\q)$ and a student $a$ such that $b:=M(a) >_a M^*(a)$. Then $s_{M^*}(b)$ exists by definition, contradicting the fact that $b$ is a sink in $D^*$ (it is possible that $s_{M^*}(b)\neq a$, as there may be other nodes that precede $a$ in $b$'s list and have the required property, but it is a contradiction regardless). 
	By what we proved above, we know that $\L=\L(G^*,<,\q)\subseteq \M(G^*,\q)$. Since legal assignments form a lattice with respect to the partial order given by $\succeq$, $M^*$ is the student-optimal legal assignment. 
\hfill{\Halmos}\endproof

Note that the previous theorem in particular implies that the output of Algorithm~\ref{alg:rar} is unique, regardless of how we choose between Step~\ref{step:sink-found} and Step~\ref{step:rotation-found} at each iteration, when multiple possibilities are present. 

%
%
% ------------------------------------------
%
%

\subsection{Time Complexity}\label{sec:rrr-time-complexity}

In this section, we show how to implement Algorithm~\ref{alg:rar} so that it runs in time $O(|E|)$.  We start by observing that Gale-Shapley's algorithm for stable assignment problems can be implemented as to run with the same asymptotic time complexity as the one for stable marriage problems. This does not follow from the mapping $\pi$ defined in Section~\ref{sec:basic}, which may increase the number of vertices and edges by a factor $|V|$. The proof can be found in Appendix~\ref*{app:proof:GS-complexity}.

\begin{lemma} \label{lem:GS-complexity}
	Gale-Shapley's algorithm with students or schools proposing can be implemented to run in time $O(|E|)$.
\end{lemma}

Note that \texttt{$X$-rotate-remove} seems to require the complete rotation digraph at each step. However, this is too expensive to obtain and forbids us from achieving the same time complexity bound as in Lemma~\ref{lem:GS-complexity}. Instead, in our implementation, we will only \emph{locally} build and update a directed path of the rotation digraph until a cycle or an illegal edge is found.

\begin{theorem}\label{thr:time-complexity-rotate-remove}
	Algorithm~\ref{alg:rar} can be implemented as to run in time $O(|E|)$.
\end{theorem}

The proof of Theorem~\ref{thr:time-complexity-rotate-remove} can be found in Appendix~\ref*{app:proof:time-complex-rrr}. Here instead, we show how to implement \texttt{school-rotate-remove} on a specific instance, see Example~\ref{ex:rrr-fast}.  We remark that our implementation only requires simple data structures, such as arrays and linked lists.

\begin{example} \label{ex:rrr-fast}
	Consider the following instance with $5$ students and $5$ schools, where each school has quota $1$. The student-optimal stable assignment is $\{a_1b_4, a_2b_3, a_3b_2, a_4b_1, a_5b_5\}$, denoted succinctly by $(4,3,2,1,5)$ (ordered list of school to which each student is matched). 
	$$\begin{array}{clccl}
	a_1: & b_1 > b_2 > b_3 > b_4 > b_5 & & b_1: &  a_4 > a_5 > a_3 > a_2 > a_1 \\
	a_2: & b_2 > b_1 > b_4 > b_3 > b_5 & & b_2: & a_3 > a_5 > a_4 > a_1 > a_2 \\
	a_3: & b_3 > b_4 > b_1 > b_2 > b_5 & & b_3: &a_2 > a_5 > a_1 > a_4 > a_3 \\
	a_4: & b_4 > b_3 > b_2 > b_1 > b_5 & & b_4: & a_1 > a_5 > a_2 > a_3 > a_4 \\
	a_5: & b_4 > b_3 > b_2 > b_1 > b_5 & \qquad \qquad& b_5: & a_5 > a_1 > a_2 > a_3 > a_4
	\end{array}$$
	
	For the fast implementation of \texttt{school-rotate-remove}, at each iteration $i$, together with $M^i$, we will additionally keep the following items:
	
	\begin{enumerate}[label=(\roman*), topsep=1ex, itemsep=-.5ex]
	    \item a directed path $P^i$ of the school-rotation digraph $D^i$ stored as a doubly-linked list which we will constructed step-by-step until a sink is reached or a cycle is closed;
    	\item for each $b \in B$, a position $p_b$ for which the algorithm maintains the invariant that $M^i(a) \ge_a b$ for all $a\ge_b b(p_b)$, where $b(p_b)$ is the student at position $p_b$ on $b$'s preference list;
	    \item a Boolean array $W^i$ of dimension $|B|$, recording whether each school is in $P^i$;
		\item a list $T^i$ of identified sinks\footnote{Note that $T^i$ can be easily deduced from $p_b$, given that a school $b$ is in $T^i$ if and only if $p_b(b)=\deg_G(b)+1$. However, we keep $T^i$ in our illustration to elucidate the steps.} of $D^i$, stored as a Boolean array of dimension $|B|$;
		\item a position $f$ such that $b_{f}$ is the first school that is not in $T^i$.
	\end{enumerate}
	
	\begin{table}[h!t]
	\centering
	\setlength{\tabcolsep}{8pt}
	\renewcommand\arraystretch{.85}
	\begin{tabular}{llllll}
		$(i.j)$ & $P^i$ & $\{p_b\}_{b\in B}$ & $M^i$ & $T^i$ & $f$ \\
		\hline
		$(0.0)$ & $[] \rightarrow b_1$ & $[1,1,1,1,1]$ & $(4,3,2,1,5)$ & $\emptyset$ & 1 \\
		$(0.1)$ & $b_1, a_5, {b_5}$ & $[2,1,1,1,1]$ & & & \\
		\hline
		$(1.0)$ & $b_1$ & $[2,1,1,1,6]$ & & $b_5$ & \\
		$(1.1)$ & $b_1, a_3, b_2$ & $[3,1,1,1,6]$ & & & \\
		$(1.2)$ & $b_1, a_3, b_2, a_5, b_5$ & $[3,2,1,1,6]$ & & & \\
		\hline
		$(2.0)$ & $b_1, a_3, b_2$ & & & & \\
		$(2.1)$ & $b_1, a_3, b_2, a_4, (b_1)$ & $[3,3,1,1,6]$ & & & \\
		\hline
		$(3.0)$ & $[] \rightarrow b_1$ & & $(4,3,1,2,5)$ & & \\
		$(3.1\text{-}2)$ & $b_1, a_2, b_3, a_5, b_5$ & $[4,3,2,1,6]$ & & & \\
		\hline
		$(4.0)$ & $b_1, a_2, b_3$ & & & & \\
		$(4.1\text{-}2)$ & $b_1, a_2, b_3, a_1, b_4, a_5, b_5$ & $[4,3,3,2,6]$ & & & \\
		\hline
		$(5.0)$ & $b_1, a_2, b_3, a_1, b_4$ & & & & \\
		$(5.1)$ & $b_1, a_2, b_3, a_1, b_4, (a_2)$ & $[4,3,3,3,6]$ & & & \\
		\hline
		$(6.0)$ & $b_1$ & $[3,3,3,3,6]$ & $(3,4,1,2,5)$ & & \\
		$(6.1\text{-}2)$ & $b_1, a_2, b_4, a_3, (b_1)$ & $[4,3,3,4,6]$ & & & \\
		\hline
		$(7.0)$ & $[] \rightarrow b_1$ & & $(3,1,4,2,5)$ & & \\
		$(7.1\text{-}3)$ & $b_1, a_1, b_3, a_4, b_2, (a_1)$ & $[5,4,4,4,6]$ & & & \\
		\hline
		$(8.0)$ & $b_1$ & $[4,4,4,4,6]$ & $(2,1,4,3,5)$ & & \\
		$(8.1\text{-}2)$ & $b_1, a_1, b_2, a_2, (b_1)$ & $[5,5,4,4,6]$ & & & \\
		\hline
		$(9.0)$ & $[]\rightarrow b_1$ & & $(1,2,4,3,5)$ & & \\
		\hline
		$(10.0)$ & $[]\rightarrow b_2$ & $[6,5,4,4,6]$ & & $b_1, b_5$ & $2$ \\
		\hline
		$(11.0)$ & $[]\rightarrow b_3$ & $[6,6,4,4,6]$ & & $b_1, b_2, b_5$ & $3$ \\
		$(11.1\text{-}2)$ & $b_3, a_3, b_4, a_4, (b_3)$ & $[6,6,5,5,6]$ & & & \\
		\hline
		$(12.0)$ & $[]\rightarrow b_3$ & & $(1,2,3,4,5)$ & & \\
		\hline
		$(13.0)$ & $[]\rightarrow b_4$ & $[6,6,6,5,6]$ & & $b_1, b_2, b_3, b_5$ & $4$ \\
		\hline
		$(14.0)$ & $[]$ & $[6,6,6,6,6]$ & & $b_1, b_2, b_3, b_4, b_5$ & $\infty$ \\
	\end{tabular}
	\caption{Iterations of \texttt{school-rotate-remove} of Example~\ref{ex:rrr-fast}.}
	\label{tab:rrr-fast}
	\end{table}
	
	In Table~\ref{tab:rrr-fast}, we outline the updates occurred at all steps (denoted by $j$) of all iterations (denoted by $i$) during the fast execution of \texttt{school-rotate-remove}. A cell is left blank if no update happens. The steps of iteration $i$ illustrate the steps in building the directed path $P^i$. $W^i$ can be easily deduced from $P^i$ and is therefore not included in the table.
    
    The main idea of the construction is that, in order to find a directed cycle or a sink in a digraph, it suffices to follow a path. (i) allows us to carry over information on such paths from one iteration of the algorithm to the next. (ii) and (v) allow us to extend such path quickly, without going through the full preference lists of agents again. (iii) and (iv) allow a quick detection when a sink or a cycle has been found when following a path.

	When extending the directed path $P^i$, if $P^i=[]$, as in $(0.0)$ and $(10.0)$, we add the first school not in $T^i$ to the directed path, which is achieved by repeatedly checking if $b_f\in T^i$ and while so, updating $f:=f+1$. If $P^i$ is non-empty with $b$ at the tail, and $b$ is not the node corresponding to $\emptyset$, we rely on $p_b$ to find $s_{M^i}(b)$. That is, we repeatedly update $p_b:=p_b+1$ until either $p_b>5$ or $a:=b(p_b)$ satisfies $b>_a M^i(a)$. So $p_b$ strictly increases every time an extension happens with $b$ at the tail. The only time that $p_b$ will decrease is when $b$ points to a directed cycle, as the school $b_1$ in $(5.1)$ and $(7.3)$. In such case, $p_b$ is decremented by $1$ after the rotation elimination, as seen in $(6.0)$ and $(8.0)$. This is because it is possible to have $s_{M^{i+1}}(b) = s_{M^i}(b)$. There are two scenarios, corresponding to Step~\ref{step:sink-found} and Step~\ref{step:rotation-found} in Algorithm~\ref{alg:rar}, where we stop extending the directed path $P^i$: one is when the tail $b$ is a sink, implied by having $p_b>5$; the other is when the additional node is already in the directed path, which can be checked against $W_i$. In the latter case, such nodes are written as $(\text{node})$ in Table~\ref{tab:rrr-fast}.  
	\hfill {{$\diamondsuit$}}  
\end{example}

%
%
% ------------------------------------------
%
%

\section{An $O(|E|)$ Algorithm for Computing $G_L$} \label{sec:G-L-efficient}

Throughout the section, we fix an instance $(G,<,\q)$ with $G=(A\cup B, E)$ and abbreviate $\SS:=\SS(G,<,\q)$. We start with a preliminary fact. Recall that we denote by $\Rot(G,<,\q)$ (resp. $\SR(G,<,\q)$) the set of student-rotations (resp. school-rotations) exposed in some stable assignment of $(G,<,\q)$. Let $G_L$ be the subgraph of $G$ as defined in Theorem~\ref{thr:main1}.

\begin{lemma}\label{lem:rotations}
	Let $e$ be an illegal edge of $(G,<,\q)$, and $\widetilde G=G[E \setminus \{e\}]$. Then $\Rot(G,<,\q)\subseteq \Rot(\widetilde G,<,\q)$ and $\SR(G,<,\q)\subseteq \SR(\widetilde G,<,\q)$. 
\end{lemma}

\proof{Proof.}
    Fix $M \in \SS$. Since $\SS\subseteq \SS(\widetilde G,<,\q)$, $M$ is also a stable assignment of $(\widetilde G,<,\q)$. First consider any student-rotation $\rho \in \Rot(G, <, \q)$ exposed in $M$. We want to show that $\rho$ is also exposed in $M$ in $(\widetilde G,<,\q)$. Assume $\rho=b_0, a_0, b_1, a_1, \cdots, b_{r-1}, a_{r-1}$. By Lemma~\ref{thr:256-ext}, edges $a_ib_{i+1}$ and $a_{i+1}b_{i+1}$ for all $i=0, 1, \cdots, r-1$, are stable and therefore legal. Hence, all such edges are in $E(\widetilde{G})$, implying that $b_{i+1} = s_M(a_i)$ and $next_M(a_i) = a_{i+1}$ hold in $(\widetilde G, <, \q)$ as well. Thus, $\rho \in \Rot(\widetilde G,<,\q)$. A similar argument  shows $\SR(G,<,\q)\subseteq \SR(\widetilde G,<,\q)$.
\hfill{\Halmos}\endproof

\begin{theorem}\label{thr:main2}
	$G_L$ can be found in time $O(|E|)$.
\end{theorem}

\proof{Proof.} 
    By Theorem~\ref{thr:main1} and Lemma~\ref{thr:256-ext}, $E(G_L)$ is given by edges in $M_0^{\L}$, plus all pairs $a_i b_{i+1}$ for some student-rotation $\rho=b_0,a_0,\dots,a_k \in \Rot(G_L,<,\q)$. By Lemma \ref{thm:255-ext}, there exists exactly one set $\Rot_1$ of student-rotations whose elimination leads from $M^{\L}_0$ to $M_0$; one set $\Rot_2$ leading from $M_0$ to $M_z$; and one set $\Rot_3$ leading from $M_z$ to $M_z^{\L}$; and their union gives $\Rot(G_L,<,\q)$. We argue that $\Rot_3$ is computed during the execution of \texttt{student-rotate-remove}. Indeed, throughout the algorithm, a sequence of rotations is found and eliminated, leading from $M_z$ to $M^{\cal L}_z$. Each of these is exposed in some stable assignment in an instance that contains all legal edges. Hence, by repeated application of Lemma~\ref{lem:rotations}, those rotations form set ${\cal R}_3$. They can be computed in time $O(|E|)$ by Theorem \ref{thr:time-complexity-rotate-remove}. By Lemma~\ref{thm:255-ext} and repeated applications of Lemma \ref{lem:rotations}, $\Rot_2$ coincides with the set $\Rot(G,<,\q)$, which can be computed in time $O(|E|)$ by classical algorithms, see, e.g., \cite{GI}. \texttt{school-rotate-remove} computes in time $O(|E|)$, again by Theorem~\ref{thr:time-complexity-rotate-remove}, the set of school-rotations $\SR_1$ whose sequential elimination starting from $M_0$ leads to $M_0^{\L}$. By Lemma~\ref{lem:students-rotations-and-school-rotation}, the set $\Rot_1$ can be obtained from $\SR_1$ via the bijection $\sigma$. Since $\sigma$ maps $\rho:=b_0,a_0, b_1,a_1, \cdots,a_{r-1} \in \Rot_1$ to $\sigma(\rho):= a_0,b_1, a_1,\cdots, a_{r-1},b_0 \in \SR_1$, computing $\Rot_1$ from $\SR_1$ takes time $O(|E|)$, concluding the proof.
\hfill{\Halmos}\endproof

%
%
% ------------------------------------------
%
%

\section{An $O(|E|)$ Algorithm for EADAM with Consent} \label{SEC:CONNECT-TO-EADAM}

In this section, we first formally introduce EADAM with consent~\cite{Kesten}. Then in Section~\ref{sec:rrr-modified} we show that a fast implementation of EADAM can be achieved by a suitable modification of our \texttt{school-rotate-remove} algorithm. The proof relies on a simplified and outcome-equivalent version of EADAM introduced by~\citet{Tang.Yu}. Thus, we defer the proof as well as a formal introduction of simplified EADAM to Appendix~\ref*{app:seadam-outcomeequiv}. Together with Theorem \ref{thr:time-complexity-rotate-remove}, this implies the following. 

\begin{theorem}\label{thr:main3}
	EADAM with consent on a stable assignment instance $(G(A\cup B,E),<,\q)$ can be implemented as to run in time $O(|E|)$.
\end{theorem}

We also compare our algorithm with previous versions of EADAM through computational experiments. In Section~\ref{sec:computation-compare}, the theoretical advantage of \texttt{student-rotate-remove} is verified computationally on random instances.

%
%
% ------------------------------------------
%
%

\subsection{Kesten's EADAM}	

Recall that Gale-Shapley's algorithm (with students proposing) is executed in successive steps. During each step, every student that is currently unmatched applies to the first school in his preference list that he has not yet applied to, and gets either temporarily accepted or rejected. A student $a$ is called an \emph{interrupter} (\emph{for school $b$}, \emph{at step $k'$}) if: $a$ is temporarily accepted by school $b$ at some step $k<k'$; $a$ is rejected by school $b$ at step $k'$; and there exists a student that is rejected by school $b$ during steps $k, k+1, \cdots, k'-1$. In such case, we will also call $ab$ an \emph{interrupting pair} (\emph{at step $k'$}). Informally speaking, an interrupter is a student who, by applying to school $b$, interrupts a desirable assignment between school $b$ and another student at no gain to himself. Removing such interruptions is crucial in neutralizing their adverse effects on the outcome. 

Kesten's EADAM algorithm takes as input an instance $(G,<,\mathbf{q})$ with $G=(A\cup B, E)$ and a set $\overline{A}\subseteq A$ of students which we call \emph{consenting}. Each iteration of EADAM starts by running Gale-Shapley's algorithm from scratch. It then removes from the graph certain interrupting pairs involving consenting interrupters. The algorithm terminates when there are no interrupting pairs whose corresponding interrupters are consenting students.

Details of Kesten's algorithm can be found in Algorithm~\ref{alg:Kesten-EADAM} and an example of EADAM can be found later in Example~\ref{ex:eadam-seadam}.

\begin{algorithm}
	\caption{Kesten's EADAM\label{alg:Kesten-EADAM}}
	\begin{algorithmic}[1]
		\REQUIRE $(G(A\cup B,E),<,\q)$, consenting students $\overline{A}\subseteq A$
		\STATE Let $G^0=G$, $i=0$.
		\STATE Run student-proposing Gale-Shapley's algorithm on $(G^i,<,\q)$ to obtain assignment $M^i$. 
		\WHILE {there is a consenting interrupter}
		\STATE Identify the maximum $k'$ such that there exists a consenting interrupter at step $k'$. \label{step:last-consent-intr} % rejected by a school for which $a$ is an interrupter. 
		\STATE Let $E'$ be the set of all interrupting pairs $ab$ at step $k'$ such that $a$ is consenting. \label{step:last-consent-intr-pairs}
		\STATE Define $G^{i+1}$ from $G^i$ by removing edges in $E'$. Set $i=i+1$.
		\STATE Run student-proposing Gale-Shapley's algorithm on $(G^i,<,\q)$ to obtain assignment $M^i$. 
		\ENDWHILE
		\STATE{Output $M^{i}$.}
	\end{algorithmic}
\end{algorithm}

The following theorem collects some results from~\citet{Kesten} and~\citet{Tang.Yu}, demonstrating the transparency of the consenting incentives and some attractive properties of EADAM's output. Recall that an assignment $M$ is \emph{constrained efficient} if it does not violate any nonconsenting students' priorities\footnote{A student $a$'s priority is violated at assignment $M$ if there is a school $b$ such that $ab$ is a blocking pair of $M$.}, but any other assignment $M'\succeq M$ does. 

\begin{theorem} \label{thm:sEADAM}
    Under Kesten's EADAM mechanism:
    \begin{enumerate}
        \item \label{lem:underdemanded-fixed} The assignment of a student does not change whether he consents or not. That is, for any student $a \in A$ and any set of consenting students $\bar A\subseteq A$, if $M$ and $M'$ are the outputs of EADAM on inputs $\big\{(G,<,\mathbf{q}), \overline{A}\setminus \{a\}\big\}$ and $\big\{(G,<,\mathbf{q}), \overline{A}\big\}$ respectively, then $M(a)=M'(a)$.
        \item \label{lem:sEADAM-Pareto} The output is Pareto-efficient when all students consent and is constrained efficient otherwise.
    \end{enumerate}
\end{theorem}

\begin{example}\label{ex:eadam-seadam} 
	Each school in this example has a quota of $1$. Their preference lists are given below. All students are consenting except for $a_3$. $$\begin{array}{clccl}
	a_1: & b_1 >b_2>b_3>b_4 & \qquad \qquad & b_1: &  a_4 >a_2>a_1>a_3 \\
	a_2: & b_1 >b_2>b_3>b_4 & & b_2: & a_2 >a_3>a_1>a_4 \\
	a_3: & b_3 >b_2>b_4>b_1 & & b_3: & a_1 >a_4>a_3>a_2 \\
	a_4: & b_3 >b_1>b_2>b_4 & & b_4: & a_3 >a_1>a_2>a_4 
	\end{array}$$
	
	\noindent \textbf{Gale-Shapley's algorithm:} The	student-proposing Gale-Shapley's algorithm outputs the assignment $M^0 =\{a_1b_3, a_2b_2, a_3b_4, a_4b_1\}$. Steps of the algorithm are given below:
	$$\begin{array}{cccclclcl}
	\text{step} & & b_1 & & b_2 & & b_3 & & b_4 \\
	\hline
	1 & \quad \qquad & \xcancel{a_1}, \mathbf{a_2} & \quad \qquad &  & \quad \qquad & \xcancel{a_3}, \mathbf{a_4} & \quad \qquad  &\\
	2 & &  & & \xcancel{a_1}, \mathbf{a_3} & & & & \\
	3 & &  & &  & & a_1, \xcancel{\mathbf{a_4}}& & \\
	4 & & \xcancel{\mathbf{a_2}}, a_4 & & & & & & \\
	5 & &  & & a_2, \xcancel{\mathbf{a_3}} & & & & \\
	6 & &  & &  & &  & & a_3 \\
	\end{array}$$
	
	\noindent \textbf{Iteration \#1:} From the steps of Gale-Shapley's algorithm, one can identify all interrupting pairs. For instance, $a_2$ proposes to $b_1$ at step 1. This causes $a_1$ to be rejected by $b_1$. However, $a_2$ is later rejected by $b_1$ at step $4$. Thus, by definition, $a_2b_1$ is an interrupting pair at step 4. 
	
	In total, there are three interrupting pairs, $a_3 b_2, a_2 b_1, a_4 b_3$, from the last step to the first. The last interrupting pair of a consenting interrupter is $a_2b_1$, given that $a_3$ is not a consenting student. Thus, $k'=4$. Since there is only one interrupting pair at step $k'=4$, EADAM algorithm simply removes $a_2 b_1$ from the instance. On the new instance, EADAM re-runs Gale-Shapley's algorithm:
	$$\begin{array}{cccclclcl}
	\text{step} & & b_1 & & b_2 & & b_3 & & b_4 \\
	\hline
	1 & \quad \qquad & a_1 & \quad \qquad & a_2 & \quad \qquad & \xcancel{a_3}, a_4 & \quad \qquad &\\
	2 & &  & & \xcancel{a_3}, a_2 & & & & \\
	3 & &  & &  & &  & & a_3 
	\end{array}$$
	The resulting assignment is $M^1=\{a_1b_1, a_2b_2, a_3b_4, a_4b_3\}$.
	
	\smallskip \noindent \textbf{Iteration \#2:} There are no interrupting pairs, thus no consenting interrupters. Hence, EADAM terminates and outputs assignment $M^1$.
	
	Note that using tools developed in previous sections, one can show that $a_2b_1$, the first edge that is removed by EADAM, is actually a legal edge and the assignment output of EADAM, $M^1$, is not a legal assignment. 
	\hfill {$\diamondsuit$}
\end{example}

%
%
% ------------------------------------------
%
%

\subsection{\texttt{School-Rotate-Remove with Consent}} \label{sec:rrr-modified}

\citet{Morr} showed that, when all students consent, the output of EADAM is the student-optimal legal assignment. Hence, \texttt{school-rotate-remove} can be employed to find this assignment in time $O(|E|)$ (see Theorem~\ref{thr:time-complexity-rotate-remove}). However, as Example~\ref{ex:eadam-seadam} shows, when only some students consent, EADAM may output an assignment that is not legal. We show in this section how to suitably modify \texttt{school-rotate-remove} in order to obtain the assignment output of EADAM for any given set of consenting students, without sacrificing the running time.

In \texttt{school-rotate-remove}, the key idea is to reroute arcs that point to students who are assigned to sinks in the rotation digraph. This allows us to identify school-rotations in the underlying legalized instance. Assume for instance that $(b',a), (a,b)\in A(D_B)$, and $b$ is a sink. Upon such rerouting, $a$'s priority might be violated. In particular, if $b'$ successfully participates in a school-rotation after the rerouting, then $ab'$ will be a blocking pair for the new assignment. Hence, under the EADAM framework, if $a$ is not consenting, we can no longer freely reroute arcs pointing to $a$. In fact, in order to respect $a$'s priority (i.e., to avoid $ab'$ becoming a blocking pair), $b'$ cannot be assigned to any student $a'$ such that $a>_{b'} a'$. This means that the arc coming out of $b'$ cannot be rerouted to any other student, essentially marking $b'$ a sink. 

A detailed description of our algorithm is presented in Algorithm~\ref{alg:rrr-consent}. Throughout the rest of the section, we call school-rotations simply rotations. As in Algorithm~\ref{alg:rar}, when both cases (i) and (ii) are present at Step~\ref{step:rot-or-sink} of some iteration, we are free to choose between Step~\ref{clause:illegal-edge} and Step~\ref{clause:rotation}.  These choices do not affect the final assignment output, as shown in Theorem~\ref{thm:rrr-consent-unique}. A fast implementation is provided later in Section~\ref{sec:fast-rrr-consent}. A step-by-step application of our algorithm on the instance from Example~\ref{ex:eadam-seadam} is outlined in Example~\ref{ex:rrr-consent}.

\begin{algorithm}[h!t]
\caption{\texttt{school-rotate-remove with consent}} \label{alg:rrr-consent}
\begin{algorithmic}[1] 
	\REQUIRE $(G(A\cup B,E),<,\q)$, consenting students $\overline{A}\subseteq A$
	\STATE{Find the student-optimal stable assignment $M_0$ of $(G,<,\q)$ via Gale-Shapley's algorithm.}  \label{step:find-optimal}
	\STATE{Let $G^0:=G$ and $M^0:=M_0$.}
	\STATE{Set $i=0$ and let $D^0$ to be the school-rotation digraph of $M^0$ in $(G^0,<,\q)$.}
	\WHILE{$D^i$ still has an arc}	\label{step:while}
	\STATE Find (i) arcs $(b',a)$ and $(a,b) \in A(D^i)$ where $b$ is a sink in $D^i$, or  (ii) a cycle $C^i$ of $D^i$. \label{step:rot-or-sink}
	\IF{(i) is found}\label{clause:illegal-edge}
	\STATE Define $G^{i+1}$ from $G^{i}$ by removing $ab'$, and set $M^{i+1} = M^i$.  \label{step:remove-illegal-edge}
	\IF {$a\notin \overline{A}$}
	\STATE Remove from $G^{i+1}$ edges $a'b'$ for all $a'$ such that $a>_{b'} a'$.\label{step:remove-nonconsent-edge}
	\ENDIF
	\ELSIF{(ii) is found}\label{clause:rotation}
	\STATE Let $\rho^i$ be the corresponding school-rotation. Set $M^{i+1}=M^i / \rho^i$, and $G^{i+1}=G^i$.
	\ENDIF 
	\STATE{Set $i=i+1$ and let $D^i$ to be the school-rotation digraph of $M^i$ in $(G^i,<,\q)$.}
	\ENDWHILE
	\STATE{Output $M^i$.}
\end{algorithmic}
\end{algorithm}

\begin{example}\label{ex:rrr-consent} 
	Consider the instance given in Example~\ref{ex:eadam-seadam}. From the student-optimal stable assignment $M^0:=\{a_1b_3, a_2b_2, a_3b_4, a_4b_1\}$, we can construct the rotation digraph as below. Note that in this graph and in the following, some isolated nodes are not included.
	
	\begin{center}
		\centering
		\begin{tikzpicture}[scale = 1]
		\node[] () at (0,0.5) {};
		\node[circle, draw] (1) at (0,0) {$b_3$};
		\node[rectangle, draw] (2) at (1.5, 0) {$a_4$};
		\node[circle, draw] (3) at (2.5,0) {$b_1$};
		\node[rectangle, draw] (4) at (4, 0) {$a_2$};
		\node[circle, draw] (5) at (5,0) {$b_2$};
		\node[rectangle, draw] (6) at (6.5, 0) {$a_3$};
		\node[circle, draw] (7) at (7.5,0) {$b_4$};
		\draw[solid, thick, ->] (1) -- (2);
		\draw[solid, thick, ->] (2) -- (3);
		\draw[solid, thick, ->] (3) -- (4);
		\draw[solid, thick, ->] (4) -- (5);
		\draw[solid, thick, ->] (5) -- (6);
		\draw[solid, thick, ->] (6) -- (7);
		\end{tikzpicture}
	\end{center}
    
    \noindent\textbf{Iteration \#1:} Since $b_4$ is a sink, we will remove edge $a_3b_2$ as in Step~\ref{step:remove-illegal-edge}, in the hope of rerouting the arc coming out of $b_2$. However, because $a_3$ is not consenting, we have to additionally remove edges $a_1b_2$ and $a_4b_2$ as in Step~\ref{step:remove-nonconsent-edge}. This completely removes the possibilities of rerouting, essentially making $b_2$ a sink, as seen in the rotation digraph of the updated instance:
	
	\begin{center}
		\centering
		\begin{tikzpicture}[scale = 1]
		\node[] () at (0,0.5) {};
		\node[circle, draw] (1) at (0,0) {$b_3$};
		\node[rectangle, draw] (2) at (1.5, 0) {$a_4$};
		\node[circle, draw] (3) at (2.5,0) {$b_1$};
		\node[rectangle, draw] (4) at (4, 0) {$a_2$};
		\node[circle, draw] (5) at (5,0) {$b_2$};
		\node[circle, draw] (7) at (7.5,0) {$b_4$};
		\draw[solid, thick, ->] (1) -- (2);
		\draw[solid, thick, ->] (2) -- (3);
		\draw[solid, thick, ->] (3) -- (4);
		\draw[solid, thick, ->] (4) -- (5);
		\end{tikzpicture}
	\end{center}

    \noindent\textbf{Iteration \#2:} Now, $b_2$ is a sink. Since its assigned student $a_2$ is consenting, the algorithm simply removes edge $a_2b_1$ in Step~\ref{step:remove-illegal-edge}, resulting in the following updated rotation digraph:
	
	\begin{center}
		\centering
		\begin{tikzpicture}[scale = 1]
		\node[] () at (0,0.5) {};
		\node[circle, draw] (1) at (0,0) {$b_3$};
		\node[rectangle, draw] (2) at (1.5, 0) {$a_4$};
		\node[circle, draw] (3) at (2.5,0) {$b_1$};
		\node[rectangle,draw](8) at (1,-1) {$a_1$};
		\node[circle, draw] (5) at (5,0) {$b_2$};
		\node[circle, draw] (7) at (7.5,0) {$b_4$};
		\draw[solid, thick, ->] (1) -- (2);
		\draw[solid, thick, ->] (2) -- (3);
		\draw[solid, thick, ->] (3) -- (8);
		\draw[solid, thick, ->] (8) -- (1);
		\end{tikzpicture}
	\end{center}

    \noindent\textbf{Iteration \#3:} We can now eliminate the school-rotation (i.e., trading schools between $a_1$ and $a_4$), and update the assignment to be $\{a_1b_1, a_2b_2, a_3b_4, a_4b_3\}$. After the assignment update, the new rotation digraph only contains sinks, and thus the algorithm terminates. 
    This final assignment coincides with the assignment output of EADAM.
	\hfill {$\diamondsuit$}	
\end{example}

In our rotation-based algorithm, the students from whom we seek consent are those who are assigned to schools corresponding to sinks, and thus they are not in any directed cycles in the current and subsequent rotation digraphs. Therefore, there is a clear separation between the students from whom we ask for consent and those participating in Pareto-improvement cycles (i.e., school-rotations). This is consistent with the result in Theorem~\ref{thm:sEADAM}, part~\ref{lem:underdemanded-fixed} that students have no incentive to not consent.

The proof of the following statement can be found in Appendix~\ref*{app:seadam-outcomeequiv}. 

\begin{theorem}\label{thm:rrr-eadam-equiv}
	For any given input, the outputs of Algorithm~\ref{alg:rrr-consent} and Algorithm~\ref{alg:Kesten-EADAM} coincide. 
\end{theorem}

We remark that the proof of Theorem~\ref{thm:rrr-eadam-equiv} is different (and quite harder) than the proof of Theorem~\ref{thr:rotate-remove-finds-legal}. Indeed, in the latter case we can build on the fact that legal assignments form a lattice, while in the former we do not have such a well-behaved structural result at our disposal. Hence, a more careful analysis of the algorithms is needed.

%
%
% ------------------------------------------
%
%

\subsection{Fast Implementation of \texttt{School-Rotate-Remove with Consent}}\label{sec:fast-rrr-consent}

The fast implementation is heavily based on the implementation presented in Section~\ref{sec:rrr-time-complexity}. Therefore, we defer the proof of Lemma~\ref{lem:rrr-conset-impl} to Appendix~\ref*{app:proof:rrr-consent-impl}, but demonstrate it in Example~\ref{ex:rrr-consent-fast}.

\begin{lemma}\label{lem:rrr-conset-impl}
	Algorithm~\ref{alg:rrr-consent} can be implemented as to run in time $O(|E|)$.
\end{lemma}

\begin{example} \label{ex:rrr-consent-fast}
	Consider the instance in Example~\ref{ex:rrr-fast}. Assume $a_5$ is not consenting. In Table~\ref{tab:rrr-consent-fast}, we outline the updates, similar to those in Example~\ref{ex:rrr-fast}. When school $b$ points to the nonconsenting student $a_5$ (whose partner $b_5$ is a sink) in the rotation digraph, in addition to remove $a_5$ and $b_5$ from the directed path $P^i$, we also remove $b$ from $P^i$, set $T^i:=T^i\cup \{b\}$, and update $p_b := 6$ in lieu of the edge removals in Step~\ref{step:remove-nonconsent-edge}. Such updates can be seen in $(1.0)$, $(2.0)$, $(3.0)$, and $(4.0)$. 
    \hfill {$\diamondsuit$}	
\end{example}
	
\begin{table}[h!t]
	\centering
	\setlength{\tabcolsep}{8pt}
	\begin{tabular}{llllll}
		$(i.j)$ & $P^i$ & $\{p_b\}_{b\in B}$ & $M^i$ & $T^i$ & $f$ \\
		\hline
		$(0.0)$ & $[] \rightarrow b_1$ & $[1,1,1,1,1]$ & $(4,3,2,1,5)$ & $\emptyset$ & 1 \\
		$(0.1)$ & $b_1, a_5, b_5$ & $[2,1,1,1,1]$ & & & \\
		\hline
		$(1.0)$ & $[] \rightarrow b_2$ & $[6,1,1,1,6]$ & & $b_1, b_5$ & 2 \\
		$(1.1)$ & $b_2, a_5, b_5$ & $[6,2,1,1,6]$ & & & \\
		\hline
		$(2.0)$ & $[] \rightarrow b_3$ & $[6,6,1,1,6]$ & & $b_1, b_2, b_5$ & 3 \\
		$(2.1)$ & $b_3, a_5, b_5$ & $[6,6,2,1,6]$ & & & \\
		\hline
		$(3.0)$ & $[] \rightarrow b_4$ & $[6,6,6,1,6]$ & & $b_1, b_2, b_3, b_5$ & 4 \\
		$(3.1)$ & $b_4, a_5, b_5$ & $[6,6,6,2,6]$ & & & \\
		\hline
		$(4.0)$ & $[]$ & $[6,6,6,6,6]$ & & $b_1, b_2, b_3, b_4, b_5$ & $\infty$ \\
	\end{tabular}
	\caption{Iterations of \texttt{school-rotate-remove with consent} of Example~\ref{ex:rrr-consent-fast}}
	\label{tab:rrr-consent-fast}
\end{table}

\proof{Proof of Theorem~\ref{thr:main3}.} 
    It follows immediately from Theorem~\ref{thm:rrr-eadam-equiv} and Lemma~\ref{lem:rrr-conset-impl}.
\hfill{\Halmos}\endproof

%
%
% ------------------------------------------
%
%

\subsection{Computational Experiments} \label{sec:computation-compare}

\begin{figure}[!t]
\centering
\begin{subfigure}{.48\textwidth}
	\centering
	\includegraphics[width=\linewidth]{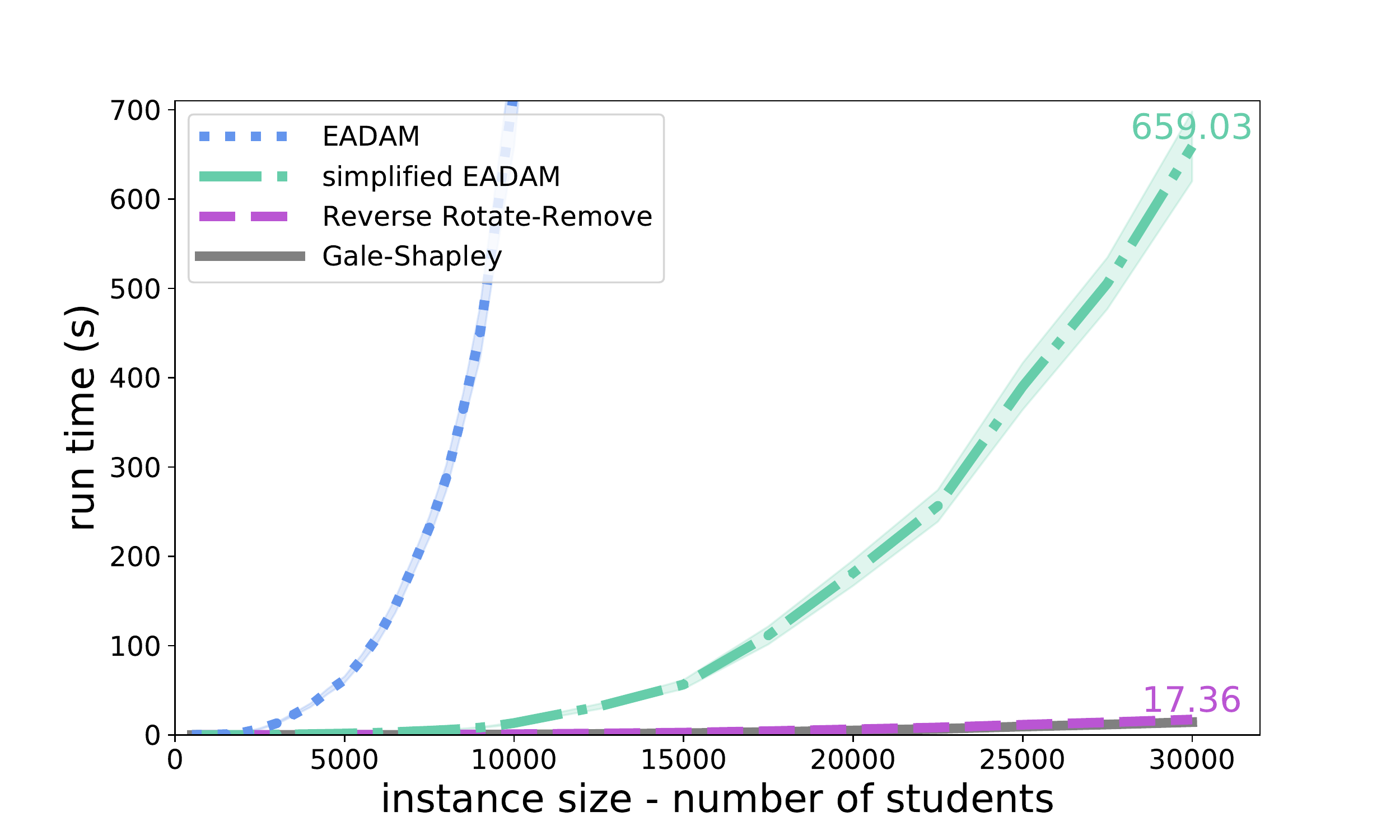}
	\caption{all students consent}
\end{subfigure}%
\hspace*{\fill}
\begin{subfigure}{.48\textwidth}
	\centering
	\includegraphics[width=\linewidth]{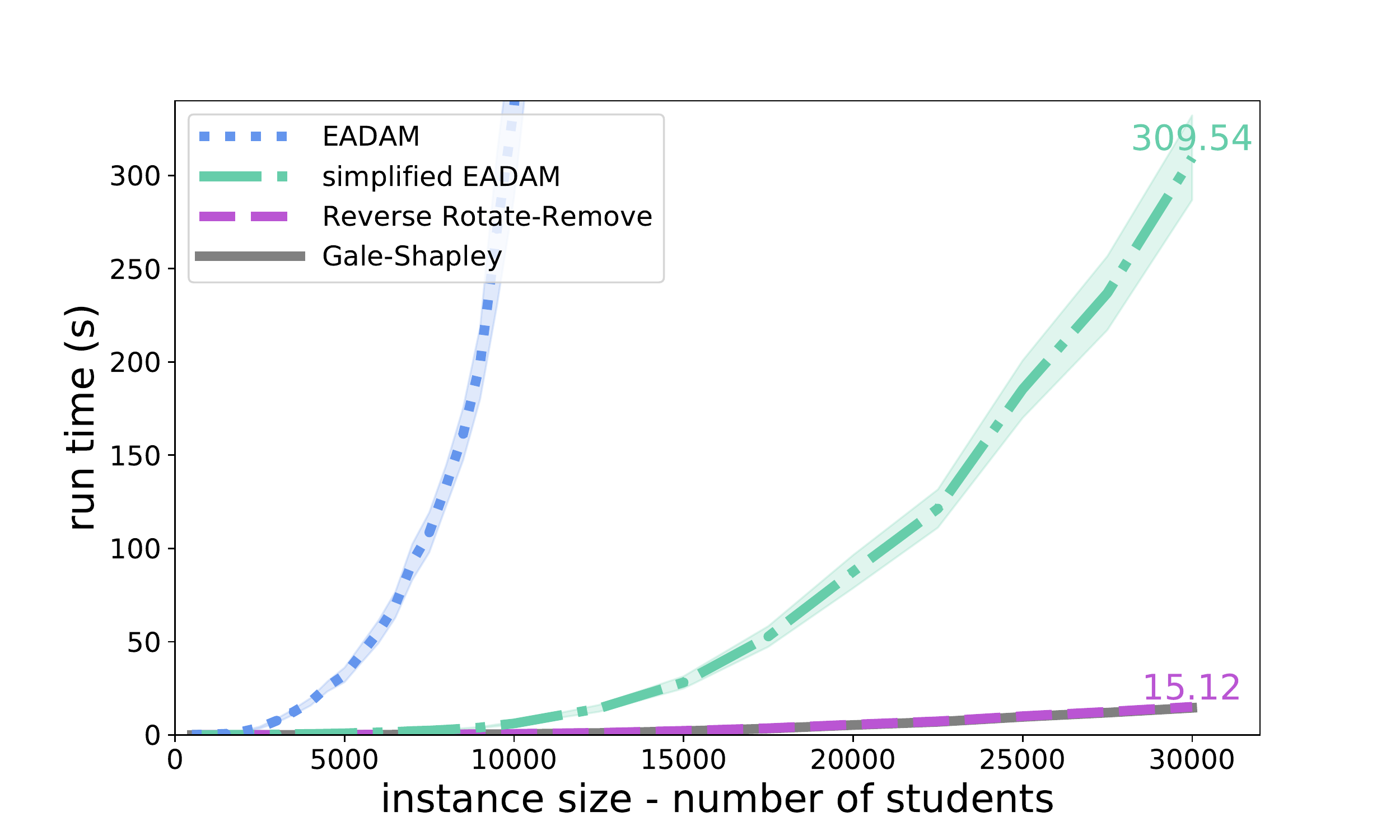}
	\caption{$80\%$ students consent}
\end{subfigure}%

\begin{subfigure}{.48\textwidth}
	\centering
	\includegraphics[width=\linewidth]{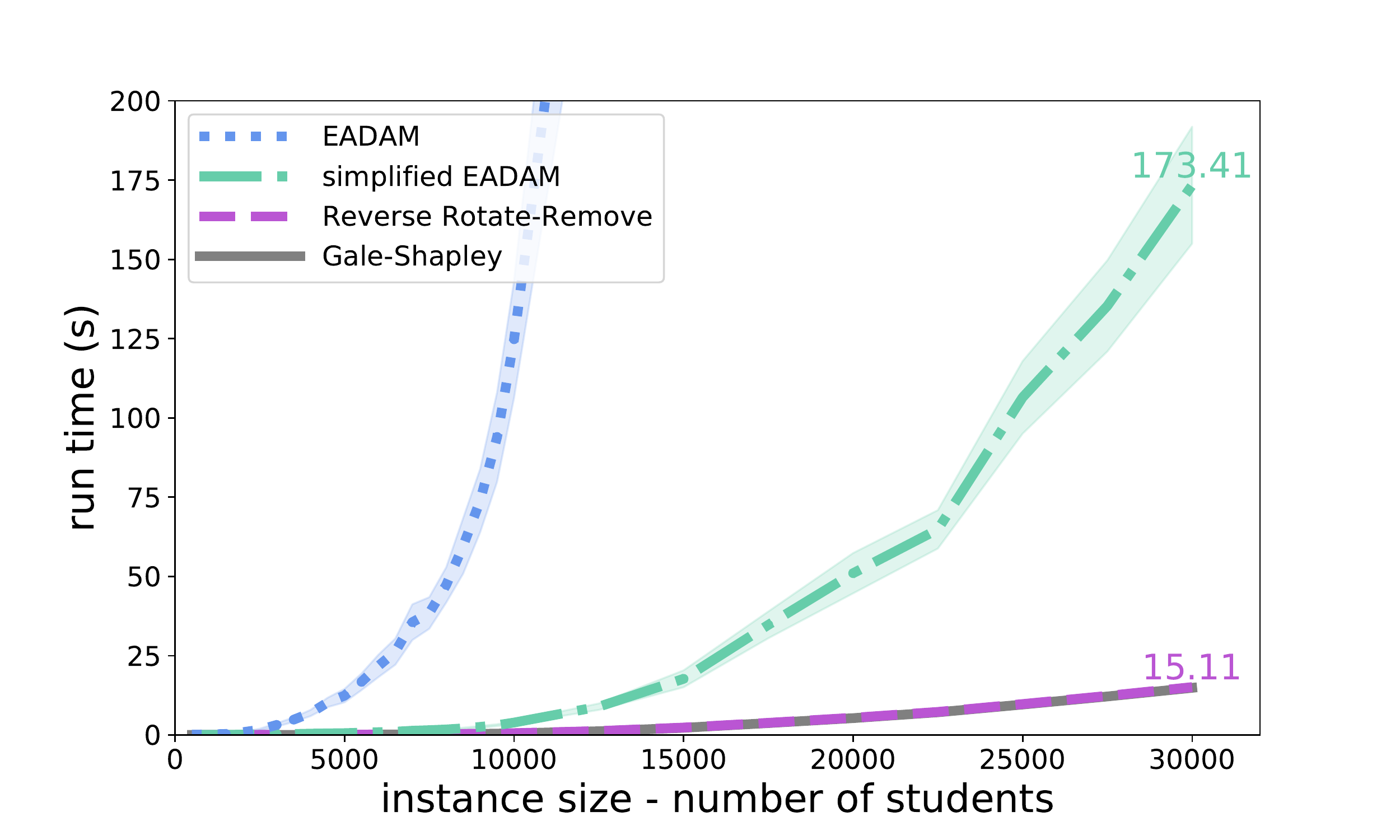}
	\caption{$50\%$ students consent}
\end{subfigure}%
\hspace*{\fill}
\begin{subfigure}{.48\textwidth}
	\centering
	\includegraphics[width=\linewidth]{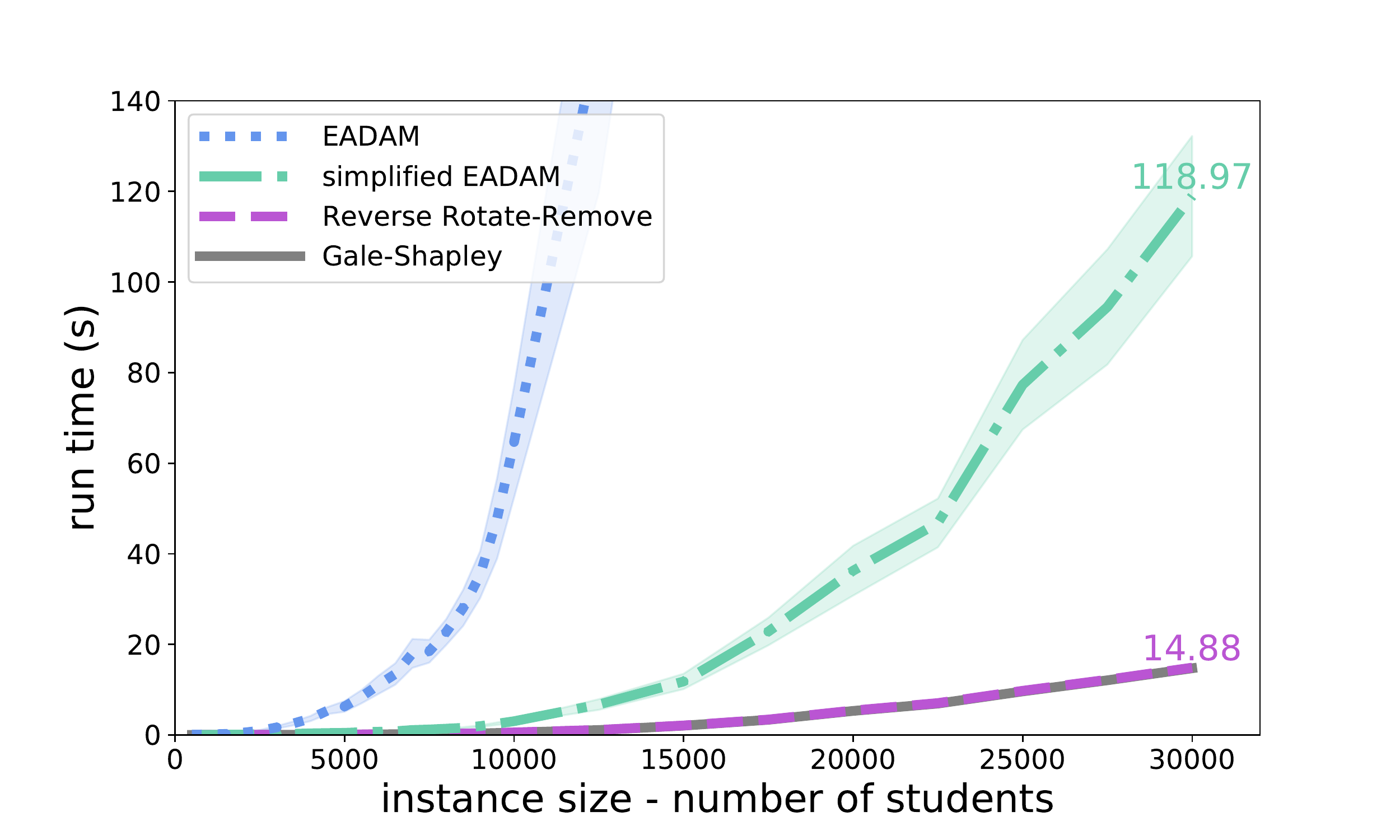}
	\caption{$30\%$ students consent}
\end{subfigure}%

\begin{subfigure}{.5\textwidth}
	\centering
	\includegraphics[width=\linewidth]{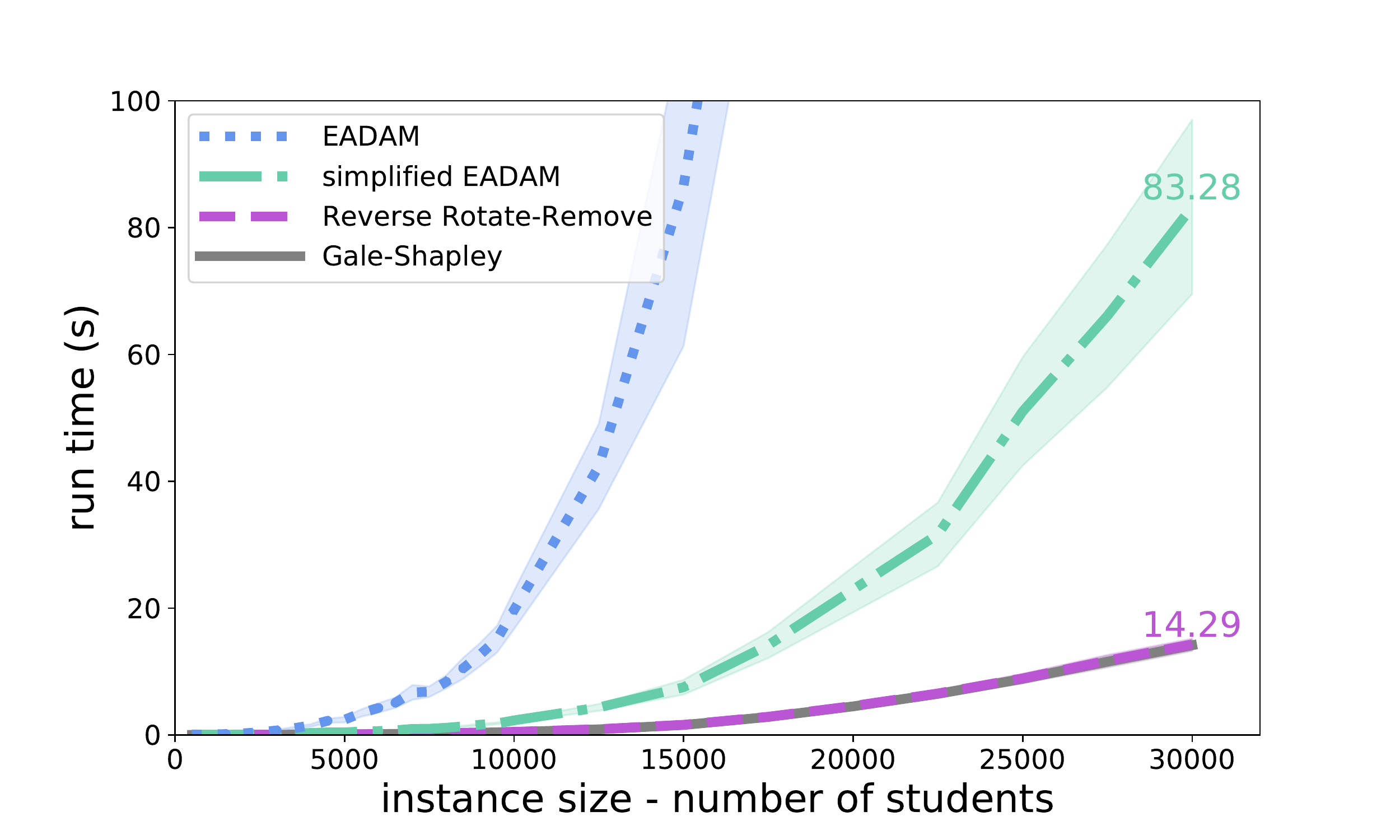}
	\caption{$10\%$ students consent}
\end{subfigure}
\caption{\small Comparing EADAM, simplified EADAM, and \texttt{school-rotate-remove with consent} in the one-to-many setting on random instances of varying sizes. Average run time of simplified EADAM and \texttt{school-rotate-remove with consent} included for the largest instance in our experiment.} \label{fig:algo-compare}
\end{figure}

Since Gale-Shapley's algorithm on stable assignment instances can be implemented to run in time $O(|E|)$ by Lemma~\ref{lem:GS-complexity}, the original EADAM algorithm~\citep{Kesten} runs in time $O(|E|^2)$ because it runs Gale-Shapley's routine iteratively for at most $|E|$ times. Moreover, a simplified version of EADAM~\cite{Tang.Yu}, for which the details are presented in the Appendix, runs in time $O(|E||V|)$ because it runs Gale-Shapley's routine iteratively for at most $|V|$ times. We remark that although mechanism design, rather than computational complexity, is the primary interest of Kesten's paper, computational efficiency is nevertheless crucial in putting the mechanism into practice, especially for large markets such as the New York school system. In fact, \citet{Tang.Yu} mentioned computational tractability as one of their contributions.

\begin{figure}[!t]
\centering
\begin{subfigure}{.48\textwidth}
	\centering
	\includegraphics[width=\linewidth]{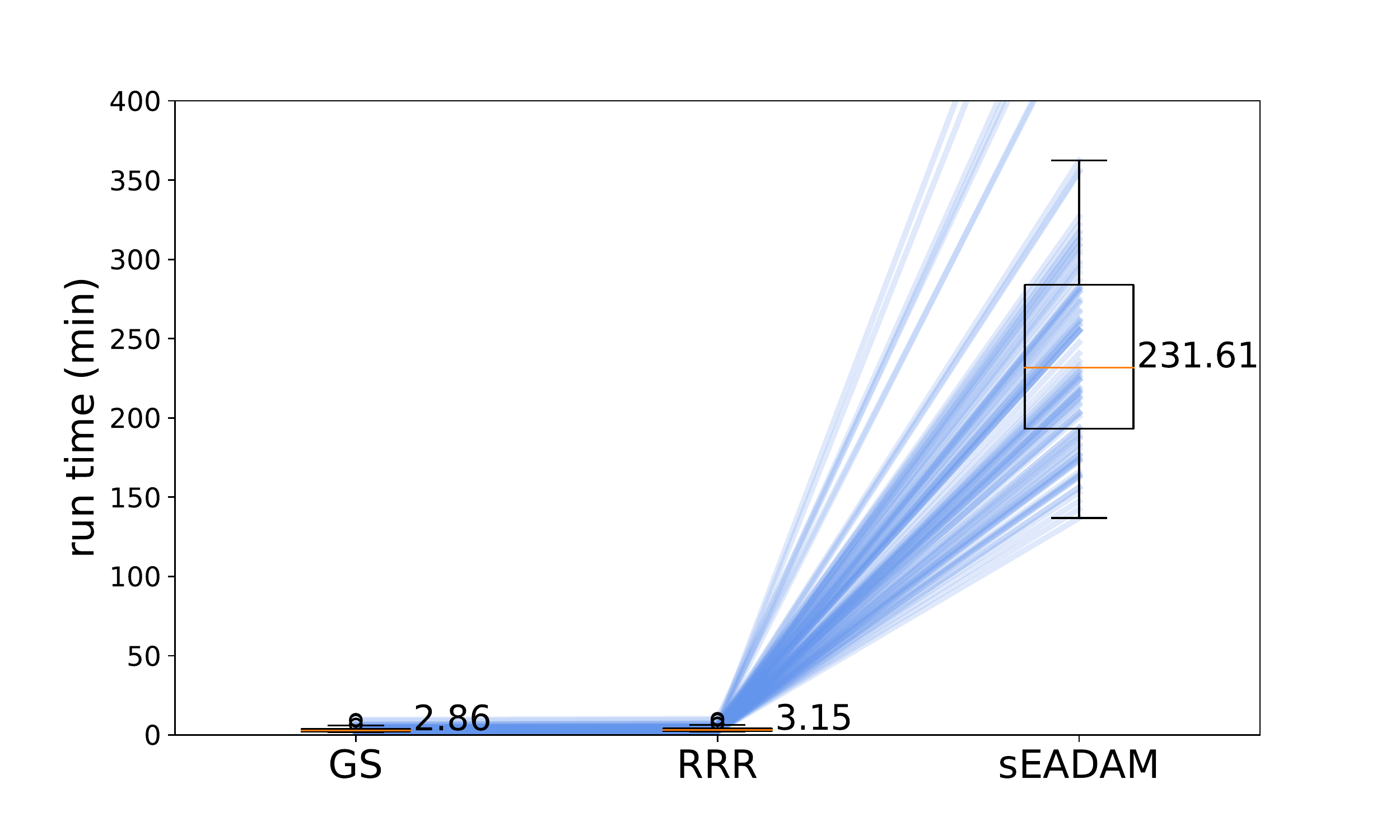}
	\caption{all students consent}
\end{subfigure}%
\hspace*{\fill}
\begin{subfigure}{.48\textwidth}
	\centering
	\includegraphics[width=\linewidth]{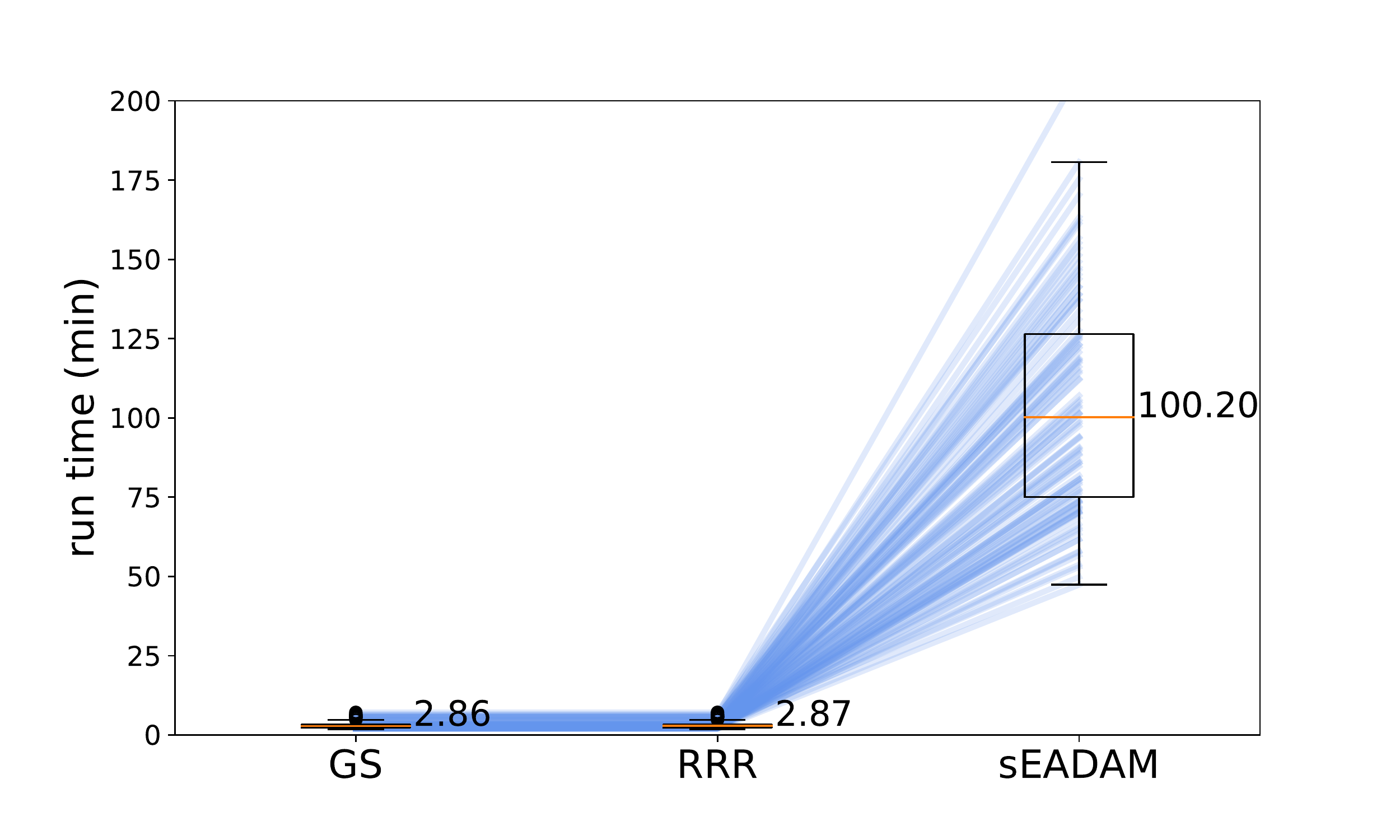}
	\caption{$80\%$ students consent}
\end{subfigure}%

\begin{subfigure}{.48\textwidth}
	\centering
	\includegraphics[width=\linewidth]{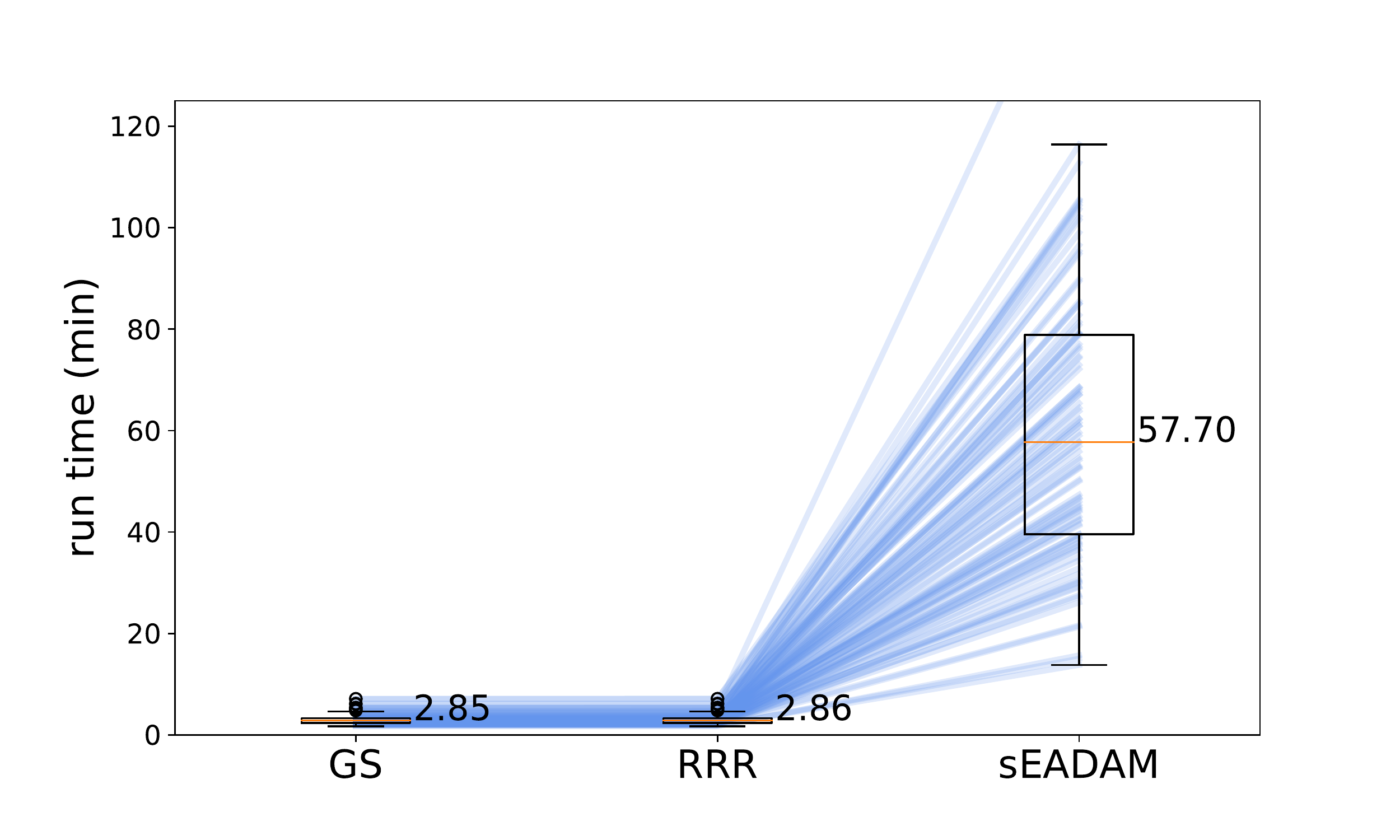}
	\caption{$50\%$ students consent}
\end{subfigure}%
\hspace*{\fill}
\begin{subfigure}{.48\textwidth}
	\centering
	\includegraphics[width=\linewidth]{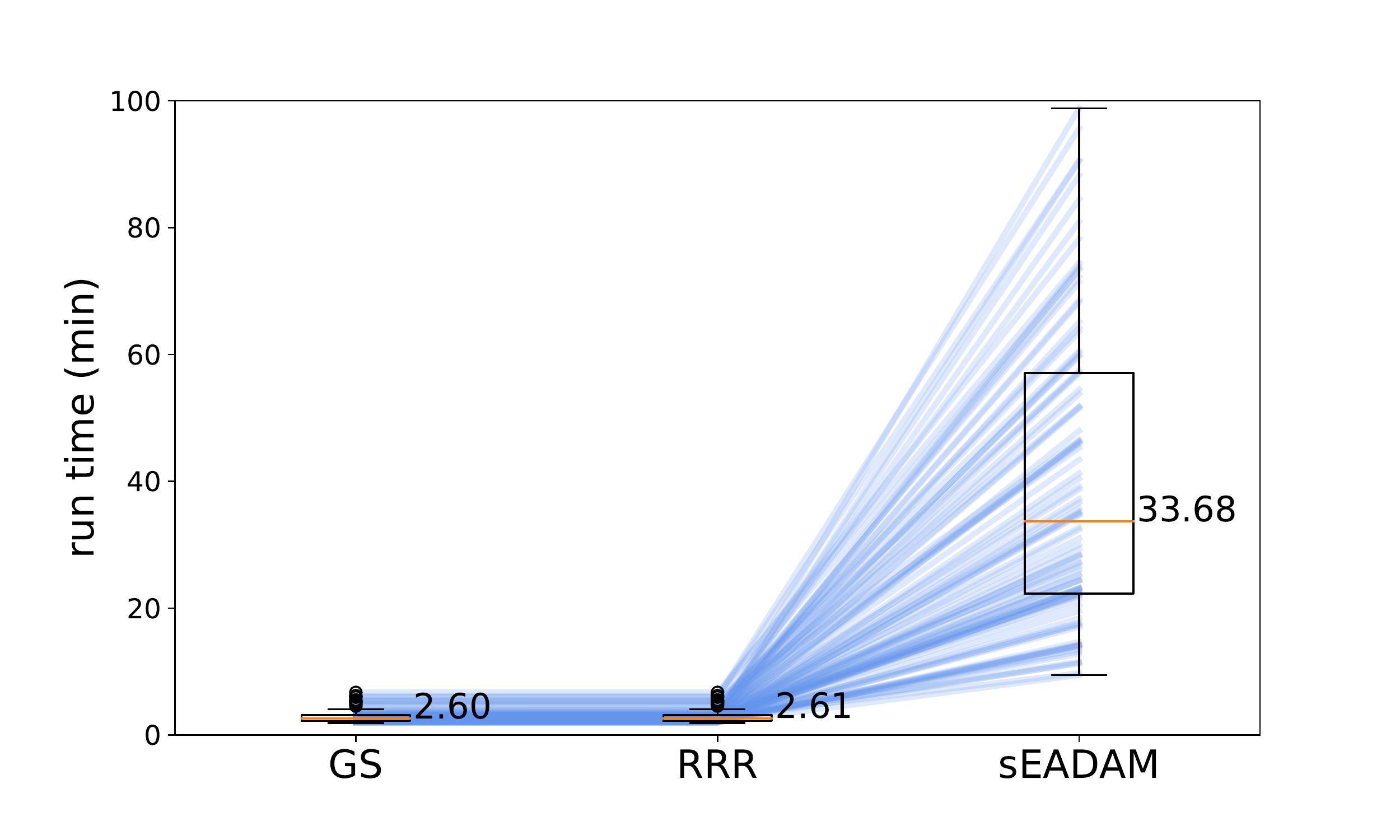}
	\caption{$30\%$ students consent}
\end{subfigure}%

\begin{subfigure}{.5\textwidth}
	\centering
	\includegraphics[width=\linewidth]{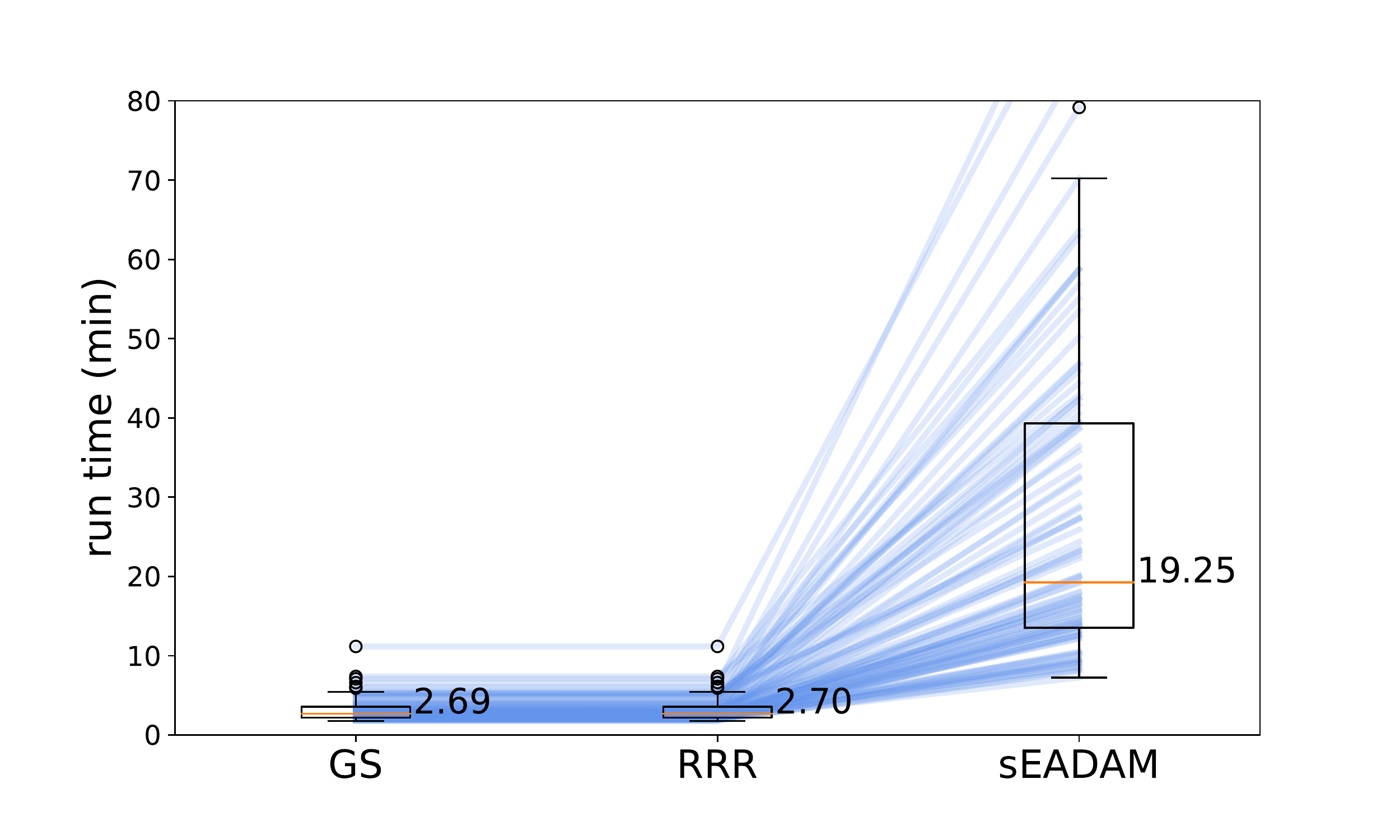}
	\caption{$10\%$ students consent}
\end{subfigure}
\caption{\small Comparing simplified EADAM (sEADAM) and \texttt{school-rotate-remove with consent} (SchRR) on random instances whose sizes are similar to those of the New York City school system. Run times of Gale-Shapley's algorithm (GS) are included as a benchmark. Run time of Kesten's original EADAM algorithm is not included because most instances fail to finish within 24 hours. Each line represents one instance. Box plots and averages of run times are included for each algorithm.} \label{fig:algo-compare-large}.
\end{figure}

One major advantage of our \texttt{school-rotate-remove with consent} is that instead of repeatedly running Gale-Shapley's algorithm, we update the assignment locally using the structural results (lattice structure and rotations) of stable assignments. Our algorithm runs in time $O(|E|)$ as shown in Lemma~\ref{lem:rrr-conset-impl}.

To further demonstrate the computational advantage of our algorithm, we randomly generated instances of varying sizes, and recorded the running time of all three algorithms. The running time of Gale-Shapley's algorithm is also recorded as a benchmark. The number of students in our instances ranges from $500$ to $30,000$, and the corresponding number of schools ranges from $5$ to $300$. For each instance size, $100$ instances $(G,<,\q)$ are obtained by randomly generating $<$ and $\q$. For each student $a$ (resp. school $b$), $<_a$ (resp. $<_b$) is defined by a random permutation of $B$ (resp. $A$). The quota of each schools is randomly selected between $50$ and $150$ uniformly. Note that in this set of simulations, students and schools have complete preference ranking of the opposite side. That is, in all instances, $G$ is a complete bipartite graph. We also conduct another set of simulation (details later) with incomplete preference list. We tested scenarios where each student is randomly determined to be consenting with proability $10\%$, $30\%$, $50\%$, $80\%$, and $100\%$. The experiments were carried out on a computing node with $1$ core and $4$GB RAM.

A visual representation of  the running times of different algorithms can be found in Figure~\ref{fig:algo-compare}. The shaded areas are $95\%$ confidence intervals of each algorithm for given instance sizes. Our algorithm performs significantly faster than the simplified EADAM~\citep{Tang.Yu} and dramatically faster than the original EADAM~\citep{Kesten}, with the differences being especially pronounced when all students consent. 

The New York City school district has approximately $90,000$ students applying to $700$ public high school programs every year, where students can list up to $12$ schools in their application~\citep{narita2016match}. We further conducted computational experiments whose instance size are similar to those of New York City. We compared our algorithm with simplified EADAM on random instances generated similarly as previously described. However, in this set of the simulations, we fix instance size with $|A|=90,000$ and $|B|=700$. Moreover, the quota of each school is uniformly randomly selected from integers between $\lceil 0.5\times\mu \rceil$ and $\lceil 1.5\times\mu \rceil$ where $\mu=\lceil\frac{|A|}{|B|} \rceil$. In generating $<$, for every student $a$, $<_a$ is obtained by truncating the random permutation such that only the top 12 schools are listed; for every school $b$, $<_b$ is obtained by restricting the random permutation to students who have $b$ in their preference lists. Graph $G$ can be deduced from the preference lists. Results of our experiments are summarized in Figure~\ref{fig:algo-compare-large}. The difference in computational time is noticeably different with all levels of consenting percentages. In particular, when all students consent, \texttt{school-rotate-remove} takes approximately $3$ minutes, whereas simplified EADAM takes on average $4$ hours and its run time has a much higher variance.

%
%
% ------------------------------------------
%
%

\section{Legal matchings and Latin marriages} \label{SEC:LATIN}

In this section, we restrict to one-to-one instances. For an instance $(G,<)$ of the stable marriage problem, let $\SS(G,<)$ and $\L(G,<)$ denote the set of stable matchings and legal matchings respectively. In addition, we call $(G_L, <)$ the \emph{legalized instance} of $(G,<)$, where $G_L$ is the subgraph of $G$ defined as in Theorem~\ref{thr:main1}. We say an instance $(G,<)$ is \emph{legal} if $G_L=G$.

%
%
% ------------------------------------------
%
%

An $n\times n$ matrix is a \emph{Latin square} if each row and each column is a permutation of numbers $1, 2, \cdots, n$. Given an instance $(G,<)$ of the stable marriage problem with complete lists, we call the position of $a$ in the preference list of $b$ the \emph{rank} of $a$ in $b$'s list. Following the work of~\citet{Ben.et.al}, we say an instance $(G,<)$ with $|A|=|B|=n$ is \emph{Latin} if there exists a Latin square $Q$  with $n$ rows indexed by elements of $A$ and $n$ columns indexed by elements of $B$ such that, for each row $a$ and column $b$, $Q(a,b)$ is the rank of $b$ in $a$'s list, and $n+1-Q(a,b)$ is the rank of $a$ in $b$'s list. We call such $Q$ the \emph{Latin ranking matrix}. See Example~\ref{ex:latin-example} for an example of a Latin ranking matrix and its associated stable marriage instance. In this section, we prove the following.

\begin{theorem}\label{thr:main4}
	Let $(G,<)$ be a Latin instance. Then $G_L=G$. Moreover, assume $(G,<)$ have $n$ men and $n$ women,  then there exists an instance $(G',<')$ with an additional man $\tilde{a}$ and an additional woman $\tilde{b}$ such that $|\SS(G',<')|=1$ and $\L(G',<')=\{M\cup \{\widetilde{a} \widetilde{b}\}: M\in \SS(G,<) \}$.  
\end{theorem}

\citet{Ben.et.al} provide, for each even $n$, a Latin instance $(G,<)$ with $n$ men and $n$ women such that $|\SS(G,<)|=\omega(2^n)$ and in the man-optimal stable matching, each man is given his favorite partner. Hence, Theorem~\ref{thr:main4} implies that for each odd $n$, there is an instance $(G',<')$ with $n$ men and $n$ women such that $|\SS(G',<')|=1$ and $|\L(G',<')|=\omega(2^n)$. Moreover, proofs of our construction for $(G',<')$ shows that the man-optimal legal matching in $\L(G',<')$ assigns to each man from $G$ his favorite partner, while the stable matching in $\SS(G',<')$ assigns to each man from $G$ his second least favorite partner (see Lemma \ref{lem:latin-aux-one-stable}). Note that, up to a different constant in the basis, the asymptotic ratio between the quantities $|\L(G,<)|$ and $|\SS(G,<)|$ cannot be increased, as it has been recently shown that there exists an absolute constant $c>1$ such that every instance of the stable marriage problem with $n$ men and $n$ women has $O(c^n)$ stable matchings \citep{Karl}. 

We believe that future investigations of the relationship between Latin instances and legal matchings may provide further advancement on a question by~\citet{Knuth}. In particular, in his seminal work,~\citet{Knuth} asks for a characterization of instances that maximize $|\SS(G,<)|$ for each value of $|A|=|B|=n \in \N$. While an asymptotic upper bound follows from the work cited above~\cite{Karl}, the characterization of these instances is unsolved even for reasonable small sizes. Note that, for each $n\in \N$, there is always a legal instance achieving the maximum, as for any $(G,<)$ we have $|\SS(G_L,<)|=|\L(G_L,<)|\geq|\SS(G,<)|$. 

The theorem below~\citep{Ben.et.al} gives a necessary and sufficient condition for a matching to be stable in a Latin instance. 

\begin{theorem}\label{thm:latin-stability-condition}
	Let $M$ be a matching of the instance defined by a Latin ranking matrix $Q$. $M$ is stable if and only if there do not exist row $a$ and column $b$ such that $Q(M(b),b) >Q(a,b) >Q(a,M(a))$ or $Q(M(b),b) <Q(a,b) <Q(a,M(a))$.
\end{theorem}

The following lemma shows that in a Latin instance, the set of legal matchings is exactly the set of stable matchings. 

\begin{lemma}\label{lem:latin-legal}
	Let $(G,<)$ be a Latin instance. Then $G_L=G$.
\end{lemma}

\proof{Proof.}
	Assume $Q$ is the Latin ranking matrix of instance $(G,<)$ and $Q\in \Z^{n\times n}$. For $i\in [n]$, let $M^i= \{ab: Q(a,b)=i\}$, and then by definition of Latin squares, $M^i$ is a matching. By construction, for any row $a$ and column $b$, $Q(M^i(b),b) =i= Q(a,M^i(a))$. Therefore, $M^i$ must be stable and thus legal due to Theorem~\ref{thm:latin-stability-condition}. Since $\bigcup_{i\in [n]} M^i = E(G)$, by Theorem~\ref{thr:main1}, $G_L = G$.
\hfill{\Halmos}\endproof

As we will show next, the set of stable matchings of a Latin instance can be ``masked'' into the set of legal matchings of an auxiliary instance  with only one more man and one more woman, such that the auxiliary instance has only one stable matching. The construction is as follows: given a Latin instance $(G(A\cup B, E),<)$, construct an auxiliary instance $(G'(A'\cup B', E'), <')$, where $A'= A\cup \{ \widetilde{a} \}, B'= B\cup \{ \widetilde{b} \}, E'=A'\times B'$, and $<'$ is defined as follows:
\begin{enumerate} [label=(\roman*)]
	\item every $a\in A$ ranks $\widetilde{b}$ in the last position, and $<'_a$ restricted to $B$ is exactly $<_a$.
	\item $\widetilde{a}$ has an arbitrary ranking of $B'$ as long as $\widetilde{b}$ is the least preferred.
	\item every $b\in B$ ranks $\widetilde{a}$ in the second place, and $<'_{b}$ restricted to $A$ is exactly $<_b$.
	\item $\widetilde{b}$ has an arbitrary ranking of $A'$ as long as $\widetilde{a}$ is ranked the first.
\end{enumerate}
An example of our construction can be found in Example~\ref{ex:latin-example}.  

\begin{example}\label{ex:latin-example} 
	Consider the following Latin ranking matrix $Q$ and the associated instance $(G,<)$.
	$$\begin{blockarray}{ccccc}
	& b_1 & b_2 & b_3 & b_4 \\
	\begin{block}{c(cccc)}
	a_1 & \framebox{1} & 2 & 3 & 4 \\
	a_2 & 2 & 1 & \framebox{4} & 3 \\
	a_3 & 3 & \framebox{4} & 1 & 2 \\
	a_4 & 4 & 3 & 2 & \framebox{1} \\
	\end{block}
	\end{blockarray} \qquad \qquad \begin{array}{clccl}
	a_1: & b_1 > b_2 > b_3 > b_4 & & b_1: &  a_4 > a_3 > a_2 > a_1 \\
	a_2: & b_2 > b_1 > b_4 > b_3 & & b_2: & a_3 > a_4 > a_1 > a_2 \\
	a_3: & b_3 > b_4 > b_1 > b_2 & & b_3: &a_2 > a_1 > a_4 > a_3 \\
	a_4: & b_4 > b_3 > b_2 > b_1 & \qquad \quad & b_4: & a_1 > a_2 > a_3 > a_4
	\end{array}$$
	
	Consider the matching $M=\{a_1b_1, a_2b_3, a_3b_2, a_4b_4\}$, which corresponds to the boxed cells in the Latin ranking matrix. Since $a_3b_1$ blocks $M$, we can conclude that $M$ is not stable. Equivalently, we can apply Theorem \ref{thm:latin-stability-condition} on the Latin ranking matrix with $a=a_3, b=b_1$. Then, we have $Q(M(b), b)=1 < Q(a,b)=3 < Q(a, M(a))=4$, also implying that $M$ is not stable. 
	
	One can check that $(G,<)$ has $10$ stable matchings. Now consider the auxiliary instance $(G', <')$, whose preference lists are exactly those given in Example \ref{ex:rrr-fast} with $a_5=\tilde{a}$ and $b_5=\tilde{b}$. $(G',<')$ has only one stable matching, which is  $\{a_1b_4$, $a_2b_3$, $a_3b_2$, $a_4b_1$, $\widetilde{a}\widetilde{b}\}$, but its legalized instance $(G'_L, <'_L)$ has 10 stable matchings.
	\hfill {$\diamondsuit$}	
\end{example}

We first show the following facts before concluding the proof of Theorem \ref{thr:main4}.

\begin{lemma}\label{lem:latin-aux-one-stable}
	Given a Latin instance $(G, <)$ with $G=(A\cup B, E)$, define $(G', <')$ as above. Then, we have $|\SS(G', <')| = 1$ and each man from $A$ is given his second least favorite partner (with respect to $<'$) in this stable matching. 
\end{lemma}

\proof{Proof.}
	Let $M \in \SS (G', <')$. We will first show $M(\widetilde{b}) =\widetilde{a}$. Assume by contradiction that $M(\widetilde{b})=a$ for some $a\in A$. Let $b$ be $a$'s least preferred partner in $B$. Then $b>'_a \widetilde{b}=M(a)$ by construction. By the symmetric nature of Latin instances, $a$ must be $b$'s most preferred partner in $A$, which means $a>'_b M(b)$. But then $ab$ is a blocking pair of $M$, contradicting stability. Next, we want to show every woman in $B$ is matched to her most preferred man. Assume by contradiction that the claim is not true for some $b\in B$. Then $\widetilde{a}>'_b M(b)$. Since $b >'_{\widetilde{a}} \widetilde{b}$ by construction, $\widetilde{a}b$ blocks $M$, which again contradicts stability. Hence, $\SS(G', <')$ contains exactly one stable matching, namely the one where every woman is matched to her most preferred man according to $<'$. That is, every man $a \in A$ is given his second least favorite partner with respect to $<'$.
\hfill{\Halmos}\endproof

\begin{lemma}\label{lem:latin-legalization}
	Let $(G,<)$ and $(G',<')$ be as before with $G=(A\cup B, E)$ and $G'=(A'\cup B', E')$. Then, $\L(G', <') = \{ M\cup \{\widetilde{a} \widetilde{b}\}: M \in \SS(G, <) \}$.
\end{lemma}

\proof{Proof.}
	Let $M_0$ be the only stable matching of $(G', <')$. Since every woman in $B'$ is matched to her most preferred man in $A'$ as shown in the proof of Lemma~\ref{lem:latin-aux-one-stable}, $M_0$ is also the woman-optimal legal matching of $\L(G', <')$. In addition, since $\widetilde{b}$ is the least preferred woman of every man by construction of $G'$, $\widetilde{b}$ is a sink in the woman-rotation digraph of $M_0$ and remains a sink throughout the execution of \texttt{woman-rotate-remove}. Thus, $\widetilde{a}$ is matched to $\widetilde{b}$ in the man-optimal legal matching of $\L(G', <')$. Hence, $\widetilde{a}\widetilde{b}\in M$ for all $M\in \L(G', <')$ and according to Theorem~\ref{thr:main1}, all edges in $\widetilde{E}:=\{a\widetilde{b}: a\in A \} \cup \{\widetilde{a}b: b\in B \}$ are illegal. By Lemma~\ref{lem:remove-illegal}, we have $\L(G', <') = \L(G[E\setminus E'], <')= \{M\cup \{\widetilde a \widetilde b\}: M\in \L(G,<) \}$, where the last equality is because $E\setminus \widetilde{E} = E(G)\cup \{ \widetilde{a}\widetilde{b} \}$. Finally, by Lemma~\ref{lem:latin-legal}, we have $\L(G, <) = \SS(G, <)$ and thus, $\L(G', <') =\{M\cup \{\widetilde a \widetilde b\}: M\in \SS(G,<) \}$.
\hfill{\Halmos}\endproof

\medskip
\proof{Proof of Theorem~\ref{thr:main4}.} 
    Immediately implied  by Lemmas~\ref{lem:latin-legal},~\ref{lem:latin-aux-one-stable} and~\ref{lem:latin-legalization}.
\hfill{\Halmos}\endproof

\bigskip \noindent {\bf Acknowledgments.} We thank Yiannis Mourtos, Jay Sethuraman, and the anonymous reviewers for their insightful comments and useful suggestions on an earlier version of this draft.

\bibliography{legal} 
\bibliographystyle{plainnat}

\appendix

\section{Self-contained Proof of Theorem~\ref{thr:main1}} \label{app:self-contained}

Here, we present a proof of Theorem~\ref{thr:main1} that builds only on concepts from Section~\ref{sec:basic} and Section~\ref{sec:lattice-rotation}. Recall for $\M'\subseteq \M:=\M(G,\q)$, $\I(\M')$ is defined as the set of assignment that are blocked by some assignment form $\M'$. The proof of Theorem~\ref{thr:main1} relies on the study of the fixed points of a function $\L$ when applied to $\SS:= \SS(G,<,\q)$, where $\L(\M')$ is defined as the set of assignments that are \textit{not} blocked by \emph{any} assignment in $\M \setminus \I(\M')$. That is, $\L(\M'):=\M\setminus \I\big( \M \setminus \I(\M') \big)$\footnote[1]{We adopt the function $\L$ from a similar operator introduced in~\citet{Morr}.}. 

In order to get some intuition about the $\L$ operator, let $\L^0:=\SS$ and iteratively define $\L^{i}:=\L(\L^{i-1})$ for $i \in \N$. Also, let $\I^{i}:=\I(\L^i)$. Clearly, any set with the legal property must contain $\SS$, and as a result, such set cannot contain any assignment from $\I^0$. Thus, we can safely say that all assignments that are only blocked by assignments in $\I^0$ are also contained in every set with the legal property. These assignments, together with $\L^0$, give exactly $\L^1$. Similarly, we can enlarge $\L^1$ to $\L^2$, etc.  Hence, as ~\citet{Morr} also observed, the sequence $\L^0, \L^1, \L^2, \cdots$ converges.

\begin{lemma}\label{lem:L-grows}
	There exists $k \in \N$ such that $\L^k=\L^{k+1}$.
\end{lemma}

\proof{Proof.}
    We show by induction on $i$ that $\L^{i}\subseteq \L^{i+1}$, which concludes the proof. Clearly $\L^0={\SS}\subseteq \L^1$. Now fix $i \in \N$. Since $\L^{i-1} \subseteq \L^i$, we deduce $\I^{i}\supseteq \I^{i-1}$. Hence,
    $$\L^{i+1} =\L(\L^i) = \M \setminus \I(\M\setminus \I^i) \supseteq \M\setminus \I(\M \setminus \I^{i-1}) = \L^{i},$$
    where in the containment relation we use $\I^i \supseteq \I^{i-1}$ and therefore $\I(\M \setminus \I^{i}) \subseteq \I(\M \setminus \I^{i-1})$.
\hfill{\Halmos}\endproof

\begin{example}
    Consider the instance given in Example~\ref{ex:legal-block}. We have $\L^0:=\SS=\{M_1\}$. Since $M_1$ blocks $M_3$, $M_4$, and $M_5$, we have $\I^0 := \I(\L^0) = \{M_3, M_4, M_5\}$. Now, note that $M_2$ is only blocked by $M_3$ and $M_4$, both of which are in $\I^0$. Thus, $M_2\in \L(\L^0)$ and $\L^1=\{M_1, M_2\}$. Repeating the process again, we can see that $\I^1 := \I(\L^1) = \{M_3, M_4, M_5\}$ and $\L^2=\{M_1, M_2\}$. The sequence thereafter stabilizes.
	\hfill {$\diamondsuit$}	
\end{example}

\noindent {\bf Fixed points of ${\cal L}$ are stable assignments.} 
We want therefore to study the set where sequence $\L^0, \L^1, \L^2, \cdots$ stabilizes. One key observation is that every fixed point of $\L$ coincides with the set of stable assignments in some subinstance of the original problem. Although simple, this observation has dramatic consequences, as we can now rely on all the structural knowledge on stable assignments. This distinguishes our approach from that of~\citet{Morr}, where properties of legal assignments are developed independently from those of stable assignments. 

\begin{lemma}\label{lem:L-stops}
	Assume $\M_0\subseteq \M$ satisfies $\L(\M_0) =\M_0$. Then $\M_0=\SS(G',<,\q)$, where $G':=G[E']$ and $E':=\bigcup \{M: M\in \M\setminus \I(\M_0)\}$. 
\end{lemma}

\proof{Proof.}
    Suppose $M \in \M \setminus \M_0 =\M \setminus \L(\M_0)$. Then there is an assignment $M' \in \M\setminus \I(\M_0)$ and an edge $e \in M'$ that blocks $M$. But then $e \in \bigcup \{\bar{M} : \bar{M} \in \M\setminus \I(\M_0)\}=E'$. Hence, $M\notin \SS(G',<, \q)$. Conversely, suppose $M\in \M \setminus \SS(G',<,\q)$. If $M \in \M(G',\q)$, then $M$ is blocked by some $e \in E'$. This means an assignment in $\M\setminus \I(\M_0)$ blocks $M$, implying $M \notin \L(\M_0) =\M_0$. If $M\notin \M(G',\q)$, then $M$ contains an edge that is not in $E(G')$. This implies $M\in \I(\M_0)$ and thus $M\notin \L(\M_0)=\M_0$.
\hfill{\Halmos}\endproof

\noindent{\bf Assignments that do not block each other.} Besides properties of stable assignments, we will also use those of assignments that do not block each other. Those facts are established in the next two lemmas. They can be seen as an extension of the  ``opposition of interest'' property: if $a$ and $b$ are partners in a stable assignment $M$, then they cannot both strictly prefer another stable assignment $M'$ to $M$~\cite[Theorem 1.3.1]{GI}.

\begin{lemma}\label{lem:counting}
	Consider an instance of stable marriage problem $(G,<)$ with $G=G(A\cup B,E)$. Let $M, M' \in \M(G, \mathbf{1})$. Call an edge $ab\in M\cup M'$ \emph{irregular} if both $a$ and $b$ strictly prefer $M$ to $M'$ or both strictly prefer $M'$ to $M$. Suppose $M$ does not block $M'$ and $M'$ does not block $M$. Then:
	\begin{enumerate}
		\item[1)] there are no irregular edges;
		\item[2)] $G[M\triangle M']$ is a disjoint union of singletons and cycles;
		\item[3)] a node is matched in $M$ if and only if it is matched in $M'$. 
	\end{enumerate}
\end{lemma}

\proof{Proof.}
    1) Assume $a_1b_1$ is an irregular edge and assume wlog both endpoints strictly prefer $M$ to $M'$. Then $a_1b_1 \in M'$, because otherwise $M$ blocks $M'$. Starting from $i=2$, iteratively define $a_{i}=M(b_{i-1})$ and $b_i=M'(a_i)$. Repeatedly using the assumption that $M$ and $M'$ do not block each other, we deduce that, for all $i\ge 2$, $a_i$ strictly prefers $M'$ to $M$, and vice versa $b_i$ strictly prefers $M$ to $M'$. Moreover, $a_i\neq \emptyset$ and $b_i \neq \emptyset$. Since $M$, $M'$ are matchings, there exists $\ell \geq 2$ such that $a_\ell=a_1$. Hence, $a_1=a_\ell$ strictly prefers $M'$ to $M$, a contradiction.

    2) 	Note that the degree of each node in $G[M\triangle M']$ is at most $2$. Suppose the thesis does not hold, then $G[M\triangle M']$ contains a path, say wlog $a_1,b_1,a_2,b_2, \cdots$, whose endpoints have degree 1 in $G[M \triangle M']$. Assume wlog that $a_1b_1 \in M'$. Since $a_1$ is unmatched in $M$, $a_1$ strictly prefers $M'$ to $M$. In addition, since $a_1b_1 \in M'$ does not block $M$, $b_1$ strictly prefers $M$ to $M'$. We can iterate and conclude, similarly to part $1)$, that all nodes $a_i$ strictly prefer $M'$ to $M$, and vice versa all nodes $b_i$ strictly prefer $M$ to $M'$. Suppose first that $a_k b_k$ is the last edge of the path. Then $b_k$ strictly prefers $M'$ to $M$ as $a_k b_k \in M'$ and $M(b_k)=\emptyset$. This implies $a_k b_k$ is irregular, a contradiction to part 1). Similarly, if the last edge is $b_{k}a_{k+1}$, $a_{k+1}$ strictly prefers $M$ as $a_{k+1}$ is unmatched in $M'$. Again, this implies $b_{k}a_{k+1}$ is irregular, a contradiction to part 1).

    3) Immediately from 2).
\hfill{\Halmos}\endproof

\begin{lemma}\label{lem:magic-wand}
	Let $(G(A\cup B,E),<,\q)$ be an instance of the stable assignment problem. Let $M,M' \in \M(G,\q)$ be such that $M$ does not block $M'$ and $M'$ does not block $M$, and fix $a \in A$ that is matched in $M$. Then $a$ is matched in $M'$. Let therefore $b=M(a)$, $\overline b=M'(a)$. If $b>_a \overline b$, then there exists $\overline a \in M(\overline b)$ such that $a >_{\overline b} \overline a$ and ${\overline b} >_{\overline {a}} M'(\overline a)$.
\end{lemma}

\proof{Proof.}
    We first claim that $M'_H$ does not block $M_H$, and $M_H$ does not block $M'_H$, where $M_H=\pi(M)$ and $\pi$ is the mapping defined in Section~\ref{sec:redu}. It then follows from Lemma~\ref{lem:counting}, part 3) and the definition of mapping $\pi$ that $a$ is matched in $M'$. It is enough to show that $M'_H$ does not block $M_H$. Assume by contradiction that there exists $\bar ab^i\in M'_H$ that blocks $M_H$. That means $b^i>_{\bar a} M_H(\bar a)$ and $\bar a>_{b^i} M_H(b^i)$. If $M_H(\bar a) = b_1^j$ where $b_1\neq b$, then $b>_{\bar a} M(\bar a)$, ${\bar a}>_b a'$ for some $a'\in M(b)$, and thus $\bar ab\in M'$ blocks $M$, a contradiction. So assume $M_H(\bar a)=b^j$ for some $j\in [q_b]$. Since $b^i>_{\bar a} b^j$, we have $j>i$ by construction of $(H_G, <_G)$. Then by the definition of $\pi$, $M_H(b^i)>_b M_H(b^j)=\bar a$, a contradiction.

    Let $\overline b {}^i:=M'_{H}(a)$ and $b^{\ell}:=M_H(a)$. By Lemma~\ref{lem:counting}, part 2), there exists a cycle $C=a,b^{\ell},...,\overline b{}^i$ in $H_G[M'_{H} \triangle M_H]$, and this cycle  has no irregular edges. Since $b^\ell>_a \overline b{}^i$, (i) all nodes from $A \cap C$ strictly prefer $M_H$ to $M'_{H}$, and vice-versa (ii) all nodes from $B_H \cap C$ strictly prefer $M'_{H}$ to $M_H$. Recall that $B_H$ is the collection of seats in the reduced instance $(H_G, <_G)$. Let $\overline b{}^j \in C$ be such that all nodes of $C \cap B_H$ that precede $\overline b{}^j $ in $C$ are not seats of $\overline{b}$, while all nodes that follow $\overline b{}^j$ in $C\cap B_H$ are seats of  $\overline{b}$. Note that $\overline b{}^j$ is well-defined, since $C$ terminates with $\overline b{}^i$ (hence possibly $j=i$). Let $\overline a:=M_{H}(\overline b{}^j)$, i.e., $C=a, b^{\ell}, \cdots, \overline a, \overline b{}^j, \cdots, \overline b{}^i$. By (i) above, $\overline b{}^j = M_H(\overline a) >_{\overline a} M'_H(\overline a) $, hence $\overline b >_{\overline a} M'(\overline a)$, as required. Moreover, $\overline a = M_H(\overline b{}^j)<_{\overline b{}^j} M'_H(\overline b{}^j)\leq_{\overline b{}^j}M'_H(\overline b{}^i)=a$ (where the strict preference follows from (ii) and the non-strict one from the definition of mapping $\pi$), hence  $a>_{\overline b} \overline a$, as required. 
\hfill{\Halmos}\endproof

\noindent {\bf Edges of $G_L$.} For $k \in \N$ that satisfies Lemma~\ref{lem:L-grows}, we let ${\L}:={\L}^k$. Let $G_L$ be the subgraph of $G$ induced by edges $\bigcup\{M:M\in \M\setminus \I(\L)\}$\footnote[2]{Note that this definition of $G_L$ is different from the one given in the main body of the paper. However, results from this section imply that the two definitions coincide.}. Gale-Shapley's algorithm and structural properties deduced so far are used next to characterize edges of $G_L$ as all and only edges used by some assignment in ${\cal L}$. 

\begin{lemma}\label{lem:legal-edges}
	$E(G_L)=\bigcup\{M : M \in \L\}$. 
\end{lemma}	

\proof{Proof.}
    The containment relationship $E(G_L)\supseteq \bigcup\{M : M \in \L\}$ is clear from definition. So it suffices to show $E(G_L)\subseteq \bigcup\{M : M \in \L\}$. Assume by contradiction that there exists an edge $ab\in E(G_L)\setminus \bigcup\{M: M\in \L\}$. Let $M\in \M\setminus \I(\L)$ be an assignment such that $ab\in M$. Let $M_0$ (resp.~$M_z$) be the stable assignment output by the students (resp.~schools) proposing Gale-Shapley's algorithm in $G_L$. Since $\L = \L(\L)$, we have $M_0, M_z \in \L$ by Lemma~\ref{lem:L-stops}. By construction, $ab \notin M_0 \cup M_z$. In the following, when talking about a specific execution of Gale-Shapley's algorithm, we say that $a$ \emph{rejects} $b$ if during the execution, $a$ rejects the proposal by $b$, possibly after having temporarily accepted it. We distinguish three cases. 

    \textbf{Case a):} $b >_a M_0(a)=:\overline b$. By the choice of $M$, we know that $M_0$ and $M$ do not block each other (in either $(G,<,\q)$ or $(G_L,<,\q)$, as the preference lists are consistent),  since both are assignments in $\M\setminus \I(\L)$. Note that this case contains all and only the edges that have been rejected by some (equivalently, any) execution of the student-proposing Gale-Shapley's algorithm on $G_L$. Among all those edges, pick the one $ab$ that is last rejected by some execution of the algorithm. Apply Lemma~\ref{lem:magic-wand} (with $M=M$ and $M'=M_0$) and conclude that there exists $\overline a \in M(\overline b)$ such that $a >_{\overline b} \overline a$ and ${\overline b} >_{\overline {a}} M_0(\overline a)$. $\overline b>_{\overline a}M_0(\overline a)$ implies that $\overline b$ rejected $\overline a$ during the execution of Gale-Shapley's in consideration. Hence, when $a$ proposes to $\overline b$, either $\overline a$ still has to be rejected by $\overline b$, or it has been rejected before. In the latter case, $\overline b$ has her quota filled and rejects some other student when $a$ proposes. Hence the following events happen in this order during the execution of Gale-Shapley's algorithm: $a$ is rejected by $b$; $a$ proposes to $\overline b$; $\overline b$ rejects a student. This contradicts our assumption that $ab$ is the last rejected edge. 

    \textbf{Case b):} $M_0(a)>_a b >_a M_z(a)$. First, we want to show that there exists a stable assignment $M'$ such that $M'(a)>_a b$ and $a'>_{b} a$ for all $a' \in M'(b)$. Note that Lemma~\ref{thr:256-ext} implies $s_{M_0}(a)\neq \emptyset$. If $b>_a s_{M_0}(a)$, then we can simply take $M'=M_0$. Otherwise, we can apply a sequence of rotations from $M_0$ to obtain an assignment $M'$ such that $M'(a)>_a b>_a s_{M'}(a)\ge_a M_z(a)$. By definition of $s_{M'}(a)$, we must have $a'>_{b} a$ for all $a' \in M'(b)$. Now, by choice of $M$, $M$ and $M'$ do not block each other. We can therefore apply Lemma~\ref{lem:magic-wand} (with the roles of $M$ and $M'$ inverted) and conclude that there exists $\overline a \in M'(b)$ with $a>_{b} \overline a$, a contradiction. 

    \textbf{Case c):} $M_z(a) >_a b$. Using Lemma \ref{lem:magic-wand} (with $M=M_z$ and $M'=M$) we deduce that there exists $\bar a \in M_z(b)$ such that $a>_b \bar a$ and $b>_{\bar a} M(\bar a)=\bar b$. Hence, in some (equivalently, any) iteration of the school-proposing Gale-Shapley's algorithm, $a$ rejects $b$. Since this is the last case that still needs to be considered, we may assume edges $E(G_L)\setminus \bigcup\{M : M \in \L \}$ are exactly those rejected by some execution of the school-proposing Gale-Shapley's algorithm. Among all such edges, take $ab$ that is the last rejected by some execution. Applying Lemma~\ref{lem:magic-wand} again (with $a=\overline a$, $M=M_z$, $M'=M$), we know $\overline{a} >_{\overline b} a'$ for some $a'\in M_z(\overline{b})$. This implies that $\overline{a}$ rejected $\overline{b}$ during the execution of Gale-Shapley's in consideration. Hence, when $b$ proposes to $\overline a$, either $\overline b$ still has to be rejected by $\overline a$, or it has been rejected before. In the latter case, when $b$ proposes to $\bar a$, $\overline a$ rejects the school it temporarily accepted. Hence, the following events happen during the considered execution in this order: $a$ rejects $b$; $b$ proposes to $\overline a$; $\overline a$ rejects a school, contradicting the choice of $ab$.
\hfill{\Halmos}\endproof

\noindent {\bf Concluding the proof.} Once the previous facts have been established, concluding the proof of Theorem~\ref{thr:main1} is quite straightforward.

\begin{lemma}\label{lem:daje-roma}
	$\L$ has the legal property. That is, $\I(\L) = \M \setminus \L$.
\end{lemma}

\proof{Proof.}
    Clearly ${\I}({\L})\subseteq {\M}\setminus {\L}$. Now take $M \in \M\setminus \I(\L)$. Then $M \subseteq \bigcup \{M': M' \in \M \setminus \I(\L)\}= E(G_L)$. Hence, $M$ is an assignment of $G_L$ not blocked by any assignment from $\L=\SS(G_L,<,\q)$, where the last equality holds by Lemma~\ref{lem:L-stops}. By  Lemma~\ref{lem:legal-edges}, $M$ is not blocked by any edge in $E(G_L)$, and we conclude that $M \in \SS(G_L,<,\q) =\L$. 
\hfill{\Halmos}\endproof

Because of Lemma~\ref{lem:daje-roma}, we say that $(\L, \I:= \M \setminus \L)$ is a \emph{legal partition} of $\M$.

\begin{lemma}\label{lem:unique}
	$\L$ is the unique subset of $\M$ with the legal property.
\end{lemma}

\proof{Proof.}
    Assume by contradiction that there exists a set $\L' \subseteq \M$, $\L' \neq \L$ with the legal property. Let $\I':=\M \setminus \L'$. If $\L\subsetneq \L'$, we must have $\I' \subsetneq \I$. Take any $M\in \I\setminus \I'$, it must be blocked by some assignment in $\L\subsetneq\L'$. But we also have $M\in \L'$, which contradicts the assumption that $\L'$ has the legal property. Similarly, we cannot have $\L' \subsetneq \L$. Thus, sets $\A:=\{ M : M\in \I \cap \L'\}$ and $\B:=\{ M: M\in \L \cap \I'\}$ are both non-empty. In addition, let $\C:=\L \cap \L'$. It is also non-empty because all stable assignments are contained in any set with the legal property. In particular, $\L^0\subseteq \C$. Note that every assignment in $\B$ is blocked by some assignment from $\A$. Moreover, ($\dagger$) no assignments from $\A\cup \B$ can be blocked by any assignment from $\C$. Now take the first $i \in \N$ such that $\L^i \cap \B \neq \emptyset$, and note that $i \geq 1$. Let $M \in \L^i \cap \B$. All assignments blocking $M$ must be contained in $\I(\L^{i-1})$. Thus, we can pick $M'\in \I(\L^{i-1}) \cap\A$. Hence, $M'$ is blocked by some assignment in $\L^{i-1}\subseteq \C$ (containment relation due to the choice of $i$), contradicting ($\dagger$).
\hfill{\Halmos}\endproof

\proof{Proof of Theorem~\ref{thr:main1}.} 
    Immediately from Lemmas~\ref{lem:L-stops},~\ref{lem:legal-edges},~\ref{lem:daje-roma}, and~\ref{lem:unique}. 
\hfill{\Halmos}\endproof

%
%
% ------------------------------------------
%
%

\section{Extension of Theorem~\ref{thr:main1} to the setting of~\citet{EM}} \label{app:choice-setting}

In~\citet{EM}, each school $b$, instead of having a strict preference ordering of the students and having a quota, has a choice function $C_b:2^{A_b}\rightarrow 2^{A_b}$, where $A_b=\{a: ab\in E\}$ and for each $X\subseteq A_b$, $C_b(X)\subseteq X$, where $E$ is a subset of $A\times B$. In particular, $C_b(X)$ represents the students school $b$ chooses from $X$. \citet{EM} assume that, for every $b\in B$, the choice function $C_b$ satisfies the following two properties:
\begin{enumerate}
    \item Substitutability: For any $X\subseteq Y\subseteq A_b$, we have $C_b(Y)\cap X \subseteq C_b(X)$; and
    \item Law of aggregate demand (LAD): For any $X\subseteq Y\subseteq A_b$, we have $|C_b(X)|\le |C_b(Y)|$.
\end{enumerate}
As usual, each student $a \in A$ has a strict ordering over schools $\{b \in B : ab \in E\}$. We denote by $(G(A\cup B, E), <,\mathcal{C})$ an instance of the stable assignment problem in the setting of~\citet{EM}, where $<= \{<_v\}_{v\in A}$ and $\mathcal{C} = \{C_b\}_{b\in B}$. An assignment\footnote[3]{This is called an ``individually rational'' assignment in~\citet{EM}.} $M$ of this instance is a collection of edges of $G$ such that for each $a\in A$, at most one edge of $M$ is incident to $a$, and for each $b\in B$, $C_b(\{a: ab\in M\}) = \{a: ab\in M\}$. For an assignment $M$ and $x \in A \cup B$, we again write $M(x):=\{y : xy \in M\}$.

An edge $ab\in E$ is said to \emph{block} assignment $M$ if $b>_a M(a)$ and $a\in C_b(M(b)\cup \{a\})$. Similarly to the one-to-many setting, an assignment is stable if and only if there is no edge blocking it, and an assignment $M'$ is said to block an assignment $M$ if there is an edge $ab\in M'$ such that $ab$ blocks $M$. The set of legal assignments is defined the same way as before. Moreover, it is shown in~\citet{EM} that Theorem~\ref{thm:Morr} holds in this setting.

For a subgraph $G'\subseteq G$, we denote by $(G', <, \mathcal{C})$ the instance where students' preferences are those induced by $<$ on schools in $G'$, and schools' choice functions are defined as the restriction of $\mathcal{C}$ to students in $G'$. In particular, for every school $b\in B$, the restriction of function $C_b$ also satisfies the substitutability and LAD properties, and thus $(G', <, \mathcal{C})$ is also an instance of the stable assignment problem in the setting of~\citet{EM}.

Because of the above-mentioned properties, the three lemmas in Section~\ref{sec:disguised}, Lemma~\ref{lem:remove-illegal},~\ref{lem:legal-remain-rm-edge}, and~\ref{lem:legal-equal-stable}, go through to the setting of~\citet{EM} with exactly the same arguments, and thus so does Theorem~\ref{thr:main1}.

%
%
% ------------------------------------------
%
%

\section{Proof of Lemma~\ref{lem:find-rot2}} \label{app:proof-find-rot2}

The proof of Lemma~\ref{lem:find-rot2} builds on Lemma~\ref{lem:magic-wand} and on the following fact (see, e.g., \citet{GI}).

\begin{lemma} \label{lem:polarity}
    If $M,M' \in \SS(G,<,\q)$ and $M\succeq M'$, then for every school $b\in B$,  $a'>_b a$ for all $a \in M(b) \setminus M'(b)$ and $a' \in M'(b) \setminus M(b)$.
\end{lemma}

\proof{Proof of Lemma~\ref{lem:find-rot2}.}
    We prove the result with $X$ being the set of students, and thus let $x'=a'$, $y=b$, and $x=a$. The other case can be proved similarly. By Theorem~\ref{thr:main1}, $M\in \SS(G,<,\mathbf{q})\subseteq \L(G,<,\mathbf{q})=\SS(G_L,<,\mathbf{q})$, where recall that $G_L$ is the subgraph of $G$ whose all edges are those that appear in some legal assignment. In $(G_L, <, \q)$, consider any sequence of student-rotations, $\rho_1, \rho_2, \cdots, \rho_k$, whose elimination from $M$ gives the school-optimal legal assignment $M^\L_z$. The existence of such sequence follows from Lemma~\ref{thm:255-ext}. Let $M^i = M/\rho_1/\cdots/\rho_i$ with $M^0=M$. If $a=\emptyset$, by definition of student-rotations, we have $b\notin \rho_i$ for all $i\in [k]$. Now consider the case where $a\neq \emptyset$. Since $M^i\succeq M^j$ for all $i\le j$, by Lemma~\ref{lem:polarity}, we have $s_{M^i}(a) = \emptyset$ and $next_{M^i}(a')=a$ for $i=0,\dots,k$. We again conclude $b\notin \rho_i$ for all $i\in [k]$. Thus, we deduce $M(b)=M^\L_z(b)$. Now assume by contradiction that $a'b\in M'$ for some legal assignment $M'$. First note that $M'\succeq M^\L_z$ because $M^\L_z$ is the school-optimal legal assignment and thus the school-optimal stable assignment in $(G_L,<,\q)$ due to Theorem~\ref{thr:main1}. Also note that $M$ and $M'$ do not block each other given that both are legal assignments. Moreover, since $M(a')>_{a'} b= M'(a')$ by construction of $D_A$, we can apply Lemma~\ref{lem:magic-wand} (with $a=a'$, $\overline b = b$) and conclude that there exist $\overline{a} \in M(b)$ such that $b>_{\overline{a}} M'(\overline{a})$. However, $M'\succeq M^\L_z$ implies $M'(\overline{a}) \ge_{\overline{a}} M^\L_z(\overline{a})$ and $\overline{a}\in M(b)$ implies $M^\L_z(\overline{a}) = M(\overline{a})=b$. Hence, $M'(\overline{a}) \ge_{\overline{a}}b$, a contradiction.
\hfill{\Halmos}\endproof

%
%
% ------------------------------------------
%
%

\section{Details of implementations}

As all implementation of our algorithms run in time $O(|E|)$, we can preprocess the input in time $O(|E|)$ and assume that: for each agent, we have its preference list given as an ordered list; given $a \in A, b\in B$ and an assignment $M$, in constant time we can access $M(a)$ and the least preferred student in $M(b)$; given $x \in A \cup B$ and two neighbors $y_1$ and $y_2$ of $x$, we can, in constant time, decide if $y_1>_x y_2$. 

%
%
% ------------------------------------------
%
%

\subsection{Proof of Lemma~\ref{lem:GS-complexity}} \label{app:proof:GS-complexity} 

\proof{Proof.}
    We consider the case where students propose, the other one following in a similar fashion. Since each student proposes to each school at most once, there are $O(|E|)$ proposals. Throughout the algorithm, we keep a list of students who are not temporarily accepted and still have schools to apply to on their preference lists. At each iteration, we pick the first student on this list and he will be the proposing student at this iteration. In addition, for each student, we keep a pointer to the last school they applied to. This gives us faster access to the school he will propose to next. Since there are $O(|E|)$ rejections, these updates take $O(|E|)$ time.

    Each school $b\in B$ maintains three pieces of information throughout the algorithm. They are: a Boolean  array $\ell_b$ of size $|A|$ recording its currently assigned students; an integer $c_b$ recording the number of students it is assigned to; and the least preferred student $w_b$ it is assigned to. Now consider the iteration where $a$ proposes to $b$. If $c_b<q_b$, information update is simple and requires constant time. Assume $c_b = q_b$ where rejection happens. If $w_b>_b a$, there is nothing to update. We are left with the case where $a>_b w_b$. In such case, $b$ rejects $w_b$ and accepts $a$, so the update on $\ell_b$ is straightforward. Since the least preferred student of each school improves over the course of the algorithm, we may update $w_b$ by looking forwards on $b$'s preference list, starting from $w_b$, until reaching a student $a$ such that $\ell_b(a) = 1$. Thus, throughout the algorithm, for each school $b$, all such updates require $O(\deg_G(b))$ time. Altogether, the algorithm runs in time $O(|E|)$. 
\hfill{\Halmos}\endproof

%
%
% ------------------------------------------
%
%

\subsection{Proof of Theorem~\ref{thr:time-complexity-rotate-remove}} \label{app:proof:time-complex-rrr}

Before going into the proof, we first state the following lemma~\cite{metarot}. It implies that the number of student-rotations an instance can have is at most linear in the number of edges. 

\begin{lemma}\label{lem:a-to-b} 
    For any $xy\in E$, there is at most one $X$-rotation $y_0,x_0,y_1,\dots, y_{r-1}, x_{r-1}$ exposed in some stable assignment of $(G,<,\q)$ such that $x=x_i$ and $y=y_i$ for some $i \in \{0,1,\cdots, r-1\}$.
\end{lemma}

\proof{Proof of Theorem~\ref{thr:time-complexity-rotate-remove}.}
    We show details for Algorithm~\ref{alg:rar} with $X$ taken as the set of schools (i.e., \texttt{school-rotate-remove}), as those with $X$ taken as the set of students follow in a similar fashion. For simplicity, we call ``school-rotations'' simply ``rotations'' throughout the proof. 

    Algorithm~\ref{alg:rar} first finds the student-optimal stable assignment. This takes time $O(|E|)$ by Lemma~\ref{lem:GS-complexity}. Then the algorithm enters the \textbf{while} loop. A key fact we will resort to multiple times in our arguments is the following: $(\ddagger)$ for any pair of iterations $i,j$ such that $i>j$, we have $M^i\succeq M^j$. Given an assignment $M$, we say that we \emph{scan} an edge $ab$ when we check  $b>_a M(a)$ given the condition $M(b)>_b a$ (this is required to compute $s_{M}(b)$). From what assumed above, scanning $ab$ requires constant time.  We denote by $b(i)$ the student at the $i^{\text{th}}$ position on the preference list of $b$. Assume schools are sorted as $B=\{\overline{b}_1, \overline{b}_2, \cdots, \overline{b}_{|B|}\}$. Recall that, for all $i$, all sinks of $D^i$ are either school nodes or the node corresponding to $\emptyset$.

    At each iteration $i$, we keep the assignment $M^i$ as an $|A|$-dimensional array with the $k^{\text{th}}$ position recording the school the $k^{\text{th}}$ student is assigned to; a partial list $T^i$ of sinks of $D^i$, with one entry per node of $B$, stored as a $0/1$ Boolean array of dimension $|B|$; a position $f$ such that $\overline{b}_{f}$ is the first school that is not in $T^i$; a directed path $P^i$ of $D^i$, stored as a doubly-linked list; a Boolean array $W^i$ recording whether a school $b$ is in $P^i$; and for each $b \in B$, a position $p_b$ such that, in determining $s_{M^i}(b)$, we do not need to scan $ab$ for all $a$ such that $a\ge_b b(p_b)$. We initialize $M^0=M_0$, $T^0:=\emptyset$, $f:=1$, $P^0:= []$, $W^0:= \emptyset$, and $p_b$ to be the position of the least preferred student in $M^0(b)$ on $b$'s preference list for every $b \in B$. Clearly the initialization takes $O(|E|)$ time.

    We start by showing, for each iteration $i$, how to update the aforementioned pieces of information through two series of operations: those \ul{underlined in the text}, which require constant time, and those \uwave{wave underlined}. Second, we show the correctness of these updates. Lastly, we bound the running time of the algorithm by investigating the number of times we repeat each of underlined operations and the total time needed to perform wave underlined operations. 

    For each iteration of the \textbf{while} loop, we perform the following updates.

    \begin{itemize}
    	\item If $P^i$ is empty, we select the first school that is not in $T^i$ and add it to $P^i$. This school can be obtained by \ul{checking if $\overline{b}_{f} \in T^i$} and, while $\overline{b}_{f} \in T^i$, \ul{updating $f:= f+1$}. So we may assume $P^i$ is non-empty, and represented as $P^i=b_0, a_1, b_1, \cdots, a_k, b_k$. 
    	\item Within the iteration, we extend $P^i$ and simultaneously maintain $W^i$, by finding $a_{k+1} = s_{M^i}(b_k)$, $b_{k+1} = next_{M^i}(b_{k})$, $\cdots$ until we reach a node $b_j$ such that either (1) $b_j$ is a sink (step~\ref{step:sink-found}); or (2) $next_{M^i}(b_j) = b_\ell$ for some $\ell<j$ (step~\ref{step:rotation-found}). In particular,
    	\begin{itemize}[topsep=0ex, itemsep=0ex, leftmargin=3ex]
    		\item[a)] \ul{Check if $b_k=\emptyset$}. If so, we are in case (1). Otherwise, to obtain $s_{M^i}(b_k)$, we will repeatedly \ul{update $p_{b_k}:=p_{b_k}+1$} until \ul{$p_{b_k}>\deg_G(b_k)$} (i.e., $b_k$ is a sink and we are in case (1) above) or by \ul{scanning of $b_k(p_{b_k})b_k$} we deduce $s_{M^i}(b_k) = b_k(p_{b_k})$. 
    		\item[b)] If $s_{M^i}(b_k)$ is found, we \ul{check if $b_{k+1}:=next_{M^i}(b_k) \in W^i$}. If this happens, we are in case (2) above, otherwise we set $k:=k+1$, and go to a). 
    		
    	\end{itemize}
    	\item In case (1), $a_jb_{j-1}$ is removed from $G^i$ as an illegal edge. We achieve this by setting \uline{$P^{i+1}:= P^i\setminus \{a_j, b_j\}$}, and if $b_j \neq \emptyset$, we also set \uline{$W^{i+1}:=W^i \setminus \{b_j\}$} and \ul{update $T^{i+1}: = T^i \cup \{b_j\}$}.
    	\item In case (2), a school-rotation exposed in $M^i$ -- corresponding to the directed cycle $C^i= b_\ell, \cdots, b_j, a_{j+1}$ -- is found and eliminated, as to \uwave{construct $M^{i+1}$ from $M^i$}. We update \ul{$p_{b_{\ell-1}}:= p_{b_{\ell-1}} -1$} if $\ell> 0$, and set \uwave{$P^{i+1}:= P^i\setminus C^i$}, \uwave{$W^{i+1}:= W^i \setminus C^i$}. 
    \end{itemize}

    This shows that storing and updating $T^i, f, P^i, W^i, \{p_b\}_{b\in B}$, together with $M^i$, are sufficent for the execution of the algorithm.

    We will now argue about the correctness of these updates. In both cases (1) and (2), $P^{i+1}$ is a directed path of $D^{i+1}$ and $W^{i+1}$, $M^{i+1}$ are correctly computed. Moreover, because of ($\ddagger$), sinks of $D^i$ are also sinks in $D^{i+1}$, justifying the update on $T^i$ and $f$. Lastly, consider any node $b \in B$ whose associated position $p_b$ is updated in this iteration. There are two scenarios. The first scenario is when looking for $s_{M^i}(b)$, where $p_b$ is repeatedly updated until $p_b>\deg_G(b)$ or until $b(p_b)$ is added to the directed path $P^i$. In either case, because of ($\ddagger$) and the fact that every time $p_b$ is updated, it is incremented only by $1$, the updated $p_b$ remains a good choice for our purpose. The second scenario is when $b=b_{\ell-1}$, where $p_b$ is updated to be $p_b-1$. In this case, we found a rotation $\rho$ with $b\notin \rho$ and $next_{M^i}(b)\in \rho$. We carry out the decrement because it is possible to have $s_{M^{i+1}}(b) = s_{M^i}(b)$ and thus re-scanning of $s_{M^i}(b)b$ is required. No further decrements on $p_b$ is needed again because of $(\ddagger)$.

    Finally, we will argue about the time complexity. First note that the number of iterations is clearly bounded by the number of edges plus the number of rotations eliminated. Because of Lemma~\ref{lem:rotations}, all rotations eliminated throughout the algorithm are also rotations in $\SR(G_L,<,\q)$. Note that if $M^i$ is obtained from $M^{i-1}$ by eliminating a rotation, $M^i \succ M^{i-1}$. Hence, no rotation is eliminated twice and thus the number of rotations eliminated is $O(|E|)$ due to Lemma~\ref{lem:a-to-b}. Thus, the number of iterations is $O(|E|)$. 

    The total number of updates on \uline{$P^i$}, \uline{$W^i$} and \uline{$T^i$} in case (1) is then also $O(|E|)$. Since $f$ only increases, we \ul{update $f:=f+1$} at most $O(|V|)$ times. The number of times we \ul{check if $\overline{b}_f \in T^i$} is given by the number of positive answers (proportional to the number of updates of $f$) plus the number of negative answers (proportional to the number of iterations), hence $O(|E|)$. The number of times $\{p_b\}_{b\in B}$, is updated is given by the number of times we \underline{update $p_b:=p_b+1$} (proportional to the number of edges) plus twice the number of times we \ul{update $p_b:=p_b-1$} (proportional to the number of rotations), hence $O(|E|)$. We claim that we \ul{scan each edge} at most once, with $O(|E|)$ exceptions. From the update on $p_b$, we see that the only time an edge $ab$ is scanned more than once is when a rotation is eliminated and $b=b_{\ell-1}$, $a=a_\ell$. We call this an \emph{exception}. Since every rotation corresponds to at most one exception, the number of exceptions does not exceed the number of rotations, which is $O(|E|)$. Note that each time we check if $p_b> \deg_G(b)$, we either find a sink (which happens at most once per iteration), or we scan an edge (which has been shown to happen $O(|E|)$ times). Hence, the number of times we \ul{compare $p_b$ and $\deg_G(b)$} is $O(|E|)$. In addition, the number of times we \ul{check if $next_{M^i}(b_j) \in W^i$} is upper bounded by the number of edge scans, hence $O(|E|)$. The number of times we \uline{check if $b_k\neq \emptyset$} is upper bounded by the number of times we \ul{check if $next_{M^i}(b_j) \in W^i$}, so again $O(|E|)$. The number of individual entry updates when   \uwave{constructing $M^{i+1}$ from $M^i$}, \uwave{$P^{i+1}$ from $P^i$} and \uwave{$W^{i+1}$ from $W^i$} in case (2) is upper bounded by the number of edges in all rotations from $\SR(G_L, <, \q)$, which is $O(|E|)$ from Lemma~\ref{lem:a-to-b}, concluding the proof.
\hfill{\Halmos}\endproof

%
%
% ------------------------------------------
%
%

\subsection{Proof of Lemma~\ref{lem:rrr-conset-impl}} \label{app:proof:rrr-consent-impl}

\proof{Proof.}
    The implementation follows as in the proof of Theorem~\ref{thr:time-complexity-rotate-remove}. The only modification regards the update of $T^{i+1}$ in case (1) considered in the proof, which is when extending the directed path $P^i$, we encounter a node $b_j$ that is a sink. If $a_j$ consents, then the update on $T^{i+1}$ remains unchanged, which is to set $T^{i+1}:= T^i\cup \{b_j\}$; however, if $a_j$ is nonconsenting, we set $T^{i+1}:= T^i\cup \{b_j, b_{j-1}\}$ and update $p_{b_{j-1}}:=\deg(b_{j-1})+1$. Correctness analysis and the counting arguments used for time complexity analysis in the proof of Theorem~\ref{thr:time-complexity-rotate-remove} remain valid.
\hfill{\Halmos}\endproof

%
%
% ------------------------------------------
%
%

\section{Proof of Outcome-Equivalence} \label{app:seadam-outcomeequiv}

The goal of this section is to prove Theorem~\ref{thm:rrr-eadam-equiv}. The proof consists of three steps: first, in section~\ref{app:alg-rrr-consent-all-same}, we show that all executions of Algorithm~\ref{alg:rrr-consent} give the same output; then, in section~\ref{app:seadam}, we introduce an outcome-equivalent version of EADAM, called \emph{Simplified EADAM}, from~\citet{Tang.Yu}; and lastly, in section~\ref{sec:the-end}, we show that the output of a specific execution of Algorithm~\ref{alg:rrr-consent} and that of the Simplified EADAM coincide.

%
%
% ------------------------------------------
%
%

\subsection{Uniqueness of the output of Algorithm~\ref{alg:rrr-consent}}\label{app:alg-rrr-consent-all-same}

From now on, fix the input $(G(A\cup B, E), <, \q), \overline{A})$ of Algorithm~\ref{alg:rrr-consent}. An \emph{execution} of Algorithm~\ref{alg:rrr-consent} on the input is an ordered collection of \emph{iterations}, where iteration $i$ denotes the $i$-th repetition of the \emph{while} loop from Step \ref{step:while}. Hence, in iteration $i$, Algorithm~\ref{alg:rrr-consent} takes from the previous iteration graph $G^{i-1}$, assignment $M^{i-1}$, and rotation digraph $D^{i-1}$ and creates $G^i$, $M^i$, and $D^{i}$. We identify iteration $i$ by the cycle found in $D^{i-1}$ or the pair of arcs $(b',a),(a,b)$ with $b$ being a sink found in $D^{i-1}$. Hence, for an execution ${\mathcal{E}}$ of Algorithm~\ref{alg:rrr-consent}, we let ${\mathcal{E}}=(I^1,I^2,\dots)$, where each $I^i$ denotes iteration $i$. Note that in particular, $I^i$ is a subgraph of $D^{i-1}$. Let $G_{\mathcal{E}}^i$, $M_{\mathcal{E}}^i$, and $D_{\mathcal{E}}^i$ denote $G^i$, $M^i$, and $D^{i}$ under execution $\mathcal{E}$. The collection of all possible executions is denoted by $\mathbb{E}$. We start with several useful observations.

\begin{lemma}\label{lem:one-iteration-rrr}
    Let ${\mathcal{E}} \in \mathbb{E}$. $\mathcal E$ contains a finite number $k$ of iterations. Moreover, for every $i\in [k]$, we have $M_{\mathcal{E}}^i\succeq M_{\mathcal{E}}^{i-1}$.
\end{lemma}

\proof{Proof.}  
    At each iteration $i$, either case (i) or case (ii) is found. In case of case (i), we have $M_{\mathcal{E}}^i=M_{\mathcal{E}}^{i-1}$ and some edges are removed from $G_{\mathcal{E}}^{i-1}$. In case of case (ii), a school-rotation exposed in $M_{\mathcal{E}}^{i-1}$ is eliminated and thus, we have $M_{\mathcal{E}}^i\succ M_{\mathcal{E}}^{i-1}$. This proves immediately the second thesis. The first follows from the fact that the number of edges and the number of assignments of the instance are both finite.
\hfill{\Halmos}\endproof

Lemma~\ref{lem:one-iteration-rrr} implies that for every ${\mathcal{E}}=(I^1, I^2, \dots,I^k) \in \mathbb{E}$ and $i, j \in [k]$, if $i\neq j$, then $I^i \neq I^j$. 

\begin{lemma}\label{lem:rrr-unique}
	Let ${\mathcal{E}}_1=(I_1^1, I_1^2, \cdots, I_1^{k_1})$ and ${\mathcal{E}}_2=(I_2^1, I_2^2, \cdots, I_2^{k_2})$ be two executions such that $\{I_1^1, I_1^2, \cdots, I_1^{k}\} = \{I_2^1, I_2^2, \cdots, I_2^{k}\}$ for some $k\le \min(k_1, k_2)$. Then, $M_{\mathcal{E}_1}^{k} = M_{\mathcal{E}_2}^{k}$ and $G_{\mathcal{E}_1}^{k} = G_{\mathcal{E}_2}^{k}$.
\end{lemma}

\proof{Proof.} 
    Note that both executions start from the same assignment $M_0$. Let $\rho_1, \rho_2, \cdots, \rho_\ell$ be all the rotations eliminated in the first $k$ iterations of execution $\mathcal{E}_1$ and let $\rho'_1, \rho'_2, \cdots, \rho'_{\ell'}$ be those of execution $\mathcal{E}_2$. Then, since the first $k$ iterations of these two executions coincide, we have $\{\rho_1, \rho_2, \cdots, \rho_\ell\} = \{\rho'_1, \rho'_2, \cdots, \rho'_{\ell'}\}$ and thus, $M_{\mathcal{E}_1}^k = M_0/\rho_1/\rho_2/\cdots/\rho_\ell$ and $M_{\mathcal{E}_2}^k = M_0/\rho'_1/\rho'_2/\cdots/\rho'_{\ell'}$ coincide. Similarly, let $\widetilde E_1$ and $\widetilde E_2$ be the set of edges removed in the first $k$ iterations of $\mathcal{E}_1$ and $\mathcal{E}_2$ respectively. Again, because $\{I_1^1, I_1^2, \cdots, I_1^{k}\} = \{I_2^1, I_2^2, \cdots, I_2^{k}\}$, we have $\widetilde E_1 = \widetilde E_2$ and thus, $G_{\mathcal{E}_1}^{k} = G_{\mathcal{E}_2}^{k}$.
\hfill{\Halmos}\endproof

\begin{lemma} \label{lem:iter-next-cycle}
    Let $\mathcal{E}\in \mathbb{E}$ and assume $\mathcal{E} = (I^1, I^2, \cdots, I^k)$. Assume at some iteration $j< k$, rotation digraph $D_{\mathcal{E}}^{j-1}$ contains a cycle $I'$ and $I'\neq I^j$. Then, $I'$ must also be a subgraph of $D_{\mathcal{E}}^{j}$. Moreover, there exists a unique $i\in \{j+1, \cdots, k\}$ such that $I'=I^i$.
\end{lemma}

\proof{Proof.} 
    Since every vertex in the rotation digraph $D_{\mathcal{E}}^{j-1}$ has outdegree at most $1$, $I'$ and $I^j$ must be vertex-disjoint. Thus, for every vertex $x\in I'$, we have $M_{\mathcal{E}}^{j-1} (x) = M_{\mathcal{E}}^j (x)$ and for all $b\in I'\cap B$, we have $\{a: ab\in E( G_{\mathcal{E}}^{j-1})\} = \{a: ab\in E(G_{\mathcal{E}}^{j})\}$. Since $M_{\mathcal{E}}^{j}\succeq M_{\mathcal{E}}^{j-1}$ by Lemma~\ref{lem:one-iteration-rrr}, we can conclude that $s_{M_{\mathcal{E}}^{j-1}}(b)=s_{M_{\mathcal{E}}^{j}}(b)$ for all $b \in I'\cap B$. Hence, $I'$ is a subgraph of $D_{\mathcal{E}}^{j}$. The second thesis follows from repeated application of the first thesis and the termination criterion of Algorithm~\ref{alg:rrr-consent}, and uniqueness holds because for $\ell_1, \ell_2\in [k]$, if $\ell_1\neq\ell_2$, then $I^{\ell_1} \neq I^{\ell_2}$ as implied by Lemma~\ref{lem:one-iteration-rrr}.
\hfill{\Halmos}\endproof

\begin{lemma} \label{lem:iter-next-sink}
    Let $\mathcal{E}\in \mathbb{E}$ and assume $\mathcal{E} = (I^1, I^2, \cdots, I^k)$. Assume at some iteration $j< k$, rotation digraph $D_{\mathcal{E}}^{j-1}$ contains a pair of arcs $I'=(b',a), (a,b)$ with $b$ being a sink of $D_{\mathcal{E}}^{j-1}$ and $I'\neq I^j$. Then, $I'$ must also be a subgraph of $D_{\mathcal{E}}^{j}$ with $b$ being a sink in $D_{\mathcal{E}}^{j}$. Moreover, there must exists a unique $i\in \{j+1, \cdots, k\}$ such that $I'=I^i$.
\end{lemma}

\proof{Proof.} 
    Since every vertex in the rotation digraph $D_{\mathcal{E}}^{j-1}$ has outdegree at most $1$, we must have $b'\notin I^j$. Note that if $I^j$ is a directed cycle, then $I^j$ and $I'$ are vertex-disjoint, otherwise, it is possible to have $(a,b)\in I^j$. Nevertheless, we have that for $x\in \{b', b\}$, $M_{\mathcal{E}}^{j-1} (x) = M_{\mathcal{E}}^j (x)$ and $\{a: ax\in E( G_{\mathcal{E}}^{j-1})\} = \{a: ax\in E(G_{\mathcal{E}}^{j})\}$. Since $M_{\mathcal{E}}^{j}\succeq M_{\mathcal{E}}^{j-1}$ by Lemma~\ref{lem:one-iteration-rrr}, we can conclude that $s_{M_{\mathcal{E}}^{j-1}}(b')=s_{M_{\mathcal{E}}^{j}}(b')$ and thus, $I'$ is a subgraph of $D_{\mathcal{E}}^{j}$ and $b$ is a sink in $D_{\mathcal{E}}^{j}$. The second thesis follows as in the proof of Lemma~\ref{lem:iter-next-cycle}.
\hfill{\Halmos}\endproof

\begin{lemma} \label{lem:iter-switch}
    Let ${\mathcal{E}} = (I^1, I^2, \cdots, I^k) \in \mathbb{E}$ and assume for some iteration $j\in \{2, 3, \cdots, k\}$, $I^j$ is a subgraph of $D_{\mathcal{E}}^{j-2}$ and if $I^{j} = (b',a),(a,b)$, we also have $b$ being a sink of $D_{\mathcal{E}}^{j-2}$. Then, we have $\mathcal{E}':=(I^1, I^2, \cdots, I^{j-2}, I^{j}, I^{j-1}, I^{j+1} \cdots, I^k) \in \mathbb{E}$.
\end{lemma}

\proof{Proof.} 
    Let $\bar M^i$, $\bar G^i$, and $\bar D^i$ be the assignment, graph, and rotation digraph after the first $i$ iterations of $\mathcal{E}'$. For $i\le j-2$, they are well defined (i.e., $\bar M^i = M_{\mathcal{E}}^i$) since the first $j-2$ iterations of $\mathcal{E}$ and $\mathcal{E}'$ are exactly the same. Thus, $I^j$ is either a case (i) at Step~\ref{clause:illegal-edge} or a case (ii) at Step~\ref{clause:rotation} of $\bar D^{j-2}=D^{j-2}_{\mathcal{E}}$. Therefore, $\bar M^i$, $\bar G^i$, and $\bar D^i$ are also well-defined for $i=j-1$. Now, because of Lemma~\ref{lem:iter-next-cycle} and Lemma~\ref{lem:iter-next-sink} with $I' = I^{j-1}$, they are also well-defined for $i=j$. Lastly, since the first $j$ iterations of $\mathcal{E}$ and $\mathcal{E}'$ coincide, we know $\bar M^j = M_{\mathcal{E}}^j$ and $\bar G^j = G_{\mathcal{E}}^j$ due to Lemma~\ref{lem:rrr-unique}. Since all iterations after the $j^\text{th}$ iteration are exactly the same and in the same order in $\mathcal{E}$ and $\mathcal{E}'$, we also know that for all $i>j$, $\bar M^i = M_{\mathcal{E}}^i$ and $\bar G^i = G_{\mathcal{E}}^i$. This concludes the proof.
\hfill{\Halmos}\endproof

\begin{corollary}\label{cor:iter-forward}
    Let $\mathcal{E} = (I^1, I^2, \cdots, I^k) \in \mathbb{E}$ and assume for some iterations $j_1, j_2\in [k]$ with $j_1<j_2$, $I^{j_2}$ is a subgraph of $D_{\mathcal{E}}^{j_1-1}$, and if $I^{j_2} = (b',a),(a,b)$, we also have $b$ being a sink of $D_{\mathcal{E}}^{j_1-1}$. Then, $\mathcal{E}':=(I^1, I^2, \cdots, I^{j_1-1}, I^{j_2}, I^{j_1}, \cdots, I^{j_2-1}, I^{j_2+1} \cdots, I^k) \in \mathbb{E}$.
\end{corollary}

\proof{Proof.} 
    Because of Lemma~\ref{lem:iter-next-cycle} and Lemma~\ref{lem:iter-next-sink}, $I^{j_2}$ is a subgraph of $D_{\mathcal{E}}^{j_2-2}$and if $I^{j_2} = (b',a),(a,b)$, we also have $b$ being a sink of $D_{\mathcal{E}}^{j_2-2}$. Thus, due to Lemma~\ref{lem:iter-switch}, $\mathcal{E}_1 = (I^1, \cdots, I^{j_2-2}, I^{j_2}, I^{j_2-1}, I^{j_2+1}, \cdots, I^k)\in \mathbb{E}$. Repeatedly applying the argument and moving the iteration $I^{j_2}$ to earlier steps, we can conclude that $\mathcal{E}'\in \mathbb{E}$.
\hfill{\Halmos}\endproof

For two executions ${\mathcal{E}}_1=(I_1^1, I_1^2, \cdots, I_1^{k_1})$ and ${\mathcal{E}}_2=(I_2^1, I_2^2, \cdots, I_2^{k_2})$, let $C(\mathcal{E}_1, \mathcal{E}_2)$ denote the largest $i\le \min(k_1, k_2)$ such that $(I_1^1, I_1^2, \cdots, I_1^{i}) = (I_2^1, I_2^2, \cdots, I_2^{i})$ in the same order. We can now prove the following theorem. 

\begin{theorem}\label{thm:rrr-consent-unique}
    The output of Algorithm~\ref{alg:rrr-consent} is unique.
\end{theorem}

\proof{Proof.} 
    Assume by contradiction that there are two executions $\mathcal{E}_1=(I_1^1, I_1^2, \cdots, I_1^{k_1})\in \mathbb{E}$ and $\mathcal{E}_2=(I_2^1, I_2^2, \cdots, I_2^{k_2})\in \mathbb{E}$ such that $M_{\mathcal{E}_1}^{k_1} \neq M_{\mathcal{E}_2}^{k_2}$.  Also assume that among all executions that output distinct assignments, $\mathcal{E}_1$ and $\mathcal{E}_2$ are the ones with the largest value $C(\mathcal{E}_1, \mathcal{E}_2)$. Let $j:=C(\mathcal{E}_1, \mathcal{E}_2)+1$. That is, we assume $I_1^i=I_2^i$ for all $i \in [j-1]$ but $I_1^j\neq I_2^j$. 
    
    Because of Lemma~\ref{lem:rrr-unique}, we know $G_{\mathcal{E}_1}^{j-1} = G_{\mathcal{E}_2}^{j-1}$ and $M_{\mathcal{E}_1}^{j-1} = M_{\mathcal{E}_2}^{j-1}$. Thus, $I_2^j$ is also a subgraph of $D_{\mathcal{E}_1}^{j-1}$ and if $I_2^j = (b',a),(a,b)$, we also have $b$ being a sink of $D_{\mathcal{E}_1}^{j-1}$. Thus, due to Lemma~\ref{lem:iter-next-cycle} and Lemma~\ref{lem:iter-next-sink}, there must exist a unique $\ell>j$ such that $I_2^j = I_1^\ell$. Therefore, we can apply Corollary~\ref{cor:iter-forward} on $\mathcal{E}_1$ with $j_2 = \ell$ and $j_1 = j$ and conclude that $\mathcal{E}_1' := (I_1^1, I_1^2, \cdots, I_1^{j-1}, I_1^{\ell}, I_1^{j}, \cdots, I_1^{\ell-1}, I_1^{\ell+1} \cdots, I_1^{k_1}) \in \mathbb{E}$. Because of Lemma~\ref{lem:rrr-unique}, we know $M_{\mathcal{E}_1}^{k_1} = M_{\mathcal{E}'_1}^{k_1}$ and thus, $M_{\mathcal{E}'_1}^{k_1}\neq M_{\mathcal{E}_2}^{k_2}$. However, $C(\mathcal{E}'_1, \mathcal{E}_2) = j > C(\mathcal{E}_1, \mathcal{E}_2)$ contradicts the choice of $\mathcal{E}_1$ and $\mathcal{E}_2$.
\hfill{\Halmos}\endproof

%
%
% ------------------------------------------
%
%

\subsection{Simplified EADAM}\label{app:seadam}

In this section, we introduce a simplified and outcome-equivalent version of EADAM by \citet{Tang.Yu}. The key concept exploited by~\citet{Tang.Yu} is that of \emph{underdemanded} schools at an assignment $M$. A school $b \in B$ is underdemanded in $M$ if there is no student $a$ that strictly prefers $b$ to $M(a)$. They observe that, at the student-optimal stable assignment $M_0$, a student $a$ that is assigned to an underdemanded school is not \emph{Pareto-improvable}. That is, if an assignment $M'$ dominates $M_0$, it must be that $M_0(a)=M'(a)$. With this key observation, they develop the simplified EADAM algorithm and show that it is output-equivalent to~\citet{Kesten}'s original algorithm. 

\begin{algorithm}[h]
\caption{simplified EADAM\label{alg:sEADAM}}
\begin{algorithmic}[1] 
	\REQUIRE $(G(A\cup B,E),<,\q)$, consenting students $\overline{A}\subseteq A$
	\STATE Let $G^0=G$ and $i=0$.
	\REPEAT
	\STATE Run student-proposing Gale-Shapley's algorithm on $(G^i,<,\q)$ to obtain assignment $M^i$. 
	\STATE Identify underdemanded schools $B^i$ in $M^i$ and their assigned students $A^i:= \cup_{b\in B^i} M^i(b)$.
	\STATE Set $G^{i+1}=G^i$. 
	\FOR {$a \in A^i$}
	\FOR {$b \in B$ such that $ab \in E(G^{i+1})$ and $b>_aM^i(a)$}
	\STATE remove edge $ab$ from $G^{i+1}$.\label{step:seadam-remove-edges}
	\IF{$a \notin \overline{A}$}
	\STATE{remove edge $a'b$ from $G^{i+1}$ for all $a' \in A$ such that $a>_b a'$ and $a'b \in E(G^{i+1})$.} \label{step:seadam-rm-edges-nonconsent}
	\ENDIF
	\ENDFOR
	\ENDFOR
	\STATE Set $i=i+1$.
	\UNTIL{$B^{i-1} = B$}
	\STATE{Output $M^{i-1}$.}
\end{algorithmic}
\end{algorithm}

Simplified EADAM takes as input an instance and a list of \emph{consenting} students, and similarly to Kesten's original algorithm, it iteratively re-runs Gale-Shapley's procedure. In each iteration, it identifies underdemanded schools and fixes their assignments via deletion of edges. If a non-consenting student is matched to an underdemanded school, more edges are removed from the instance in order to respect his priorities. A formal description is presented in Algorithm~\ref{alg:sEADAM} and an example is given in Example~\ref{ex:seadam} for the same instance as in Example~\ref{ex:eadam-seadam}. The following theorem is proved in~\citet{Tang.Yu}.

\begin{theorem}\label{thm:eadam-seadam-equiv} 
	For any given input, the outputs of Algorithm~\ref{alg:Kesten-EADAM} and Algorithm~\ref{alg:sEADAM} coincide.
\end{theorem}

\begin{example}\label{ex:seadam}
    Consider the instance given in Example~\ref{ex:eadam-seadam}. 
    
    \noindent \textbf{Iteration \#1:} The iteration starts with the student-proposing Gale-Shapley's algorithm, whose steps can be found in Example~\ref{ex:eadam-seadam}. Since $b_4$ never rejects any students in the execution, no student strictly prefers $b_4$ to his current assignment, and thus, $b_4$ is an underdemanded school in $M^0$. In fact, $b_4$ is the only underdemanded school. Hence, $B^0=\{b_4\}$ and $A^0=\{a_3\}$. Simplified EADAM then settles assignment $a_3b_4$ by removing edges $a_3b_3$ and $a_3b_2$ from the instance, as in Step~\ref{step:seadam-remove-edges} of Algorithm~\ref{alg:sEADAM}. In addition, since $a_3$ is not consenting, edges $b_2a_1, b_2a_4$, and $b_3a_2$ are removed to respect his priority at school $b_3$ and $b_2$, as in Step~\ref{step:seadam-rm-edges-nonconsent} of the algorithm. 
    
    \noindent \textbf{Iteration \#2:} Re-running Gale-Shapley's algorithm on the updated instance:
	$$\begin{array}{cccclclcl}
	\text{step} & & b_1 & & b_2 & & b_3 & & b_4\\
	\hline
	1 & \quad \qquad & \xcancel{a_1}, a_2 & \quad \qquad &  & \quad \qquad & a_4 & & a_3\\
	2 & &  & &  & & a_1, \xcancel{a_4} & &\\
	3 & & \xcancel{a_2}, a_4 & &  & & & &\\
	4 && && a_2 && & &
	\end{array}$$
	
	The resulting assignment is $M^1 = \{a_1b_3, a_2b_2, a_3b_4, a_4b_1\}$. $b_2$ is an additional underdemanded school in $M^1$ and its assigned student $a_2$ is consenting. So simplified EADAM simply fixes assignment $a_2b_2$ by removing edge $a_2b_1$ from the instance. 
	
	\noindent \textbf{Iteration \#3:} The algorithm then runs Gale-Shapley's algorithm again on the updated instance:
	$$\begin{array}{lccclclcl}
	\text{step} & & b_1 & & b_2 & & b_3 & & b_4\\
	\hline
	1 & & a_1 & & a_2 & & a_4 & & a_3
	\end{array}$$

	The resulting assignment is $M^2=\{a_1b_1$, $a_2b_2$, $a_3b_4$, $a_4b_3\}$. Now, all schools are underdemanded. Hence, the algorithm terminates and outputs assignment $M^2$, which is equivalently to the assignment output of Kesten's EADAM that we obtained in Example~\ref{ex:eadam-seadam}. 
    \hfill{$\diamondsuit$}
\end{example}

%
%
% ------------------------------------------
%
%

\subsection{Equivalence between \texttt{School-Rotate-Remove with Consent} and Simplified EADAM}\label{sec:the-end}

The goal of this section is show that our Algorithm~\ref{alg:rrr-consent} is outcome-equivalent to this simplified version of EADAM algorithm, which together with Theorem~\ref{thm:eadam-seadam-equiv} implies Theorem~\ref{thm:rrr-eadam-equiv}. 

The following lemmas show an interesting connection between underdemanded schools and sinks in rotation digraphs (of any stable assignment).

\begin{lemma} \label{lem:sink-is-underdemanded}
	Consider the school-rotation digraph $D_B$ associated with an instance $(G,<,\q)$ at a stable assignment $M\in \SS(G,<,\q)$, and let $b \in B$. $b$ is a sink in $D_B$ if and only if it is an underdemanded school in $M$.
\end{lemma}

\proof{Proof.}
	Let $a$ be $b$'s least preferred student among $M(b)$. $b$ is a sink implies that for all students $a'$ such that $a>_b a'$, we have $M(a') >_{a'} b$. On the other hand, stability of $M$ implies that for all $a'$ such that $a'>_b a$ and $a'\notin M(b)$, we have $M(a')>_{a'} b$. These two cases conclude the proof for the ``only if'' direction. The other direction is clear from the construction of the rotation digraph.	
\hfill{\Halmos}\endproof

\begin{lemma}~\label{lem:sink-remain}
	In Algorithm~\ref{alg:rrr-consent}, let $b\in B$ be a sink in $D^i$ for some iteration $i$. Then, it remains a sink in $D^j$ for all $j\ge i$. Moreover, if $a\in M^i(b)$, $M^i(a) = M^j(a)$ for all $j\ge i$.
\end{lemma}

\proof{Proof.}
	The first part follows from the observation that $M^j\succeq M^i$ for all iterations $j\ge i$. For any $a\in M^i(b)$, since $b$ is a sink in $D^j$ for all $j\ge i$, $(a,b)$ is not part of a directed cycle of $D^j$ for any $j\ge i$. Thus, the assignment of $a$ remains for all iterations $j\ge i$.
\hfill{\Halmos}\endproof

Next, we will show that our \texttt{school-rotate-remove with consent} is outcome-equivalent to simplified EADAM. The main idea is to show that for any given input, a specific sequence of iterations of Algorithm~\ref{alg:rrr-consent} leads to an output that is the same as the one from Algorithm~\ref{alg:sEADAM}. This is sufficient because of Theorem~\ref{thm:rrr-consent-unique}. 

\begin{theorem}\label{thm:rrr-seadam-equiv}
	For any given input, the outputs of Algorithm~\ref{alg:rrr-consent} and Algorithm~\ref{alg:sEADAM} coincide.
\end{theorem}

\proof{Proof.} 
    To distinguish between notations in Algorithm~\ref{alg:sEADAM} and Algorithm~\ref{alg:rrr-consent}, we use $M_i$ (subscripts) to represent the assignment at iteration $i$ in Algorithm~\ref{alg:rrr-consent}, and $M^i$ (superscripts) for the assignment in Algorithm~\ref{alg:sEADAM}. We use $G^i$ to represent instances generated during Algorithm~\ref{alg:sEADAM}, and  $H^i$ to represent instances generated during Algorithm~\ref{alg:rrr-consent}. Underdemanded schools $B^i$'s are for Algorithm~\ref{alg:sEADAM} only, and rotation digraph $D^i$ are for Algorithm~\ref{alg:rrr-consent}. We call each while loop of Algorithm~\ref{alg:sEADAM} and Algorithm~\ref{alg:rrr-consent} an \emph{iteration} and a \emph{loop}, respectively. Note that loop $i$ takes assignment $M_{i-1}$ from the previous loop and produce $M_i$. Similarly, iteration $i$ starts with $M^{i-1}$ and outputs $M^i$.

    Because of Theorem~\ref{thm:rrr-consent-unique}, it suffices to show that there is a specific execution of Algorithm~\ref{alg:rrr-consent} whose output coincides with that of Algorithm~{\ref{alg:sEADAM}}. In particular, we consider an execution of  Algorithm~\ref{alg:rrr-consent} in which loops are partitioned in consecutive \emph{batches}, as follows. In the first batch, we eliminate any rotation that is found in the rotation digraph, but we can only enter the \textbf{if} clause in Step~\ref{clause:illegal-edge} if the sink $b$, described in case (i) of Step~\ref{step:rot-or-sink}, is a sink in $D^0$. After repeating this, we arrive at a loop $j_1$ where $D^{j_1}$ has no cycles, and none of its sinks with positive indegree is a sink in $D^0$. This is when the second batch starts. We call $j_1$ the last iteration of the first batch. Similarly, in the next batch, we can only enter the \textbf{if} clause if the sink $b$ in case (i) is a sink in $D^{j_1}$, and define $j_2,j_3,...$ analogously. Let $k$ be the last iteration of Algorithm~\ref{alg:sEADAM}. In the following paragraphs, we will show by induction on $i$ that for all $i=0,1, \cdots, k$, we have $M_{j_i}=M^i$, where $j_0=0$, and $D^{j_i}$ is the rotation digraph of $M^i$ in $(G^i, <, \q)$ ($\ddagger$) Note that by construction, $M^i$ is stable in $G^i$.

    The base case is when $i =j_i = 0$. The claim holds because $M^0=M_0$ is the student-optimal stable assignment in $(G^0, <, \q)$, and $D^0$ is the rotation digraph of $M^0$ in $(G^0, <, \q)$ since $G^0=H^0$. Next, assume the claim is true for all indices $t\le i$, and we show it for $i+1$. That is, we consider the loops in the $(i+1)^\text{th}$ \emph{batch} of Algorithm~\ref{alg:rrr-consent}. As described above, in this \emph{batch}, we can only enter the \textbf{if} clause if the sink $b$ in case (i) is a sink in $D^{j_i}$. 
    
    Fix $\overline{b}\in B$. Define the following sets: $S$ is the set of $a \in A$ such that $(\overline{b},a)\in A(D^{h})$ for some loop $h\in [j_i, j_{i+1}]$; $E^i$ (resp. $E^i_{nc}$) is the set of edges removed in some iteration $1,\dots,i$ of Algorithm~\ref{alg:sEADAM} at Step~\ref{step:seadam-remove-edges} (resp. Step~\ref{step:seadam-rm-edges-nonconsent}). We will show the following: 
    
    \noindent\textbf{(i).} \emph{If $\overline{b}\in B^i$, then $S=\emptyset$.}

    If $\overline{b}\in B^i$, it is a sink in $D^{j_i}$ by Lemma~\ref{lem:sink-is-underdemanded} and the inductive hypothesis. Moreover, it remains a sink in $D^h$ for all $h\geq j_i$ by Lemma~\ref{lem:sink-remain}. Thus, $S=\emptyset$.

    In (ii), (iii), and (iv) below, we therefore assume that $\bar b\notin B^i$ and $S\neq \emptyset$.

    \noindent\textbf{(ii).} \emph{If $(\bar{b},a)\in D^{\bar h}$ for some $\bar{h}\in [j_i, j_{i+1}]$, then $a\overline{b} \notin E_{nc}^{i+1}$}.

    Assume by contradiction that for some $a\in S$, $a\overline{b} \in E_{nc}^{i+1}$. Then, there must be an iteration $t\leq i$ and a nonconsenting student $a'$ such that $M^t(a')\in B^t\setminus B^{t-1}$, $a'>_{\overline{b}} a$, and $\overline{b}>_{a'} M^t(a')$ in $G^t$. That is, in Algorithm~\ref{alg:sEADAM}, both $a\overline{b}$ and $a'\overline{b}$ are removed when creating graph $G^{t+1}$. Since $M^t$ is stable in $(G^t, <, \q)$ and  $a'\overline{b} \in E(G^t)$, we must have $M^t(\overline{b}) >_{\overline{b}} a'$. By  inductive hypothesis, we know $M^t = M_{j_t}$; and by Lemmas~\ref{lem:sink-is-underdemanded} and~\ref{lem:sink-remain}, we know that school $M_{j_t}(a')$ is a sink in $D^{j_t}$, and $M_{t'}(a')=M_{j_t}(a')$ remains a sink for all $t'\ge j_t$. Thus, there must be $\bar{\jmath} \in [j_t, \bar{h})$ such that $(\overline{b},a')\in A(D^{\bar{\jmath}})$. In particular, $M_{\bar{\jmath}}(a')$ is a sink in $D^{\bar{\jmath}}$. Hence, we can wlog assume $\bar{\jmath}$ is the loop during which Algorithm~\ref{alg:rrr-consent} removes the edge $a'\overline{b}$. Since $a'>_{\overline{b}} a$, we must have $(\overline{b},a)\notin A(D^{r})$ for all $r\le \bar{\jmath}$. In addition, since $a'$ is nonconsenting, Step~\ref{step:remove-nonconsent-edge} of Algorithm~\ref{alg:rrr-consent} removes edge $a\overline{b}$ from the instance at the same loop $\bar{\jmath}$. Hence, $(\overline{b},a)\notin A(D^h)$ for any loop $h$, which implies $a\notin S$, a contradiction.

    \noindent \textbf{(iii).} \emph{Let $a \in S$, and $(\bar{b}, a)\in D^h$ for some $h\in [j_i, j_{i+1}]$. If $M_h(a)\in B^i$, then $(\overline{b}, a)$ is not part of a directed cycle in $D^{h'}$ for all $h'\ge j_i$. If $(\overline{b},a)\in D^{j_{i+1}}$, then $M_{j_{i}}(a)\notin B^i$.} 

    The first part follows from Lemma~\ref{lem:sink-is-underdemanded} and Lemma~\ref{lem:sink-remain}. The second part follows from the definition of $j_{i+1}$. In the following, we extend the definition of $s_M(\cdot)$ to any assignment $M$.

    \noindent\textbf{(iv).} \emph{Let $a \in S$ and $(\bar{b}, a)\in D^h$ for some $h\in [j_i, j_{i+1}]$. If $M_h(a)\notin B^i$ and $M_h\subseteq E(G^{i+1})$, then $a = s_{M_h}(\overline{b})$ in $(G^{i+1}, <, \q)$}. 

    We first show that $a\overline{b}\in E(G^{i+1})$. If not, then we must have $a\overline{b}\in E^{i+1}$, as we already showed in (ii) that $a\overline{b}\notin E^{i+1}_{nc}$. However, $a\overline{b}\in E^{i+1}$ implies $M_{j_i}(a) = M^i(a)\in B^i$ and thus $M_h(a)\in B^i$ by Lemma~\ref{lem:sink-is-underdemanded} and Lemma~\ref{lem:sink-remain}, a contradiction. Since $a = s_{M_h}(\overline{b})$ in $(H^h, <, \q)$, we only need to show that $a$ is the first student, in $G^{i+1}$, on $\overline{b}$'s preference list that prefers $\overline{b}$ to his assigned school. Assume by contradiction that there exists another student $a'$ such that $a' >_{\overline{b}} a$, $\overline{b}>_{a'} M_h(a')$, and $a'\overline{b} \in E(G^{i+1})$ but $a'\overline{b} \notin E(H^h)$. If $a'\overline{b}$ is removed at Step~\ref{step:remove-nonconsent-edge} of Algorithm~\ref{alg:rrr-consent}, then $a\overline{b}$ must also be removed at the same loop, contradicting $a = s_{M_h}(\overline{b})$ in $(H^h, <, \q)$. Thus, $a'\overline{b}$ must be removed at Step~\ref{step:remove-illegal-edge} of some loop. That is, there must exist a loop $\bar h<h$, where $\bar h\in [j_{\bar \imath}, j_{\bar \imath+1})$ for some $\bar \imath \le i$, such that $(\overline{b},a')\in A(D^{\bar h})$ and $M_{\bar h}(a')\in B^{\bar \imath}$. However, this means $a'\overline{b}\in E^{\bar \imath+1}\subseteq E^{i+1}$, which contradicts the assumption that $a'\overline{b} \in E(G^{i+1})$.

    In order to conclude the proof, let us first observe that $M_{j_i} = M^i\in \SS(G^{i+1}, <, \q)$. Indeed, consider the edges removed from $G^i$ to obtain $G^{i+1}$ in Algorithm~\ref{alg:sEADAM}. It is easy to see that none of the edges removed are in $M^i$. Thus, $M^i$ is an assignment in $G^{i+1}$. Moreover, since the order in preference lists remains from iteration to iteration, $M^i$ is stable in $(G^{i+1}, <, \q)$. We show next by induction that for all $h\in [j_i, j_{i+1}]$, $M_h\in \SS(G^{i+1},<,\q)$. Due to the inductive hypothesis of ($\ddagger$), we know the claim holds for the base case when $h= j_i$. Now, assume $M_h\in \SS(G^{i+1},<,\q)$ for some $h\in [j_i, j_{i+1}]$, we want to show that so is $M_{h+1}$. If a case (i) is found in loop $h+1$, then $M_{h+1}=M_h\in \SS(G^{i+1},<,\q)$. Otherwise, if case (ii) is found, then (iii) and (iv) imply that the rotation $\rho^h$ identified is a rotation exposed in $M_h$ in $(G^{i+1},<,\q)$ and thus, $M_{h+1}= M_h/\rho^h\in \SS(G^{i+1},<,\q)$.

    Hence, in particular, the last assignment $M_{j_{i+1}}$ obtained at this \emph{batch} is a stable assignment in $(G^{i+1}, <, \q)$. Furthermore, by iteratively applying the second part of (iii) and (iv), we know $D^{j_{i+1}}$ is the rotation digraph of $M_{j_{i+1}}$ in $(G^{i+1}, <, \q)$. By choice of $j_{i+1}$, $D^{j_{i+1}}$ contains no directed cycles, which means there are no exposed rotations at $M_{j_{i+1}}$ in $(G^{i+1}, <, \q)$. Thus, $M_{j_{i+1}}$ must be the student-optimal stable assignment in $(G^{i+1}, <, \q)$, which coincides with $M^{i+1}$. This concludes the proof of the induction ($\ddagger$).

    Finally, consider the last batch of Algorithm~\ref{alg:rrr-consent}. By Lemma~\ref{lem:sink-is-underdemanded}, sinks in $D^{j_k}$ are exactly underdemanded schools in $M^k$. Moreover, by the termination criterion of Algorithm~\ref{alg:sEADAM}, all schools are underdemanded in $M^k$. Thus, $D^{j_k}$ only has sinks and Algorithm~\ref{alg:rrr-consent} terminates with $M_{j_k}$, which is the same assignment as $M^k$.
\hfill{\Halmos}\endproof

\proof{Proof of Theorem~\ref{thm:rrr-eadam-equiv}.} 
    Immediately from Theorem~\ref{thm:eadam-seadam-equiv} and Theorem~\ref{thm:rrr-seadam-equiv}. 
\hfill{\Halmos}\endproof

%%%%%%%%%%%%%%%%%
\end{document}